\RequirePackage{lineno}
\documentclass[aps,prc,amsfonts,reprint,showpacs,showkeys,nofootinbib,twocolumn,preprintnumbers,superscriptaddress]{revtex4-2}

\usepackage{xfrac}
\usepackage[colorlinks=true, urlcolor=navyblue, linkcolor=navyblue, citecolor=navyblue]{hyperref}

\usepackage{graphicx}
\usepackage{dcolumn}
\usepackage{bm}

\usepackage{xcolor}
\usepackage{lipsum}
\usepackage{ulem}

\usepackage{graphicx,amsmath,amssymb}
\usepackage{placeins}
\usepackage{multirow}
\usepackage{epstopdf}
\usepackage{times}
\usepackage{color}

\newcommand\ua{\uparrow}
\newcommand\da{\downarrow}
\newcommand\Ua{\Uparrow}
\newcommand\Da{\Downarrow}

\newcommand\q{\mathrm}

\usepackage{longtable}
\definecolor{navyblue}{rgb}{0,0.08,0.45}
\usepackage{booktabs}
\usepackage{sidecap}
\begin{document}


\title{Measurement of the nucleon spin structure functions for $0.01<Q^2<1$~GeV$^2$ using CLAS}

\newcommand*{\ANL}{Argonne National Laboratory, Argonne, Illinois 60439, USA}
\newcommand*{\ANLindex}{1}
\affiliation{\ANL}
\newcommand*{\ASU}{Arizona State University, Tempe, Arizona 85287, USA}
\newcommand*{\ASUindex}{2}
\affiliation{\ASU}
\newcommand*{\CSUDH}{California State University, Dominguez Hills, Carson, California 90747, USA}
\newcommand*{\CSUDHindex}{3}
\affiliation{\CSUDH}
\newcommand*{\CANISIUS}{Canisius College, Buffalo, New York 14208, USA}
\newcommand*{\CANISIUSindex}{4}
\affiliation{\CANISIUS}

\newcommand*{\CMU}{Carnegie Mellon University, Pittsburgh, Pennsylvania 15213, USA}
\newcommand*{\CMUindex}{4}
\affiliation{\CMU}

\newcommand*{\CUA}{Catholic University of America, Washington, D.C. 20064, USA}
\newcommand*{\CUAindex}{5}
\affiliation{\CUA}
\newcommand*{\SACLAY}{CEA, IRFU ,Universit\'{e} Paris-Saclay, F-91191 Gif-sur-Yvette, France}
\newcommand*{\SACLAYindex}{6}
\affiliation{\SACLAY}
\newcommand*{\CNU}{Christopher Newport University, Newport News, Virginia 23606, USA}
\newcommand*{\CNUindex}{7}
\affiliation{\CNU}

\newcommand*{\WM}{College of William and Mary, Williamsburg, Virginia 23187, USA}
\newcommand*{\WMindex}{7}
\affiliation{\WM}

\newcommand*{\DUQUESNE}{Duquesne University, 600 Forbes Avenue, Pittsburgh, Pennsylvania 15282, USA}
\newcommand*{\DUQUESNEindex}{9}
\affiliation{\DUQUESNE}

\newcommand*{\FU}{Fairfield University, Fairfield, CT 06824, USA}
\newcommand*{\FUindex}{11}
\affiliation{\FU}

\newcommand*{\FIU}{Florida International University, Miami, Florida 33199, USA}
\newcommand*{\FIUindex}{11}
\affiliation{\FIU}
\newcommand*{\FSU}{Florida State University, Tallahassee, Florida 32306, USA}
\newcommand*{\FSUindex}{12}
\affiliation{\FSU}
\newcommand*{\GWUI}{The George Washington University, Washington, D.C. 20052, USA}
\newcommand*{\GWUIindex}{13}
\affiliation{\GWUI}
\newcommand*{\GEORGIA}{Georgia Institute of Technology, Atlanta, Georgia 30332, USA}
\newcommand*{\GEORGIAindex}{42}
\affiliation{\GEORGIA}
\newcommand*{\INFNFE}{INFN, Sezione di Ferrara, 44100 Ferrara, Italy}
\newcommand*{\INFNFEindex}{14}
\affiliation{\INFNFE}
\newcommand*{\INFNFR}{INFN, Laboratori Nazionali di Frascati, 00044 Frascati, Italy}
\newcommand*{\INFNFRindex}{15}
\affiliation{\INFNFR}
\newcommand*{\INFNGE}{INFN, Sezione di Genova, 16146 Genova, Italy}
\newcommand*{\INFNGEindex}{16}
\affiliation{\INFNGE}
\newcommand*{\INFNPAV}{INFN, Sezione di Pavia, 27100 Pavia, Italy}
\newcommand*{\INFNPAVindex}{19}
\affiliation{\INFNPAV}
\newcommand*{\INFNRO}{INFN, Sezione di Roma Tor Vergata, 00133 Rome, Italy}
\newcommand*{\INFNROindex}{17}
\affiliation{\INFNRO}
\newcommand*{\INFNTUR}{INFN, Sezione di Torino, 10125 Torino, Italy}
\newcommand*{\INFNTURindex}{18}
\affiliation{\INFNTUR}
\newcommand*{\KNU}{Kyungpook National University, Daegu 41566, Republic of Korea}
\newcommand*{\KNUindex}{21}
\affiliation{\KNU}
\newcommand*{\LAMAR}{Lamar University, 4400 MLK Blvd, PO Box 10046, Beaumont, Texas 77710, USA}
\newcommand*{\LAMARindex}{22}
\affiliation{\LAMAR}
\newcommand*{\MIT}{Massachusetts Institute of Technology, Cambridge, Massachusetts  02139, USA}
\newcommand*{\MITindex}{23}
\affiliation{\MIT}
\newcommand*{\MISS}{Mississippi State University, Mississippi State, Mississippi 39762, USA}
\newcommand*{\MISSindex}{24}
\affiliation{\MISS}
\newcommand*{\NMSU}{New Mexico State University, PO Box 30001, Las Cruces, New Mexico 88003, USA}
\newcommand*{\NMSUindex}{26}
\affiliation{\NMSU}
\newcommand*{\NSU}{Norfolk State University, Norfolk, Virginia 23504, USA}
\newcommand*{\NSUindex}{27}
\affiliation{\NSU}

\newcommand*{\OHIOU}{Ohio University, Athens, Ohio 45701, USA}
\newcommand*{\OHIOUindex}{28}
\affiliation{\OHIOU}

\newcommand*{\ODU}{Old Dominion University, Norfolk, Virginia 23529, USA}
\newcommand*{\ODUindex}{28}
\affiliation{\ODU}
\newcommand*{\JLUGiessen}{II Physikalisches Institut der Universit\~{a}t Giessen, 35392 Giessen, Germany}
\newcommand*{\JLUGiessenindex}{29}
\affiliation{\JLUGiessen}
\newcommand*{\SEOUL}{Seoul National University, Seoul 08826, Republic of Korea}
\newcommand*{\SEOULindex}{50}
\affiliation{\SEOUL}
\newcommand*{\MSU}{Skobeltsyn Institute of Nuclear Physics, Lomonosov Moscow State University, 119234 Moscow, Russia}
\newcommand*{\MSUindex}{31}
\affiliation{\MSU}
\newcommand*{\TEMPLE}{Temple University,  Philadelphia, Pennsylvania 19122, USA}
\newcommand*{\TEMPLEindex}{33}
\affiliation{\TEMPLE}
\newcommand*{\JLAB}{Thomas Jefferson National Accelerator Facility, Newport News, Virginia 23606, USA}
\newcommand*{\JLABindex}{34}
\affiliation{\JLAB}
\newcommand*{\BRESCIA}{Universit\`{a} di Brescia, 25123 Brescia, Italy}
\newcommand*{\BRESCIAindex}{36}
\affiliation{\BRESCIA}
\newcommand*{\FERRARAU}{Universit\`{a} di Ferrara, 44121 Ferrara, Italy}
\newcommand*{\FERRARAUindex}{10}
\affiliation{\FERRARAU}
\newcommand*{\ROMAII}{Universit\`{a} di Roma Tor Vergata, 00133 Rome Italy}
\newcommand*{\ROMAIIindex}{30}
\affiliation{\ROMAII}
\newcommand*{\UTFSM}{Universidad T\'{e}cnica Federico Santa Mar\'{i}a, Casilla 110-V Valpara\'{i}so, Chile}
\newcommand*{\UTFSMindex}{35}
\affiliation{\UTFSM}
\newcommand*{\ORSAY}{Universit\'{e} Paris-Saclay, CNRS/IN2P3, IJCLab, 91405 Orsay, France}
\newcommand*{\ORSAYindex}{20}
\affiliation{\ORSAY}
\newcommand*{\UCR}{University of California Riverside, 900 University Avenue, Riverside, California 92521, USA}
\newcommand*{\UCRindex}{37}
\affiliation{\UCR}
\newcommand*{\UCONN}{University of Connecticut, Storrs, Connecticut 06269, USA}
\newcommand*{\UCONNindex}{8}
\affiliation{\UCONN}
\newcommand*{\GLASGOW}{University of Glasgow, Glasgow G12 8QQ, United Kingdom}
\newcommand*{\GLASGOWindex}{38}
\affiliation{\GLASGOW}
\newcommand*{\ULI}{University of Ljubljana, 1000 Ljubljana, Slovenia}
\newcommand*{\ULIindex}{25}
\affiliation{\ULI}
\newcommand*{\UNH}{University of New Hampshire, Durham, New Hampshire 03824, USA}
\newcommand*{\UNHindex}{25}
\affiliation{\UNH}

\newcommand*{\URICH}{University of Richmond, Richmond, Virginia 23173, USA}
\newcommand*{\URICHindex}{32}
\affiliation{\URICH}

\newcommand*{\SCAROLINA}{University of South Carolina, Columbia, South Carolina 29208, USA}
\newcommand*{\SCAROLINAindex}{32}
\affiliation{\SCAROLINA}
\newcommand*{\VIRGINIA}{University of Virginia, Charlottesville, Virginia 22904, USA}
\newcommand*{\VIRGINIAindex}{40}
\affiliation{\VIRGINIA}
\newcommand*{\YORK}{University of York, York YO10 5DD, United Kingdom}
\newcommand*{\YORKindex}{39}
\affiliation{\YORK}
\newcommand*{\YEREVAN}{Yerevan Physics Institute, 375036 Yerevan, Armenia}
\newcommand*{\YEREVANindex}{41}
\affiliation{\YEREVAN}

\newcommand*{\NOWJLAB}{Thomas Jefferson National Accelerator Facility, Newport News, Virginia 23606, USA}
\newcommand*{\NOWISU}{Idaho State University, Pocatello, Idaho 83209, USA}
\newcommand*{\NOWWM}{College of William and Mary, Williamsburg, Virginia 23187, USA}
\newcommand*{\NOWOAK}{Oak Ridge National Laboratory, 1 Bethel Valley Road, Oak Ridge, Tennessee 37830, USA}
\newcommand*{\NOWDECD}{Deceased}

\author{A.~Deur}
\affiliation{\JLAB} 
\affiliation{\VIRGINIA} 
\affiliation{\ODU}
\author{S.E.~Kuhn} 
\affiliation{\ODU}
\author{M.~Ripani}
\affiliation{\INFNGE}
\author{X.~Zheng}
\email{Contact Author: xiaochao@jlab.org}
\affiliation{\VIRGINIA}
%

\author {A.G.~Acar} 
\affiliation{\YORK}
\author {P.~Achenbach} 
\affiliation{\JLAB}
\author{K.P.~Adhikari}
\affiliation{\ODU}
\affiliation{\JLAB}
\affiliation{\MISS}
\author {J. S.~Alvarado} 
\affiliation{\ORSAY}
\author {M.J.~Amaryan} 
\affiliation{\ODU}
\author {W. R.~Armstrong} 
\affiliation{\ANL}

% ...

\collaboration{The CLAS Collaboration}
\noaffiliation

 \date{\today}

 \begin{abstract}
  The spin structure functions of the proton and the deuteron were measured during the EG4 experiment at Jefferson Lab in 2006. Data were collected for longitudinally polarized electron scattering off longitudinally polarized NH$_3$ and ND$_3$ targets, for $Q^2$ values as small as 0.012 and 0.02~GeV$^2$, respectively, using the CEBAF Large Acceptance Spectrometer (CLAS). This is the archival paper of the EG4 experiment that summaries the previously reported results of the polarized structure functions $g_1$, $A_1F_1$, and their moments $\overline \Gamma_1$, $\overline \gamma_0$, and $\overline I_{TT}$, for both the proton and the deuteron. In addition, we report on new results on the neutron $g_1$ extracted by combining proton and deuteron data and correcting for Fermi smearing, and on the neutron moments $\overline \Gamma_1$, $\overline \gamma_0$, and $\overline I_{TT}$ formed directly from those of the proton and the deuteron. Our data are in good agreement with the Gerasimov-Drell-Hearn sum rule for the proton, deuteron, and neutron. 
  Furthermore, the isovector combination was formed 
  for $g_1$ and the Bjorken integral $\overline \Gamma_1^{p-n}$, and compared to available theoretical predictions. All of our results provide for the first time extensive tests of spin observable predictions from chiral effective field theory ($\chi$EFT) 
  in a $Q^2$ range commensurate with the pion mass. 
  They motivate further improvement in $\chi$EFT calculations from other approaches such as the 
  lattice gauge method.
   
\end{abstract}


\maketitle


\section{Introduction}\label{sec:intro}

The study of nucleon structure has been an active field of research ever since the discovery that nucleons were 
composite~\cite{Frisch:1933,Esterman:1933,Hofstadter:1953zz}. 
It is now well known that the nucleon, like all hadrons, is made of partons (quarks, antiquarks, and gluons) and that its structure is dominantly ruled by the strong nuclear interaction described by quantum chromodynamics (QCD)~\cite{book:QCDhadron,book:Roberts}. 
However, the behavior of the QCD coupling constant $\alpha_s$~\cite{Prosperi:2006hx, Deur:2016tte, Deur:2023dzc} determines that our understanding of nucleon structure at long distance is not as developed as that at short distance. 
At high energy-momentum (short distance) typically well above the~GeV scale, $\alpha_s \ll 1$, which allows us to treat QCD perturbatively. Presently, the approximations of observables 
 are typically known to order $\alpha_s^3$, and some to $\alpha_s^5$. 
 On the other hand, at low energy-momentum, $\alpha_s \gtrsim 1$, and a perturbative expansion in $\alpha_s$ becomes inapplicable. The problem is further complicated by the rise of complex phenomena, {\it e.g.} the confinements of partons, the emergence of hadronic degrees of freedom or the spontaneous breaking of QCD's near $SU(3)_L\times SU(3)_R$ chiral symmetry.
In this domain, due to the lack of a mature 
analytic technique based on the QCD Lagrangian, effective theories or models are often used. 
In particular, chiral effective field theory ($\chi$EFT)~\cite{Bernard:1995dp}, based on the observed spontaneous breaking of 
the  chiral symmetry in hadronic states, has successfully predicted
hadronic observables at low energies. 

The principle of $\chi$EFT is that it describes the nucleon in terms of effective hadronic degrees of freedom instead of the fundamental quark and
gluon fields. It is completely non-perturbative in terms of $\alpha_s$ because it employs series whose expansion parameters are based {\it e.g.} on the pion mass, the nucleon-$\Delta_{1232}$ mass gap or the characteristic energy scale of the chiral symmetry breaking. 
However, while leading-order (LO) calculations of $\chi$EFT are relatively
tractable, next-to-leading-order (NLO) and higher order calculations are much more involved.
An important experimental task is therefore to provide
precise hadron structure data that can test, and potentially guide, 
$\chi$EFT calculations and other approaches to
non-perturbative QCD. Such data must cover the domain of low momentum transfer squared ($Q^2$) where $\chi$EFT
applies, typically below a few tenths of GeV$^2$.
Especially important in this context are low $Q^2$ nucleon spin structure data~\cite{Anselmino:1994gn, Bass:2004xa, Bass:2004xa, Chen:2005tda, Burkardt:2008jw, Kuhn:2008sy, Chen:2010qc, Aidala:2012mv, Deur:2018roz, Ji:2020ena}  
which are challenging to obtain experimentally but are crucial for the development of $\chi$EFT~\cite{JeffersonLabE94010:2004ekh}. 
One such experimental endeavor is that of the Jefferson Lab (JLab) Experimental Group EG4, constituted of experiments E03-006 and E05-111 that used the CEBAF Large Acceptance Spectrometer (CLAS) in Hall B~\cite{CLAS:2003umf}. EG4 was designed to
test $\chi$EFT predictions 
by precisely measuring the longitudinal spin structure function $g_1$ and its moments for the
proton and the deuteuron 
down to $Q^2 \simeq 0.01$~GeV$^2$. 
These low $Q^2$ data, reported in two
letters~\cite{CLAS:2017ozc,CLAS:2021apd}, provided benchmarks for $\chi$EFT calculations. This document is the archival article that presents in detail the EG4 experiments, data analysis procedure for the inclusive scattering channel, results on the proton and the deuteron, and followup analysis and results on the neutron spin structure functions. Additional results on the target- and double-spin asymmetries of pion electroproduction are available~\cite{CLAS:2016qxj}, but these were focused on the semi-inclusive channel and are not included here. 

The article is arranged as follows: In Section~\ref{sec:formalism} we present the formalism
of unpolarized and polarized electron scattering. We
introduce the nucleon structure functions, their moments,
and polarized sum rules, which are relevant not only for testing $\chi$EFT but also to form observables 
that have no known method of direct measurement. In Section~\ref{sec:exp} we describe EG4
 and the special features that distinguish it from other CLAS experiments. In Section~\ref{sec:analysis} we describe the analysis procedure
of extracting the polarized yield differences from the data. In Section~\ref{sec:sim}
we describe the Monte-Carlo simulation, the normalization of the data using the elastic
scattering yield, and the extraction of the structure function $g_1$.  In Section~\ref{sec:results_pd} we present results on $g_1$, $A_1F_1$ of the proton and the deuteron, followed by the procedure to extract the neutron $g_1$ results in Section~\ref{sec:results_g1n}. We present results for and discussions of all the moments in Section-\ref{sec:results_moments}.
We then summarize, conclude, and provide perspectives in Section~\ref{sec:conclusion}.

\section{Formalism}\label{sec:formalism}
All formalism presented in this section is based on Ref.~\cite{Anselmino:1994gn}
or~\cite{Kuhn:2008sy}, and the algebraic manipulations therein.

\subsection{Inclusive Electron Scattering}\label{sec:intro_escatt}
Within the approximation of one-photon exchange between the lepton and the target 
(Fig.~\ref{fig:eNscat}), the scattering process is described by two kinematic variables:
\begin{figure}[!h]
\includegraphics[width=0.45\textwidth]{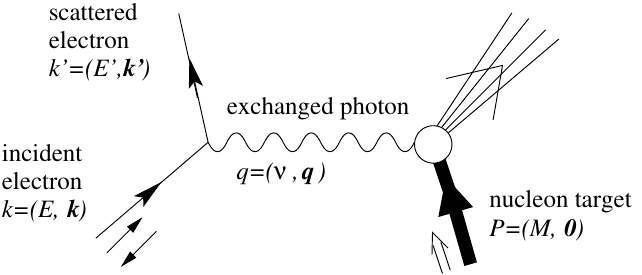}
\caption{The one-photon exchange process of polarized electron scattering off a polarized nucleon. The 4-momenta of the
  incident and the scattered electrons are $k=(E,\bm k)$ and $k'=(E',\bm {k'})$, respectively, and the helicity
  of the incident electron is indicated by the thin arrows.
 The nucleon target, if at rest, has $P=(M,\bm 0)$ and its spin is indicated by the outlined arrow. }\label{fig:eNscat}
\end{figure}
the virtuality of the exchanged photon $Q^2\equiv -q^2$
and the invariant mass of the photon-target system $W \equiv \sqrt{(q+P)^2}$.  The virtual photon
can be viewed as a probe of the substructure of the target nucleus (or nucleon) and 
$Q^2$ describes the (inverse) space-time resolution of the probe. For the fixed-target case, one has
\begin{eqnarray}
   Q^2 &=& 2E E'(1-\cos\theta), \label{eq:qsq}
\end{eqnarray}
where $\theta$ is the scattering angle of the electron, {\it i.e.},  the angle formed by $\bm k$
and $\bm {k'}$, and
  \begin{eqnarray}
    W_\mathrm{nucl} &=& \sqrt{{M_T}^2+2{M_T}\nu-Q^2}. \label{eq:Wdef_fixedtarg1}
  \end{eqnarray}
Here $M_T$ is usually the mass of the nuclear
  target in the case of nuclear scattering. However, when we study inelastic scattering from the nucleons, the nucleon
  mass $M$ is used instead:
  \begin{eqnarray}
    W &=& \sqrt{{M}^2+2{M}\nu-Q^2}~. \label{eq:Wdef_fixedtarg}
  \end{eqnarray}
  Unless indicated otherwise, Eq.~(\ref{eq:Wdef_fixedtarg}) will be used hereafter for the definition of the invariant mass.
Alternatively to $W$, one may characterize the scattering process using
the Bjorken scaling variable~\cite{Bjorken:1968dy}  
$x \equiv -q^2/(2P\cdot q) = Q^2/(W^2-M^2+Q^2)$, or for fixed targets: 
\begin{eqnarray}
  x & = & \frac{Q^2}{2M\nu}~.~\label{eq:xbj}
\end{eqnarray}

\subsection{Cross Sections and Structure Functions}\label{sec:intro_sf}
For inclusive unpolarized inelastic scattering off a spin-1/2 target, the differential cross section for detecting
the final-state lepton in the solid angle $\q{d}\Omega$ and in the
final energy range ($E^\prime$, $E^\prime+\q{d}E^\prime$) in the laboratory
frame can be written as: %

\begin{eqnarray} \label{eq:dis_xsec_f1f2}
 \frac{\mathrm{d}^2\sigma_u}{\mathrm{d}\Omega\mathrm{d}E^\prime} &=&
        \Big(\frac{\mathrm{d}^2\sigma}{\mathrm{d}\Omega}\Big)_{Mott} \times\nonumber \\
       && \Big[\frac{1}{\nu}F_2(x,Q^2)+\frac{2}{M}F_1(x,Q^2)\tan^2\frac{\theta}{2}\Big]~, \label{eq:unpolxs3}
\end{eqnarray}
where $F_1(x,Q^2)$ and $F_2(x,Q^2)$ 
are the two unpolarized structure functions characterizing the 
target structure for unpolarized inclusive lepton scattering. The Mott cross section (representing scattering off a point-like, spinless, infinitely-heavy target) 
is \begin{eqnarray}
{\Big(\frac{\mathrm{d}^2\sigma}{\mathrm{d}\Omega}\Big)}_{Mott}
 &=&\frac{\alpha^2\cos^2{\frac{\theta}{2}}}{4E^2\sin^4{\frac{\theta}{2}}}~,\label{eq:Mottxsec}
\end{eqnarray}
with $\alpha$ the quantum electrodynamics (QED) coupling.


In the polarized case where the electrons and the spin-1/2 target 
are polarized along the beam direction, the helicity-dependent cross section difference is:
\begin{eqnarray}
   &&\frac{\q{d}^2\sigma_{\ua\Ua}}{\q{d}\Omega
     \q{d}E^\prime}-\frac{\q{d}^2\sigma_{\ua\Da}}{\q{d}\Omega \q{d}E^\prime} \label{eq:polxs_long}\\
   =&& -\frac{4\alpha^2 E^\prime}{M\nu
     EQ^2}\Big [{(E+E^\prime\cos{\theta}){{g_1(x,Q^2)}} -2Mx{{g_2(x,Q^2)}}}\Big ] \nonumber\\
       = && -\frac{4\alpha^2 E^\prime}{M\nu
     EQ^2}\Big [\left(E+E^\prime\cos{\theta}+\frac{Q^2}{\nu}\right)g_1 -\frac{Q^2}{\nu} F_1A_2)\Big ]~, \nonumber
\end{eqnarray}
where the $\ua(\da)$ and $\Ua(\Da)$ represent the electron and the target spin 
directions being parallel (anti-parallel) to the beam direction, respectively. The spin-dependent properties of the target are characterized for inclusive lepton scattering 
by $g_1(x,Q^2)$ and $g_2(x,Q^2)$, the longitudinal and transverse polarized structure functions, respectively. 
The second line on the right-hand side (RHS) of Eq.~(\ref{eq:polxs_long}) indicates that
one can extract $g_1(x,Q^2)$ from this
cross section difference assuming a model
for the product $F_1(x,Q^2) A_2(x,Q^2)$,
where $A_2$ is the virtual photon asymmetry to be 
defined in the next section.

\subsection{Virtual Photon Cross Sections}\label{sec:intro_photonxsec}
The formalism provided in the previous two sections focused on the interaction
cross section between the incident electron and the target. 
It is often useful to also study cross sections
for the virtual photon being absorbed by the target, and their dependence on the photon polarization.
We follow the formalism of Ref.~\cite{book:Roberts} and Refs.~\cite{Drechsel:2000ct, Gorchtein:2004jd}. 

The photon polarization 4-vector, $e^\mu$,
includes two transverse polarization modes that satisfy
$e_T^2=-1$ and a longitudinal polarization mode that satisfies 
$e_L^2=1$. The total cross section of the photon absorption 
by the spin-1/2 target can be separated into two terms: the longitudinal and
transverse virtual
photon cross sections 
\begin{eqnarray}
  \sigma_L &=&  \frac{4\pi^2\alpha}{K}\left[\left(1+\frac{\nu^2}{Q^2}\right)\frac{F_2}{\nu}-\frac{F_1}{M}\right]~,\label{eq:sigL}\\
    \sigma_T &=&  \frac{4\pi^2\alpha}{K}\left[\frac{F_1}{M}\right]~, \label{eq:sigT}
\end{eqnarray}
that are associated with the longitudinally and 
transversely polarized virtual photons, respectively. 
The virtual photon equivalent energy $K$ (sometimes called virtual photon flux)
in Eqs.~(\ref{eq:sigL}-\ref{eq:sigT}) is not a direct observable and different definitions have been proposed:
the Hand convention $K_H\equiv \nu(1-x)$,
the Gilman convention $K_G\equiv   \sqrt{\nu^2+Q^2}$,
or using the photon energy $K_A \equiv  \nu$.
For now, we will continue to use $K$ instead of choosing
a convention. 
The ratio of the longitudinal to transverse cross sections 
$R$ is:
\begin{eqnarray}
  R &\equiv&\frac{\sigma_L}{\sigma_T} = \frac{F_2}{2xF_1}(1+\gamma^2)-1~,\label{eq:Rdef}
\end{eqnarray}
where
\begin{eqnarray}
 \gamma^2&=&\frac{Q^2}{\nu^2}=\frac{(2Mx)^2}{Q^2}~. \label{eq:gamma}
\end{eqnarray}

Using the ``relative longitudinal
polarization''~\footnote{It is
  called ``photon transverse polarization'' in Ref.~\cite{Drechsel:2000ct}. However
  since this is the weighting factor of $\sigma_L$, not $\sigma_T$, we prefer ``relative longitudinal polarization''.} 
of the virtual photon $\epsilon$, 
the total photon absorption cross section can be written as 
\begin{eqnarray}
  \frac{\q{d}^2\sigma_u}{\q{d}\Omega\q{d}E^\prime} &=& \Gamma (\sigma_T+\epsilon\sigma_L),
\end{eqnarray}
with 
\begin{eqnarray}
 \epsilon &=& \frac{1}{\left[1+2(1+1/\gamma^2)\tan^2(\theta/2) \right]} ~,\label{eq:epsilon1}
\end{eqnarray}
and
the photon flux factor $\Gamma$ 
\begin{eqnarray}
  \Gamma &=& \frac{\alpha}{2\pi^2}\frac{E'}{E}\frac{K}{Q^2}\frac{1}{1-\epsilon}~.\label{eq:Gamma}
\end{eqnarray}

For the polarized case, one can similarly divide the virtual photon absorption cross section
by a spin-1/2 target into several components based on the polarization of the photon, and the total spin of the photon and the target projected
along the direction of the photon momentum $\bm q$ ($1/2$ and $3/2$ for the photon spin antiparallel 
and parallel to the target spin, respectively),
and a component $\sigma_{I}$ resulting from the interference between transverse and
longitudinal virtual photon-nucleon amplitudes~\cite{book:Roberts}. 
From these components one further defines
\begin{eqnarray}
  \sigma_{T}&=&(\sigma_{1/2}+\sigma_{3/2})/2~,\label{eq:sigT2}\\
  \sigma_{TT}&=&(\sigma_{1/2}-\sigma_{3/2})/2~,\label{eq:sigTT}\\
  \sigma_{LT} &=& \sigma_I~.\label{eq:sigTL}
\end{eqnarray}
These spin-dependent virtual photon cross sections relate to the nucleon
polarized structure functions as:
\begin{eqnarray}
  \sigma_{LT} &=& \frac{4\pi^2\alpha}{M K}\gamma (g_1+g_2),\label{eq:sigLT2}\\
  \sigma_{TT} &=& \frac{4\pi^2\alpha}{M K} (g_1-\gamma^2 g_2)~.\label{eq:sigTT2}
\end{eqnarray}

Finally, two virtual photon asymmetries are also commonly employed.
The first is the longitudinal asymmetry $A_1$: 
\begin{eqnarray}
 A_1 &\equiv& \frac{\sigma_{TT}}{\sigma_T}
 = \frac{g_1-\gamma^2 g_2}{F_1}~, \label{eq:A1}
\end{eqnarray}
from which we obtain:
\begin{eqnarray}
   \sigma_{TT}&=&\frac{4\pi^2\alpha}{M K }A_1F_1~.\label{eq:sigTT_A1F1}
\end{eqnarray}

The second is the longitudinal-transverse interference asymmetry:
\begin{eqnarray} A_2 &\equiv&  \frac{\sigma_{LT}}{\sigma_T}
  =\frac{\gamma(g_1+g_2)}{F_1} ~.\label{eq:A2}
\end{eqnarray}
The asymmetries $A_1$, $A_2$, and $g_1/F_1$ are related as
\begin{eqnarray} 
 A_1 + \gamma A_2 &=& (1+\gamma^2)\frac{g_1}{F_1} ~.\label{eq:a1a2=g1f1}
\end{eqnarray} 
Note that one can use either $g_{1,2}$ or $A_{1,2}F_1$ to
describe the polarized cross section, Eq.~(\ref{eq:polxs_long}).

\subsection{Sum Rules and Moments of Structure Functions}\label{sec:intro_moments}

In this section we present the Bjorken, GDH, generalized GDH, and generalized polarizability sum rules, which
all involve moments of polarized structure functions or, equivalently, 
$\sigma_{TT}$ and $\sigma_{LT}$. 
For reviews of nucleon spin structure and additional sum rules, see Refs.~\cite{Chen:2005tda, Kuhn:2008sy, Chen:2010qc, Deur:2018roz}.

\subsubsection{Bjorken Sum Rule \label{bjorken SR}}
The Bjorken sum rule, derived for $Q^2 \to \infty$ using current algebra and isospin symmetry~\cite{Bjorken:1966jh,Bjorken:1969mm}, predates QCD and is not a QCD prediction.
However, its generalization to include the $Q^2$-dependence appearing at finite $Q^2$~\cite{Kataev:1994gd,Kataev:2005hv,Baikov:2010je} stems from pQCD. Studying the sum rule in the pQCD domain ($Q^2 \gtrsim 1$~GeV$^2$) therefore tests whether QCD correctly describes the strong force when spin degrees-of-freedom are explicit. The Bjorken sum has been actively measured at CERN~\cite{SpinMuon:1993gcv, SpinMuonSMC:1997voo,
SpinMuonSMC:1994met,SpinMuon:1995svc,SpinMuonSMC:1997ixm,SpinMuonSMC:1997mkb, 
COMPASS:2010wkz, COMPASS:2015mhb, COMPASS:2016jwv}, 
DESY~\cite{
HERMES:2002gmr}, JLab~\cite{Deur:2004ti, Deur:2008ej, Deur:2014vea}, and SLAC~\cite{
E143:1995rkd,E143:1995clm,E154:1997xfa,E143:1998hbs,E154:1997ysl,E155:2000qdr},
with measurement agreeing to better than 10\% with the sum rule expectation~\cite{Deur:2018roz}.
Measurements performed with the higher-energy accelerators, CERN and SLAC, provide data at larger $Q^2$ values, offering extensive coverage of the low-$x$ part of the Bjorken sum. Measurements at the lower-energy facilities, DESY and JLab, covered the smaller $Q^2$ range while overlapping with the SLAC data. Due to their lower energy, DESY and JLab have a limited low-$x$ reach, which is supplemented by extrapolating the low-$x$ data obtained from CERN and SLAC.

The original Bjorken sum rule ($Q^2 \to \infty$) reads:
\begin{eqnarray}
  \Gamma_1^{p-n}(Q^2)\vert_{Q^2\to\infty} = \int_0^1 \left[g_1^p(x,Q^2)-g_1^n(x,Q^2)\right]dx = \frac{g_a}{6}, \nonumber\\
  \label{eq:bjorken_sum}
\end{eqnarray}
where $g_a$ is the axial coupling constant measured in neutron beta decay.
Equation~(\ref{eq:bjorken_sum}) can be generalized 
 for finite $Q^2$ by accounting for gluon radiation  and higher-twist (HT) effects:
\begin{eqnarray}
  \Gamma_1^{p-n}(Q^2) &=& \frac{g_a}{6}f(Q^2)+\q{HT},
  \label{eq:generalized Bjorken SR}
\end{eqnarray}
where~\cite{Kataev:1994gd}
\begin{eqnarray}
  f(Q^2) &=& 1-\frac{\alpha_s(Q^2)}{\pi}
  -3.58\left(\frac{\alpha_s(Q^2)}{\pi}\right)^2\nonumber\\
 && -20.2\left(\frac{\alpha_s(Q^2)}{\pi}\right)^3 
 -175.7\left(\frac{\alpha_s(Q^2)}{\pi}\right)^{4}\nonumber\\
 && +\mathcal O\left(\alpha_s^5\right)
\label{eq:DGLAP Bjorken SR}
\end{eqnarray}
and HT contains higher-twist corrections proportional to powers of $1/Q^2$.

Beside the pQCD domain, the Bjorken sum rule is also important
at low $Q^2$ due to its close connection to the GDH sum rule (see next section). The behavior of the Bjorken sum 
at $Q^2<1$~GeV$^2$ was precisely mapped by several experiments
at JLab~\cite{Deur:2004ti,Deur:2008ej,ResonanceSpinStructure:2008ceg,Deur:2021klh}. Note that 
Eqs.~(\ref{eq:bjorken_sum}-\ref{eq:DGLAP Bjorken SR}) are not valid in the low $Q^2$ domain and Bjorken sum rule predictions are provided 
by either models using non-perturbative approaches~\cite{Burkert:1992tg,Burkert:1993ya,Soffer:2004ip,Pasechnik:2010fg,Pasechnik:2009yc,Brodsky:2010ur} 
or $\chi$EFT calculations~\cite{Bernard:2012hb,  Alarcon:2020icz}.

\subsubsection{Gerasimov-Drell-Hearn (GDH) Sum Rule}\label{sec:GDH_real}
The GDH sum rule~\cite{Gerasimov:1965et,Drell:1966jv, Helbing:2006zp} is based on dispersion relations, unitarity, relativity, and gauge invariance 
and is derived for real photoproduction ($Q^2=0$). 
Like the Bjorken sum rule, the GDH sum rule is derived in a more general context than that of QCD and also predates it. 
In particular, the GDH sum rule applies to any type of target. For
targets whose internal structure is governed by the strong interaction, such as nucleons and nuclei, the sum rule provides a path to study 
QCD~\cite{Deur:2018ntc}.
For real photons, the GDH sum rule gives
\begin{eqnarray}
  \int_{\nu_{thr}}^\infty \frac{\sigma_{A}-\sigma_{P}}{\nu}d\nu
   = -4\pi^2\alpha\frac{\kappa^2}{M_T^2}S~,~\label{eq:GDH_real}
\end{eqnarray}
where $\sigma_{A}$ and $\sigma_{P}$ are the photo-absorption
cross sections for target and photon spins anti-parallel (A) and
parallel (P). (For spin-1/2 targets, these are the
same as $\sigma_{1/2}$ and $\sigma_{3/2}$
 in Eqs.~(\ref{eq:sigT2}-\ref{eq:sigTT}).)
Here, $M_T$ and $S$ are the mass and the spin of the target, and $\nu_{thr}$ is the minimal photon energy for inelastic excitation
of the target.
In the context of (virtual) photon absorption
on a nucleon, this threshold energy corresponds to the pion production threshold:
\begin{eqnarray}
 \nu_{thr}&=&m_\pi+\frac{m_\pi^2+Q^2}{2M},~\label{eq:nu_thr} 
\end{eqnarray}
with $m_\pi$ the pion mass. 
The anomalous magnetic moment of the target $\kappa$ is 
defined by the total magnetic moment of the particle $\bm\mu=\frac{e(Q/e+\kappa)}{M_T}\bm S$ 
with $Q/e$ the charge of the target in units of the elementary charge,
and can be related to the gyromagnetic ratio $g$ as $\kappa=g/2-Q/e$. 
The RHS of Eq.~(\ref{eq:GDH_real}) is
$-204~\mu$b for the proton ($S=\frac{1}{2}$, $\kappa=1.793$ and, $g/2=2.793$), 
$-234~\mu$b for the neutron ($S=\frac{1}{2}$, $\kappa=-1.913$, and $g/2=-1.913$),
and $0.65~\mu$b for the deuteron ($S=1$, $\kappa=-0.143$, and $g/2=0.857$). 
Whereas the overall consensus~\cite{Helbing:2006zp} is 
that the GDH sum rule is theoretically very solid, the question of its validity has been debated, mainly regarding whether $\sigma_{TT}/\nu$ decreases fast enough with $\nu$ for the integral to converge. Concurrently, its experimental verification has been the focus of several 
experiments at MAMI and ELSA~\cite{GDH:2001zzk,Ahrens:2006yx,GDH:2003xhc,GDH:2005noz, Dutz:2004zz}, BNL~\cite{LSC:2008wiu}, JLab~\cite{CLAS:2017ozc,CLAS:2017ozc,JeffersonLabE97-110:2019fsc}, GRAAL (ESRF)~\cite{Renard:2000jc}, and  HIGS (TUNL)~\cite{Weller:2002aa} providing results 
that show, given reasonable assumptions for the large $\nu$ behavior of its integrand, the GDH sum rule is valid for the proton to within the 10\% precision of the world data~\cite{Helbing:2006zp, CLAS:2017ozc, CLAS:2021apd,  Strakovsky:2022tvu}.

\label{sec:GDH_gen}
The GDH integral for the nucleon, when expressed with spin structure functions using
Eq.~(\ref{eq:sigTT2}), is an integral over $g_1$~\cite{Anselmino:1988hn}. Like the Bjorken sum, the small correction from $g_2$ vanishes when $Q^2 \to 0$, see Eq.~(\ref{eq:gamma}). 
Therefore, the formal expression of the (isovector) GDH and Bjorken integrals are proportional, whereas the domains of applicability of the sum rules are disjoint at high $Q^2$ and $Q^2=0$, respectively.
Several methods have been proposed to bridge the two sum rules. Generalization to finite $Q^2$ values for
the left-hand side (LHS) of Eq.~(\ref{eq:GDH_real}) is straightforward since $\sigma_{TT}$ or equivalently $g_1$ and $g_2$ exist also at non-zero $Q^2$. 
A commonly used definition is~\cite{Gorchtein:2004jd, Anselmino:1994gn}:
\begin{eqnarray}
\overline  I_{TT}(Q^2) \equiv  \frac{M^2}{8\pi^2\alpha} \int_{\nu_{thr}}^\infty \frac{K}{\nu}\frac{\sigma_{TT}}{\nu}d\nu~.\label{eq:ITT}
\end{eqnarray}
Note that hereafter in this article, a line above a symbol signifies that it does not include the contribution from elastic scattering. 
At finite $Q^2$, Eq.~(\ref{eq:ITT}) can be expanded using Eq.~(\ref{eq:sigTT_A1F1}):
\begin{eqnarray}
 \overline I_{TT}(Q^2) &=& \frac{M^2}{8\pi^2\alpha} \int_{\nu_{thr}}^\infty \frac{K}{\nu}
  2 \frac{\frac{4\pi^2\alpha}{M K }A_1F_1}{\nu}d\nu\nonumber\\
   &=& \frac{2M^2}{Q^2}\int_0^{x_{th}} A_1F_1 dx~, 
   \label{eq:I_TT_2}
\end{eqnarray}
where $dx=(\sfrac{Q^2}{2M \nu^2})d\nu$ was used in the last step. 
The integral $\overline I_{TT}$ can thus be treated as a generalized form of the GDH integral and
can be determined from the measured values of $A_1F_1$. 

The $Q^2\to 0$ limit for the integral of Eq.~(\ref{eq:ITT}) can be obtained by multiplying the RHS of Eq.~(\ref{eq:GDH_real}) by the extra factor $(\frac{1}{2})M^2/(8\pi^2\alpha)$. 
Since at $Q^2=0$, $\frac{K}{\nu}=1$, it reads: 
\begin{eqnarray}
 \overline I_{TT}(Q^2=0) &=& -\frac{\kappa^2}{4}~.
\end{eqnarray}
Therefore, measuring $\overline I_{TT}$ at very low $Q^2$ also allows one to test the GDH sum rule. 

Using $\overline I_{TT}$, the Bjorken sum can be extrapolated to the real photon point as: 
\begin{equation}
\overline\Gamma_1^{p-n}(Q^2)|_{_{Q^2  \to 0}}  = 
\frac{Q^2}{2M^2}\overline I_{TT}^{p-n} =
\frac{Q^2}{8}\bigg(\frac{\kappa_n^2}{M_n^2}-\frac{\kappa_p^2}{M_p^2}\bigg)~, 
\label{eq:bj-GDH}
\end{equation}
which should have a positive slope given $\kappa_n^2/M_n^2 > \kappa_p^2/M_p^2$.

The above procedure generalizes only the GDH integral but not the GDH sum rule, {\it i.e.}, the RHS of Eq.~(\ref{eq:GDH_real}). 
Other possible generalizations of the GDH integral as reviewed in Ref.~\cite{Drechsel:2000ct} share the same caveat.
To generalize the sum rule itself, one needs to note that the integral of $g_1$ can be related to the first forward virtual Compton scattering amplitude $S_1(\nu,Q^2)$~\cite{Ji:1999mr, Kuhn:2008sy, Deur:2018roz} using the low energy theorem~\cite{Low:1954kd}.
\label{sec:GDH_gen_sr}
Accordingly, a full generalization of the GDH sum rule is~\cite{Ji:1999mr}:
\begin{eqnarray}
 \overline \Gamma_1(Q^2) &=& \int_0^{1^-} g_1(x,Q^2)dx = \frac{Q^2}{8}\bar S_1(0,Q^2),
 \label{eq:GDH_gen_ji}
\end{eqnarray}
where the bars and $1^-$ indicate that the $x=1$ elastic contribution (quasi-elastic for nuclear targets) is excluded. 
In the quark-parton model, this integral gives the quark polarization within the nucleon 
weighted by their electric charge squared, 
and provides one of the ingredients for Ellis-Jaffe
sum rule~\cite{Ellis:1973kp} and Bjorken sum rule.  For the generalized form of the GDH sum rule, Eq.~(\ref{eq:GDH_gen_ji}), the amplitude $\bar S_1(0,Q^2)$ can be calculated in $\chi$EFT 
if $Q^2$ is sufficiently low~\cite{Ji:1999pd,Ji:1999sv, Bernard:2002bs,Bernard:2002pw, Kao:2002cp, Bernard:2012hb, Lensky:2014dda,Lensky:2016nui,Alarcon:2020icz}. 
The generalized GDH integrals have been measured over an extensive $Q^2$ range and on proton, neutron, deuteron and $^3$He by 
CERN~\cite{EuropeanMuon:1987isl, SpinMuon:1993gcv, SpinMuonSMC:1997voo,
SpinMuonSMC:1994met,SpinMuon:1995svc,SpinMuonSMC:1997ixm,SpinMuonSMC:1997mkb, COMPASS:2005xxc, COMPASS:2006mhr, COMPASS:2010wkz, COMPASS:2015mhb, COMPASS:2016jwv, COMPASS:2017hef}, 
DESY~\cite{HERMES:2000apm, HERMES:1998pau, 
HERMES:2002gmr}, JLab~\cite{Amarian:2002ar,JeffersonLabE94-010:2003dvv,CLAS:2002knl,CLAS:2003rjt, RSS:2006tbm,ResonanceSpinStructure:2008ceg, CLAS:2006ozz,CLAS:2008xos,JeffersonLabE01-012:2008kgk,E01-012:2013fky,CLAS:2014qtg,CLAS:2017qga,CLAS:2015otq,CLAS:2017ozc,CLAS:2021apd,JeffersonLabE97-110:2019fsc,E97-110:2021mxm}, and SLAC~\cite{E142:1996thl,E143:1994vcg,E143:1996vck,E143:1995rkd,E143:1995clm,E154:1997xfa,E155:1999pwm,E143:1998hbs,E154:1997ysl,E155:2000qdr}. 
CERN and SLAC provide the higher $Q^2$ coverage, with extensive contributions to the moments from the low-$x$ region. JLab data cover the lowest $Q^2$ domain. All accelerators provide data points in the intermediate $Q^2$ region of a few GeV or less, including the COMPASS experiment~\cite{COMPASS:2005xxc, COMPASS:2006mhr, COMPASS:2010wkz, COMPASS:2015mhb, COMPASS:2016jwv, COMPASS:2017hef} for CERN, E143~\cite{E142:1996thl,E143:1994vcg,E143:1996vck,E143:1995rkd,E143:1995clm} for SLAC and HERMES for DESY~\cite{HERMES:2000apm, HERMES:1998pau, 
HERMES:2002gmr}.

\subsubsection{Forward Spin Polarizability $\overline \gamma_0$}\label{sec:gamma0}
The interaction of a particle with an electromagnetic field, in our case the reaction of the target particle to the photons exchanged with the electron beam, is dictated at first order in the photon energy by the particle electric charge. If the beam and target are polarized, the target particle's anomalous magnetic moment 
also comes into play. 
Both the electric charge and the magnetic moment reflect the pure elastic reaction expected from a pointlike or perfectly rigid object. 
At second order, the particle compositeness and deformability must be considered. The deformation of the particle, {\it viz} its internal rearrangement under the electromagnetic field, are described by its electromagnetic polarizabilities~\cite{Hagelstein:2015egb}. 
The forward spin polarizability $\overline\gamma_0$ arises for the case when the beam and target are polarized longitudinally to the beam direction. Polarizabilities were initially defined for 
real photons but, just like the GDH sum, they can be generalized to finite $Q^2$, becoming $Q^2$-dependent {\it generalized polarizabilities}.

There is no practical method known that allows us to directly measure generalized polarizabilities. Instead, sum rules are used to access them.
For $\overline \gamma_0(Q^2)$, the relevant sum rule involves a higher moment of $g_1$ and $g_2$~\cite{Gell-Mann:1954ttj,Guichon:1995pu, Drechsel:2002ar}:
\begin{eqnarray}
\overline  \gamma_0(Q^2) &=& \frac{16\alpha M^2}{Q^6}\int_0^{1^-} x^2\left[g_1-\gamma^2 g_2\right]dx~,
\end{eqnarray}
which can also be calculated in $\chi$EFT~\cite{Ji:1999pd,Ji:1999sv, Bernard:2002bs,Bernard:2002pw, Kao:2002cp, Bernard:2012hb, Lensky:2014dda,Lensky:2016nui, Alarcon:2020icz}.
Equivalently, we can write this integral as:
\begin{eqnarray}
 \overline \gamma_0(Q^2) &=& \frac{16\alpha M^2}{Q^6}\int_0^{1^-} x^2A_1F_1dx~.
  \label{eq:gamma0}
\end{eqnarray}
Measurement of $\overline \gamma_0$ has been carried out for the proton at the real photon point at MAMI~\cite{Dutz:2004zz}. Measurements at small and intermediate $Q^2$ have been carried out at JLab on the proton~\cite{CLAS:2008xos, CLAS:2017qga, CLAS:2021apd} and neutron~\cite{JeffersonLabE94010:2004ekh, CLAS:2008xos, CLAS:2015otq, E97-110:2021mxm}. A short review of the experimental data and predictions from $\chi$EFT,  MAID~\cite{Drechsel:1998hk}, and SAID~\cite{Workman:2012jf} models can be found in Ref.~\cite{Strakovsky:2022tvu}.

\subsection{Deuteron Nuclear Model}\label{sec:intro_deuteron}

As discussed above, the GDH sum rule
holds for any target with arbitrary spin, once we
replace the expressions $\sigma_{3/2}$ and 
$\sigma_{1/2}$ with the cross sections for target
and photon spins aligned vs. anti-aligned. 

At the same time, experimental information on free neutrons is not
realistically accessible, necessitating measurements on 
(neutrons bound in) nuclei to access the various structure
functions and integrals defined above. For the EG4 experiment, we used the
deuteron as a source of information on (bound) neutrons. In the
most na\"ive model, the rather loosely bound deuteron can be 
thought of as a combination of one neutron and one proton, 
allowing us in principle to extract neutron structure functions
after subtracting the contribution from the proton. In reality, 
this interpretation is complicated by various nuclear binding
effects, including the Fermi motion of both bound nucleons, the
depolarization of both nucleons through the D-state admixture in
the deuteron ground state wave function, and potential off-shell
corrections to the (virtual-) photon-nucleon interaction. In addition,
the measured cross section (differences) also contain contributions
not present for single free nucleons, including two-body currents, final state interactions 
and coherent production as well as the two-body breakup of the deuteron, D$(e,e^\prime p n)$. 

Ideally, a rigorous test of theoretical models like
$\chi$EFT  would compare observables measured on the deuteron
with calculations taking into account all of the above effects. In
practice, such calculations do not (yet) exist, and experimentally,
covering the entire kinematic range from two-body breakup 
(2.2~MeV for real photons) to the
limit of large $W > 2$~GeV is also extremely challenging. For example, the
GDH sum rule on the deuteron predicts a very small value for the
integral of Eq.~(\ref{eq:GDH_real}) because the anomalous magnetic
moment of the deuteron is much smaller than that of the proton
or neutron (0.14 vs. 1.8-1.9). This is due to a very
strong cancellation between the contribution from the two-body
($pn$) breakup of the deuteron, 
dominated by low photon
energy of order a few MeV, and the incoherent sum from inelastic channels on both
nucleons ({\it e.g.}, $\pi$ production with a threshold of over 140~MeV). 

In lieu of a full theoretical
description of the scattering process on the deuteron,
we use an approximation to 
express structure functions of the
deuteron as a sum of structure functions
of the proton and the neutron, convoluted
with their momentum and spin distribution inside the deuteron. Here we follow the prescription of Ref.~\cite{Kahn:2008nq}.
Our approach
corrects for the effects of Fermi motion and the effective
nucleon polarization due to the deuteron D-state contribution. At
the same time, experimentally, we avoid the kinematic region where
coherent and two-body breakup effects are expected to be most significant,
 {\it i.e.},  low energy transfer. For this reason, we extract
neutron (and deuteron) structure function results only in the
region $W > 1.15$~GeV from our data, corresponding to 
virtual photon energies in excess of 240~MeV even at our lowest
$Q^2$ point. This also minimizes the contribution from quasi-elastic 
scattering 
and its radiative tail to our data, so that we should be dominated by the
sum of contributions from the two individual nucleons. Since our lowest $Q^2$ bin, with a mean value of $0.017$~GeV$^2$, corresponds to a resolution of $\hbar/\sqrt{Q^2} = 1.5$~fm, we can also assume that these contributions largely add incoherently,
given the root mean square (RMS) separation of the two nucleons in deuterium of nearly 4~fm.
The region 1.073~GeV $< W <$ 1.15~GeV, when needed, can be augmented by a parametrization of single-pion production off protons and neutrons from the MAID partial wave analysis~\cite{Drechsel:1998hk}, where multi-particle final states do not contribute and the MAID parametrization is very reliable.

For moments of spin structure functions, the most conspicuous nuclear effect, {\it viz.} smearing of the resonances by Fermi motion, is integrated out. It vanishes exactly for first moments and to a good approximation for higher moments.  
We studied this assertion using a model for both smeared and 
unsmeared spin structure functions and found no difference in the first moments, 
and only minute differences in the higher moments ($\overline \gamma_0$) (up to 4.8\%, which is small compared to our statistical and systematic uncertainties).
Thus, no corrections were applied to the data; 
we only account for the effective polarization of the nucleons in deuterium when relating integrals for protons, neutrons, and deuterons, while supplementing the region 1.073~GeV $< W <$ 1.15~GeV with the MAID parametrization:
\begin{eqnarray}
\overline\Gamma_1^d&=& \left(\overline\Gamma_1^{p}+\overline\Gamma_1^n\right)\left(1-1.5\omega_{D}\right);\label{eq:Gamma1d=p+n}\\
\overline I_{TT}^d&=&\left(\overline I_{TT}^p+\overline I_{TT}^{n}\right)\left(1-1.5\omega_{D}\right);\label{eq:ITTd=p+n}\\
\overline\gamma_0^d&=&\left(\overline\gamma_0^p+\overline\gamma_0^n\right)\left(1-1.5\omega_{D}\right),
\label{eq:gamma0d=p+n}
\end{eqnarray}
with $\omega_{D} = 0.056 \pm0.01$~\cite{Lacombe:1980dr, Machleidt:1987hj, Zuilhof:1980ae, Kotthoff:1976fg, Desplanques:1988mp} accounting for the nucleon depolarization caused by the deuteron D-state. 
\section{The EG4 Experiment}\label{sec:exp}

The CLAS EG4 experiment~\cite{PR03006,PR05111} was carried out at JLab's experimental Hall B in 2006. Data in the range  $0.9 \lessapprox W \lessapprox 2.2$~GeV were collected using six beam energies and on longitudinally polarized NH$_3$ and ND$_3$ targets.  
Since the physics goal of EG4 was focused on the $\chi$EFT domain, data were collected with high 
statistics at very low $Q^2$ values, which enabled us to form moments down to $0.012$~GeV$^2$ 
for the NH$_3$ target.
As a comparison, the previous CLAS spin structure experiment EG1b~\cite{CLAS:2017qga} measured  
down to $Q^2=0.05$~GeV$^2$. Therefore, the data below $Q^2=0.1$~GeV$^2$ were improved 
significantly from this experiment. 

\begin{figure}[!htp]
\begin{center}
  \includegraphics[width=0.4\textwidth]{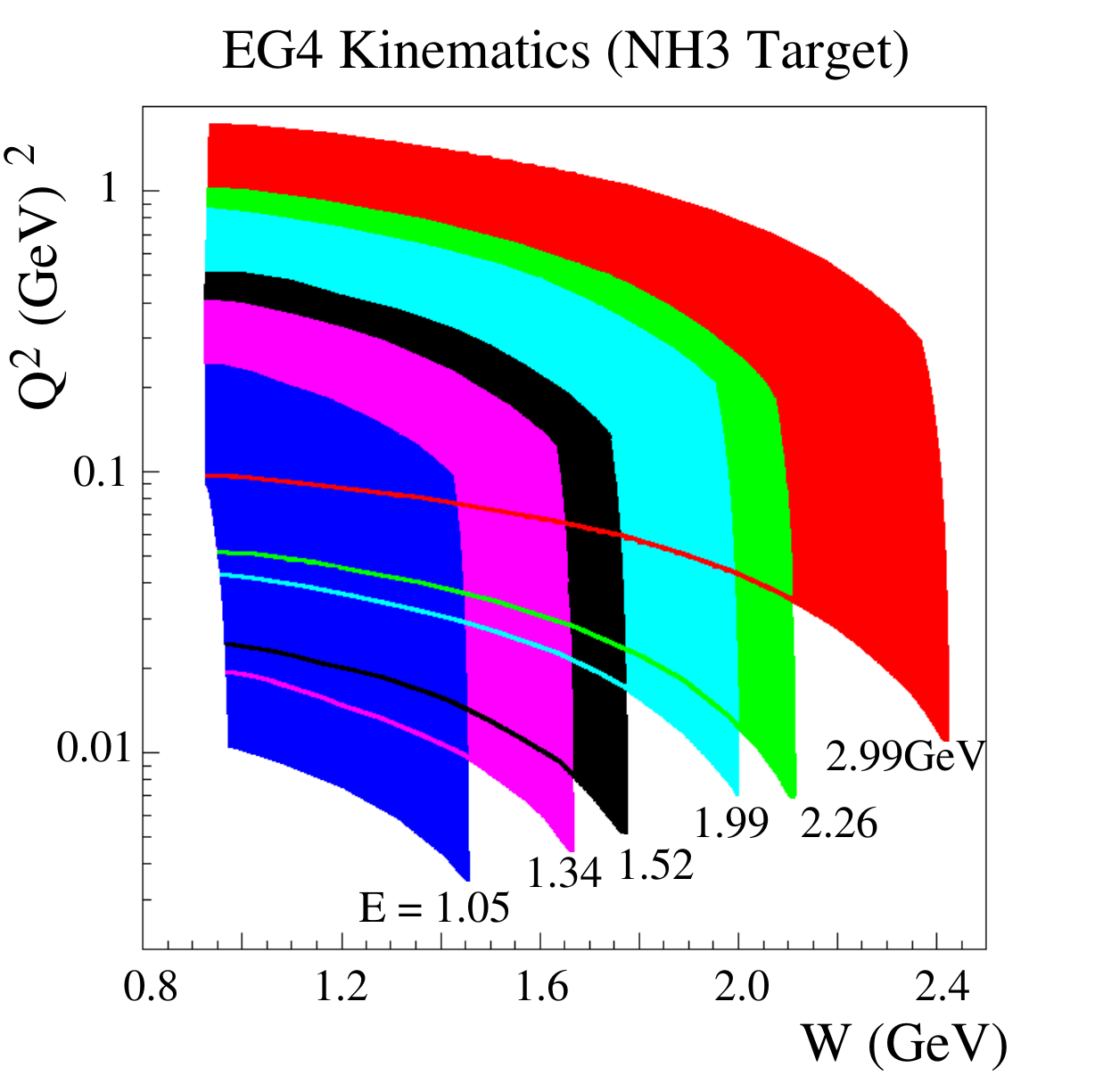}
  \includegraphics[width=0.4\textwidth]{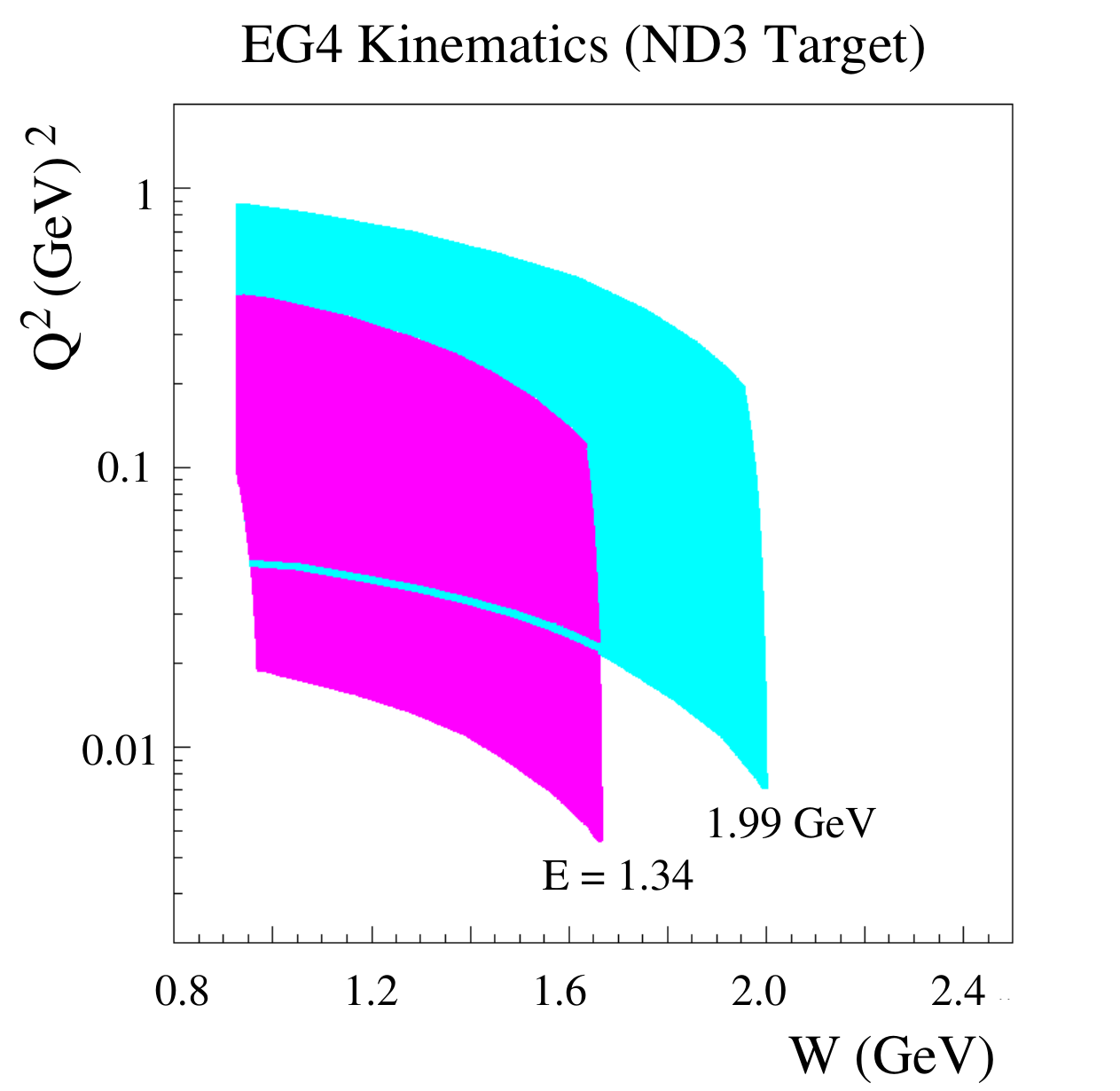}
  \end{center}
 \caption{[Color online] EG4 kinematic coverage ($Q^2$ vs. $W$) for the NH$_3$ (top) and ND$_3$ (bottom) targets.
 }\label{fig:kine}
\end{figure}

Figure~\ref{fig:kine} shows the EG4 kinematic coverage for polarized NH$_3$ and ND$_3$ targets. The beam energies used during the experiment were 1.054, 1.338, 1.989, 2.260, and 2.999~GeV. For the ND$_3$ target, only two beam energies 1.34 and 1.99~GeV were used. The beam polarization was on average 85\% throughout the experiment, as measured by the Hall B M\"oller polarimeter. The ranges in the target polarization were $(75-90)\%$ and $(30-45)\%$ for NH$_3$ and ND$_3$, respectively, measured by the NMR polarimetry.

\subsection{The Polarized Electron Beam and Beam Monitoring}\label{sec:exp_beam}

For each of the three experimental halls during the JLab 6~GeV era, a 499 MHz beam was generated with the desired beam current and polarization at the source. They were then intercalated to form a 1497 MHz beam and accelerated to about 45~MeV by superconducting radio-frequency (SRF) cavities. 
The beam was then injected into the two 0.4-km LINACs consisting of 20 SRF cavities each. The two LINACs are connected by recirculation arcs, forming a racetrack-shaped accelerator. Each revolution increased the beam energy by up to 1.2~GeV, determined by the settings of the SRF cavities. The beam could be circulated up to 5 times before being separated and dispatched to the three halls.

The EG4 experiment employed beams circulated between 1 and 3 times, with the beam current ranging from 1 to 3.5~nA. The Hall B beamline is instrumented to measure and monitor the beam properties, namely, position, transverse distribution (beam profile), current, polarization, and charge asymmetry~\cite{CLAS:2003umf}. 
Nine RF cavities grouped in three beam position monitor (BPM) sets, located 36.0, 24.6, and 8.2~m upstream from the CLAS center, measured both the beam position and current. The latter was also monitored by synchrotron light monitors (SLM) and measured absolutely by a Faraday cup positioned at the end of the beamline, 29.0~m downstream from the CLAS center. The Faraday cup consisted of a 15-cm diameter Pb cylinder of 75 radiation lengths thickness and measured the accumulated charge. Time-derivation of the accumulated charge provided the beam current. 
The beam profile was periodically verified by running thin wires (``harps'') through the beam and detecting scattered electrons using photomultipliers (PMTs) located 10~cm from the beamline. The harps also served to calibrate the BPMs.

The EG4 experiment used a longitudinally polarized beam. 
The beam helicity was flipped at 30~Hz, taking the quartet structure of either + - - + or - + + -, with the first helicity window selected from a pseudo-random sequence. 
The beam polarization was regularly measured by a M\"oller polarimeter set at the entrance of the Hall. It consisted of detectors, transport and polarizing magnets, and a 25 $\mu$m thick Permendur foil (49\% Fe, 49\% Co, 2\% Va) acting as a polarized electron target and oriented at $\pm 20^\circ$ with respect to the beamline. A pair of 120-G Helmholtz coils thermally polarized the foil longitudinally to 7.5\%. 
For each beam polarization measurement, the foil was inserted in the beam for about 30 minutes. 
The scattered (recoil) electrons were transported to the detectors (two lead-glass blocks) by two quadrupoles. The M\"oller asymmetry was measured by detecting the two electrons in coincidence and compared to the well-known theoretical expectation to obtain the beam polarization. 
A 30-min measurement provided a $\approx$1\% statistical precision. The beam polarization for EG4 was typically around 86\%~\cite{CLAS:2016qxj}, determined with a $\approx$2.5\% systematic uncertainty. Two of the M\"oller measurements were cross-checked with a Mott polarimeter set in the CEBAF injector that measured the polarization of 5~MeV electrons~\cite{Price:1998xd}. These direct measurements were used to check the stability of the beam polarization, while for the data analysis the product of beam and target polarization $P_bP_t$ was directly determined in $ep$ elastic or $ed$ quasi-elastic scattering, see Sections~\ref{sec:elsim} and~\ref{sec:elsim_comp1}.

The beam charge asymmetry, {\it viz} the relative difference in the number of electrons per helicity bunch, was monitored by  the BPMs, SLM, and Faraday cup and minimized by feedback to the electron polarized source. Most individual runs had beam charge asymmetry lower than $10^{-3}$, with a maximum of several $10^{-3}$ for some runs.

The Hall B beamline was not equipped with a beam energy measurement device. Instead, the beam energy was calculated from the setting of the CEBAF magnets together with the beam position~\cite{Krafft:2000nm}. The relative accuracy reached on the beam energy measurement was a few $10^{-4}$.
In addition to its instrumentation, the Hall B beamline had two pairs of magnetic coils to raster the beam over the target. This was to spread the heat uniformly over the target to avoid depolarization, see next section. The raster used a spiral pattern of 1.2 cm outer diameter.

\subsection{The Polarized Targets}\label{sec:target}
During EG4 the polarized target was placed 1.01~m upstream from the CLAS 
center to increase the acceptance at low $Q^2$ by reducing the minimum angle 
for scattered electrons.
The targets included frozen $^{15}$NH$_3$ and $^{15}$ND$_3$ dynamically polarized at 1~K with a 5-T field, as well as two carbon targets and an empty cell for reference. These were the same as the targets used for previous 
CLAS double-polarization measurements~\cite{Crabb:1995xi}.
The beam was rastered to protect the target from overheating and to minimize depolarization. The target 
density was 0.917 and 1.056 g/cm$^3$ for NH$_3$ and ND$_3$, respectively.

\begin{figure}[!h]
 \begin{center}
 \includegraphics[width=0.48\textwidth]{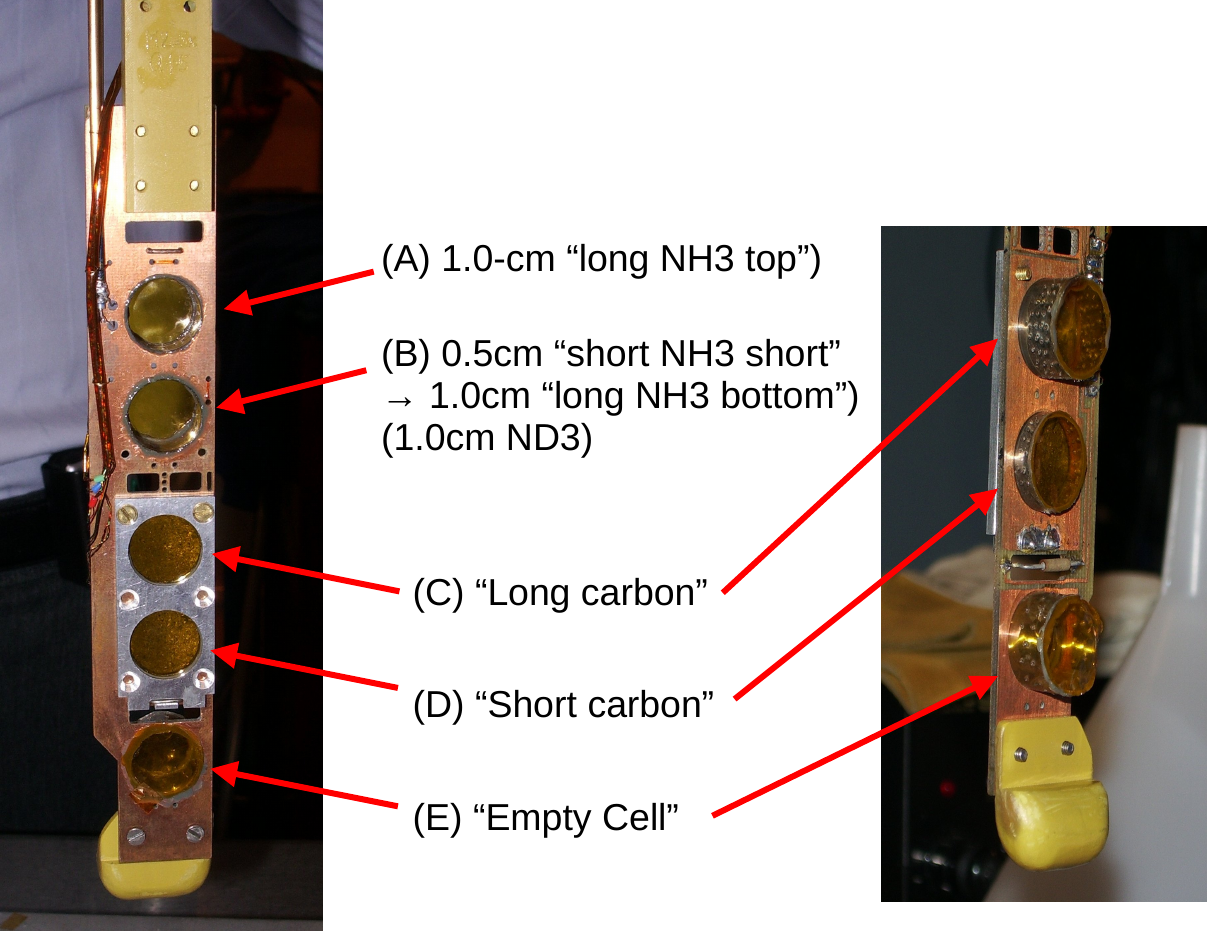}
 \end{center}
 \caption{[Color online]  
Target insert used during the EG4 experiment, viewed from the front (left) and back (right). The five target positions are labeled
A, B, C, D, and E, as shown, that accommodated polarized cells, and two carbon targets and an empty cell for calibration purposes.
Two NH$_3$ cells, of lengths 1.0~cm  and 0.5~cm, were installed 
in positions (A) and (B) during the first half of the NH$_3$ run period, and were called the ``long NH$_3$ top'' and the ``short NH$_3$'' targets, respectively. During the second half of the NH$_3$ run, two 1.0-cm NH$_3$ targets were installed in positions (A) and (B), and were called the ``long NH$_3$ top'' and the ``long NH$_3$ 
bottom'' targets, respectively. For the ND$_3$ run period only a single 1.0-cm ND$_3$ target was installed in position (B). }\label{fig:targinsert}
\end{figure}

The structure of the target insert for EG4 is shown in Fig.~\ref{fig:targinsert}. The target cells were either 1.0 cm or 0.5 cm in length, for systematic studies, and were referred to as the long and short targets, respectively. 
The insert was placed into the target ``banjo'', an approximately 1-liter vessel of 1-K liquid helium at the center of a 5-T superconducting split coil magnet.  A complete description of the polarized target can be found in Ref.~\cite{Keith:2003ca}.

\subsection{The CLAS Spectrometer and the Cherenkov Detector}

\begin{figure}[!ht]
 \includegraphics[width=0.5\textwidth]{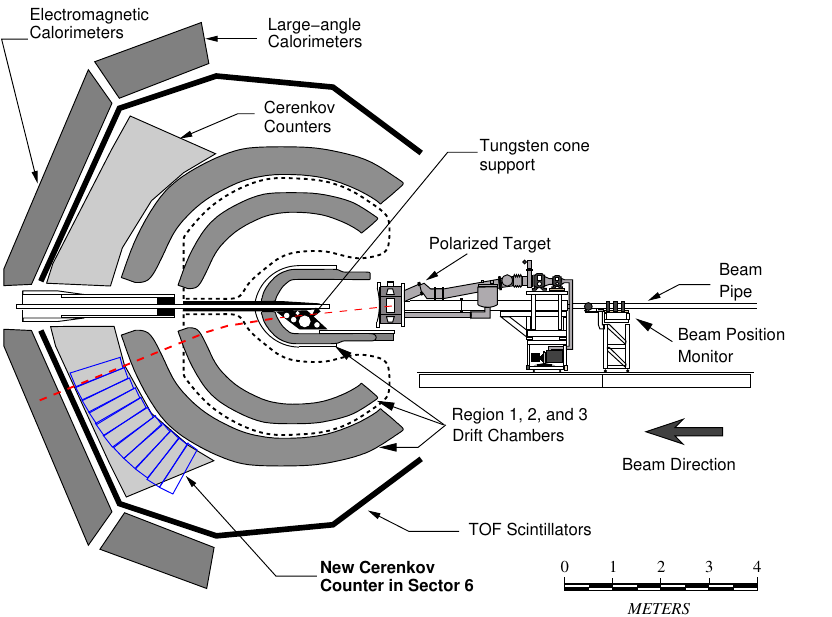}
 \caption{[Color online] Side view of CLAS with the outline of the new Cherenkov detector's segments 
shown in blue. A typical electron trajectory for the electron outbending setting is shown 
as the red dashed line.
}
 \label{fig:clas}
\end{figure}

\begin{figure}[!h]
 \begin{center}
 \includegraphics[width=0.45\textwidth]{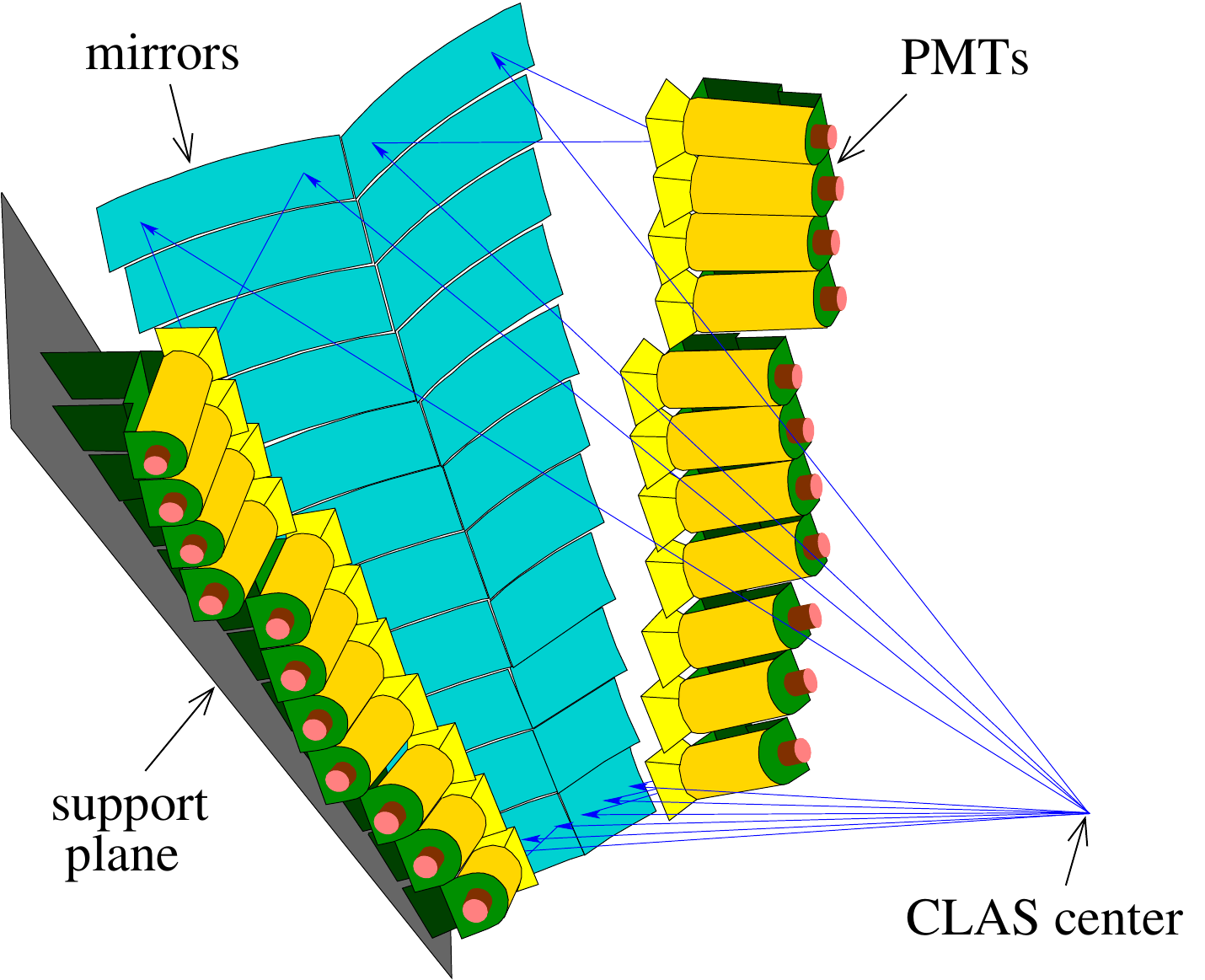}
 \end{center}
 \caption{[Color online] The new Cherenkov detector used by the EG4 experiment. This detector
consists 11 pairs of spherical mirrors that 
reflect the Cherenkov light to corresponding photomultiplier tubes (PMTs).}
 \label{fig:newcer}
\end{figure}

The basic structure of CLAS is shown in Fig.~\ref{fig:clas}.
The design of the CLAS detector was based on a toroidal magnetic field that was generated by six superconducting coils arranged around the beamline. In the EG4 experiment, the torus field setting was to bend electrons away from the beamline (outbending configuration).
The magnet coils naturally separated the detector into six “sectors”, each functioning as an independent magnetic spectrometer. Each sector included four sub-detectors: drift chambers (DC), Cherenkov counters (CC), time-of-flight scintillator counters (SC), and electromagnetic calorimeters (EC).
The drift chambers were located before, within, and after the torus magnetic field
and performed charged particle tracking, allowing for the determination of the particle momentum from the curvature of their trajectories. The other sub-detectors were located outside the magnetic field region.
In order to cover the very low $Q^2$ region with the high detector efficiency 
necessary for the absolute cross section measurements, a new CC
was built by the INFN-Genova group and 
was installed in Sector 6. The main elements of this detector are shown in
Fig.~\ref{fig:newcer}. Basic information on the new CC can be found in Ref.~\cite{CLAS:2016qxj}.

Another special feature adopted by EG4 was a new 
tungsten cone for the M\"oller shield, shown in Fig.~\ref{fig:clas}, 
which allowed direct line-of-sight from the target at smaller scattering angles down to 5$^\circ$.

\subsection{Trigger and Data Acquisition System}

The main electron trigger for EG4 was formed using
a coincidence between the signals from the EC and the new CC; 
consequently only 1/6 of the full azimuthal acceptance of CLAS was used. For calibration of the new CC performance, data were also taken using EC-only triggers.  

All photomultiplier-tube (PMT) time-to-digital-converter (TDC) and analog-to-digital-converter (ADC) signals ({\it i.e.}, SC, EC, and CC signals) generated within 90 ns of the trigger were recorded,
along with DC TDC signals [11]. A trigger supervisor (TS) directed all the signals to the data acquisition system. 
The offline physics event reconstruction code
used geometric parameters and calibration constants to convert the TDC and ADC data into kinematic and particle identification data. The code cycled through particles in the event to search for a single trigger electron -- a negatively charged particle that produced a shower in the EC. If more than one candidate was found, the one with the highest momentum was selected. This particle was traced along its geometric path back to its intersection in the target to determine the path length, which, with the assumption that its velocity $v = c$, determined the event start time. From this start time, the TOF of other particles could then be determined from the SC TDC values. The TDC values from the EC were used when SC values were not available for a given particle.

\section{Data Analysis}\label{sec:analysis}

The analysis presented in this section lead to the extraction of both the proton and the deuteron spin structure functions $g_1$, $A_1F_1$, and their moments, previously reported in Refs.~\cite{CLAS:2017ozc,CLAS:2021apd}.  We will present the general method (Section~\ref{sec:ana_method}), event selection criteria and kinematic corrections (Section~\ref{sec:ana_selection}), particle identification performance and efficiency analysis of the new CC (Section~\ref{sec:ana_pid}), 
and background (Section~\ref{sec:ana_bg}). 

\subsection{Overview of the Analysis Methods}\label{sec:ana_method}

In this work, we first extracted the event yield difference between the positive and negative beam helicity states.
The polarized cross section differences,  Eq.~(\ref{eq:polxs_long}), 
were extracted from the yield differences as follows:
\begin{eqnarray}
  \Delta\sigma_{||} (\Delta W, \Delta Q^2) &=&\hspace*{-0.4cm}
\underset{\Delta W \Delta Q^2}{\int} \left( \frac{\q{d}^2\sigma_{\ua\Ua}}{\q{d}\Omega
    \q{d}E^\prime}-\frac{\q{d}^2\sigma_{\ua\Da}}{\q{d}\Omega    \q{d}E^\prime} \right) dE' d\Omega\nonumber\\ 
  &&\hspace*{-1cm} = \left[\frac{N^+}{N_e^+}-\frac{N^-}{N_e^-}\right]\frac{1}{N_\q{targ}}\frac{1}{P_bP_t}
  \frac{1}{\eta_\q{detector}},\label{eq:analysis0}
\end{eqnarray}
where $N_e^\pm$ is the number of incident electrons with helicity $\pm$, respectively, as recorded by the Faraday cup or beam charge monitors; 
$N^\pm$ is the number of scattered electrons with incident helicity $\pm$, 
within the bin $\Delta W \Delta Q^2$, 
that pass through all analysis cuts; $N_\q{targ}$ is the number of polarized target nucleons per cm$^2$; $P_bP_t$ is the product of the beam and the target polarizations; 
and finally, $\eta_\q{detector}$ is the product of the acceptance and efficiency of the detector and the trigger. 
In practice, we calculated the yield difference in each
$(W,Q^2)$ bin and compared to the simulated value that
accounts for efficiency and acceptance, and with the same acceptance and fiducial cuts as data.

One important advantage of our method is that contributions from unpolarized material cancel in the yield differences. This is important for experiments utilizing polarized NH$_3$ and ND$_3$ targets because of the presence of significant amounts of unpolarized material, including target windows, support, insulation, liquid helium for cooling, and the nitrogen in the ammonia beads (the nitrogen can
be slightly polarized and is treated
as a background contribution). 
The remaining unknowns on the RHS of Eq.~(\ref{eq:analysis0}), $P_bP_tN_\q{targ}$, were obtained indirectly by normalizing the yield differences of elastic scattering of the data
to simulation, which provided more accurate information than combining beam and target polarimetry data ($P_b$, $P_t$) along with information on the polarized material thickness of the target. 

After event selection and kinematic correction (see next two sections), the data were divided into the $Q^2$ bins shown in Table~\ref{tab:q2bint}, with bin limits determined logarithmically (the upper edge of each bin $Q^2_\q{max}$ is $10^{1/13}$ of the lower edge $Q^2_\q{min}$). 
In practice, proton data exist for all 25 $Q^2$ bins, and deuteron data from bin 3 to 24. Our neutron results (as presented in Section~\ref{sec:results_g1n}) were extracted from combining proton with deuteron data, and thus also exist from bins 3 to 24 only.  
\begin{table}[!ht]
\begin{center}
\begin{tabular}{|c|c|c|c|c|c|c|c|c|c|} \hline\hline
 
  $Q^2$ bin ID      &  1       &  2       &  3       &  4       &  5  \\  
  min (GeV/$c$)$^2$ & 0.0110   & 0.0131  & 0.0156  & 0.0187   & 0.0223   \\
  mean (GeV/$c$)$^2$& 0.0120   & 0.0143  & 0.0171  & 0.0204   & 0.0244   \\
  max (GeV/$c$)$^2$ & 0.0131   & 0.0156  & 0.0187  & 0.0223   & 0.0266   \\\hline\hline
 
  $Q^2$ bin ID      &  6       &  7     &  8       &  9       &  10  \\  
  min (GeV/$c$)$^2$ & 0.0266   & 0.0317  & 0.0379   & 0.0452  & 0.0540    \\ 
  mean (GeV/$c$)$^2$& 0.0290   & 0.0347  & 0.0414   & 0.0494  & 0.0590    \\
  max (GeV/$c$)$^2$ & 0.0317   & 0.0379  & 0.0452   & 0.0540  & 0.0645    \\\hline \hline

  $Q^2$ bin ID      &  11      &  12      &  13     &  14     &  15   \\ 
  min (GeV/$c$)$^2$ & 0.0645   & 0.0770  & 0.0919   & 0.110    & 0.131     \\ 
  mean (GeV/$c$)$^2$& 0.0705   & 0.0841   & 0.1005  & 0.1200   & 0.143     \\
  max (GeV/$c$)$^2$ & 0.0770   & 0.0919  & 0.110    & 0.131    & 0.156     \\\hline \hline

  $Q^2$ bin ID      &  16      &  17      &  18      &  19      &  20   \\
  min (GeV/$c$)$^2$ & 0.156   & 0.187    & 0.223    & 0.266    & 0.317    \\
  mean (GeV/$c$)$^2$& 0.171    & 0.204   & 0.244    & 0.290    & 0.347    \\
  max (GeV/$c$)$^2$ & 0.187   & 0.223    & 0.266    & 0.317    & 0.379    \\\hline\hline

  $Q^2$ bin ID      &  21      &  22      &  23     &  24     &  25   \\
  min (GeV/$c$)$^2$ & 0.379    & 0.452   & 0.540   & 0.645     & 0.770     \\
  mean (GeV/$c$)$^2$& 0.414    & 0.494   & 0.590   & 0.705     & 0.841     \\
  max (GeV/$c$)$^2$ & 0.452    & 0.540   & 0.645   & 0.770     & 0.919     \\\hline\hline


\end{tabular}
\end{center}
\caption{$Q^2$ bins used for the EG4 inclusive channel analysis. For each bin, both
  minimum and maximum (lower and upper edge) are shown, with the mean $Q^2$ value calculated as
$Q^2_\q{mean}=\sqrt{Q^2_\q{min}Q^2_\q{max}}$. Note that identical binning was used in the analysis of the earlier CLAS EG1b data~\cite{CLAS:2017qga}.}\label{tab:q2bint}
\end{table}

\subsection{Event Selection and Kinematic Corrections}\label{sec:ana_selection}
The primary signal of this analysis was the scattered electrons in the inclusive mode. Multiple selection criteria were used to select good electron events, requiring the correct particle charge ($-\vert e\vert$) and valid detector signals in DC, SC, EC, and CC. The quality of the helicity signal recorded in the data stream was examined closely and any quartet sequence with inconsistency was rejected.  Kinematic corrections were then applied to each event that consisted of: incoming energy loss correction, raster correction, tracking correction, momentum correction, and outgoing ionization loss correction. These corrections were checked by the position of the $W$ peak of $ep$ elastic scattering, which is required to coincide with the proton mass perfectly.

\begin{figure}[!htp]
  \includegraphics[width=0.42\textwidth]{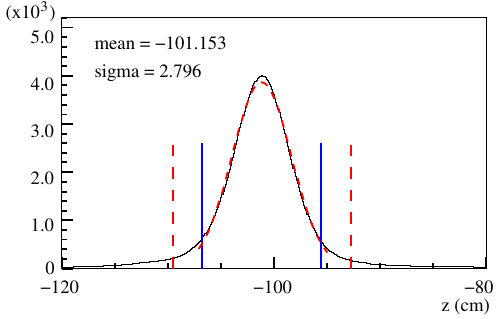}
\vspace*{1mm}
  \includegraphics[width=0.42\textwidth]{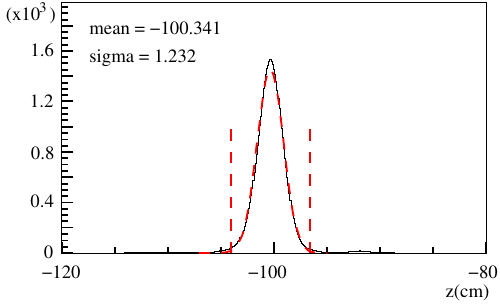}

 \caption{[Color online] Vertex $z$ distribution (solid black curves) of the 1.1~GeV NH$_3$ target data for  $Q^2=(0.011,0.0131)$~GeV$^2$ (top) and $Q^2=(0.0452,0.054)$~GeV$^2$ (bottom) bins. The dashed (red) curves show the Gaussian fit with the fitted mean and sigma values (in cm) shown in each panel. The two dashed  (red) vertical lines show the 3$\sigma_z$ widths. A  $3\sigma_z$ cut is used for the bottom figure, while a $2\sigma_z$ (solid blue vertical lines) is used for the top. When making these plots, all good electron cuts
  except the vertex $z$ cut were applied.}\label{fig:vz}
\end{figure}

After kinematic corrections, more selection criteria were applied. First, events must be detected in sector 6 of CLAS. The particle momentum must be between 20\% and 100\% of the incoming beam energy, and must be higher than $0.37$~GeV, corresponding to the trigger threshold set for the experiment. 
The distribution of the $z$ position along the beamline of the electron vertex was then studied for different $Q^2$ bins and fitted using a Gaussian function. Only events within three standard deviation ($\pm 3 \sigma_z$) of the average $z$ position of that bin were accepted, except for those 
few bins where the $z$-distribution was rather
wide and a tighter cut was used; see Fig.~\ref{fig:vz}. 
These vertex $z$ cuts ensured that data statistics were maximized while reducing backgrounds from the beamline, and that events scattered from the insulation layers of the target were rejected. 
A series of cuts were then applied on the event's momentum $p$ vs. polar angle $\theta$ as well as $\theta$ vs. the azimuthal angle $\phi$ to avoid the fiducial regions where the detector efficiencies were low. 

\subsection{Particle Identification and Cherenkov Efficiency}\label{sec:ana_pid}

Particle identification cuts were applied on the EC and CC signals. For the EC, the ratio of the total energy deposited over the particle momentum $E_\q{tot}/p$ within each $Q^2$ bin was fit with a Gaussian peak and cuts were applied to select events within $3\sigma$ from the peak center. 
Furthermore, the energy deposited in the inner layer of the EC, $E_\q{in}$, was required to be greater than 0.06~GeV. The same $3\sigma$ and $E_\q{in}$ cuts were also applied to the simulation such that a direct comparison with data could be made to account for the efficiency of these PID selection criteria.

Further PID cuts were applied to the number of photoelectrons in the CC signal, requiring $n_{\rm p.e.}>2.0$ for NH$_3$ and $n_{\rm p.e.}>2.5$ for ND$_3$ analysis.  The CC signals must also pass conditions that match their timing and geometrical location with those of the EC and DC~\cite{Osipenko}. Unlike the EC, the CC efficiency due to these PID selection criteria could not be easily obtained from detector simulation, and had to be extracted from data and applied to the simulation. We discuss below the details of the CC efficiency study.

\begin{figure}[!htp]
\begin{center}
  \includegraphics[width=0.48\textwidth]{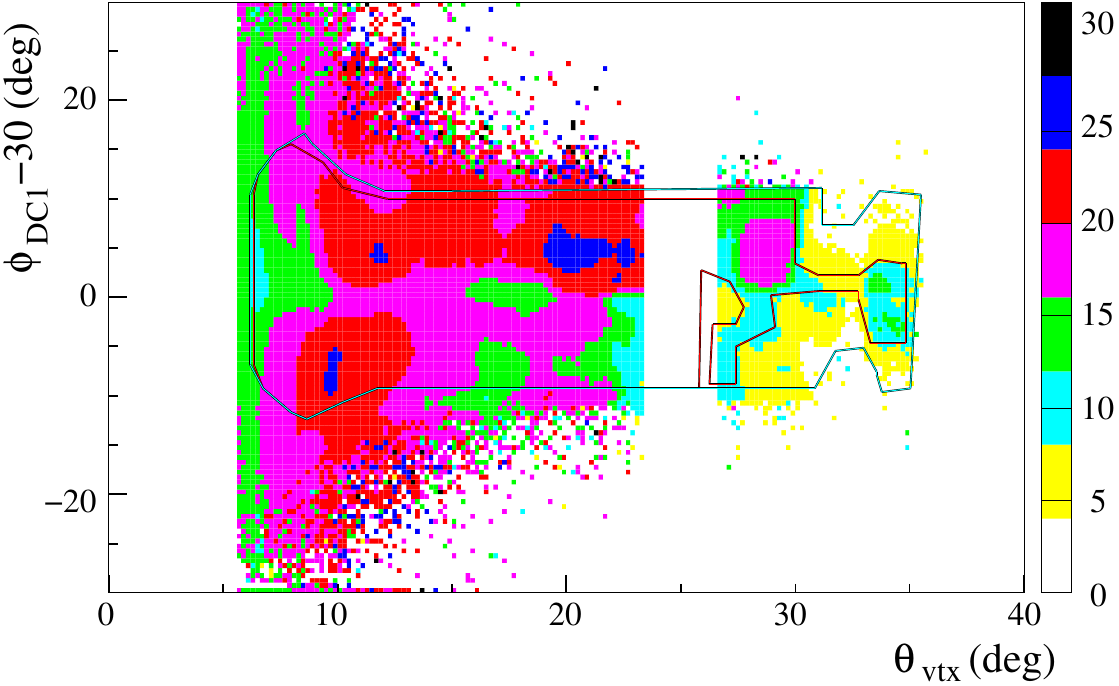}
\end{center}
\caption{[Color online] Fiducial cuts applied to $\phi_{DC1}$ (offset by $-30^\circ$) vs. $\theta_{vtx} (^\circ)$ for the $I_{torus}/2250$A$/p$(GeV)$=(0.6,0.7)$ bin for 2.3~GeV data taken on NH$_3$.
The color depth shows the average $n_{\rm p.e.}$ values for each $(\theta_{vtx},\phi_{DC1})$ bin. 
When making these plots, all good electron cuts except the $\phi_{DC1}$ vs. $\theta_{vtx}$ fiducial cut were applied.
The outer contour shows the first-round of fiducial region that traced the region of low event yield. The inner contour, which is the final CC fiducial cut, also rejects regions of low $n_{\rm p.e.}$ values. The blank region near $\theta=25^\circ$ is due to a low-efficiency region in the DC and is rejected from the analysis and the simulation. 
}\label{fig:cc_fid_nphe}
\end{figure}

The event distribution in the hit position of the CC, in terms of $\theta_{vtx}$, the polar angle at the interaction point, and $\phi_{DC1}$, the azimuthal angle in the first layer of the DC, was studied to determine both $n_{\rm p.e.}$ yield and the high efficiency region of the new CC. 
Both the event distribution and the $n_{\rm p.e.}$ depend on the event hit position in the CC, and thus vary with the ratio of the CLAS torus current over particle momentum $I_\q{torus}/p$, where $I_\q{torus}=2250$~A for beam energies 2.0, 2.3, and 3.0~GeV and $1500$~A for beam energies 1.1 and 1.3~GeV. Figure~\ref{fig:cc_fid_nphe} shows the $n_{\rm p.e.}$ distribution for one representative $I_{torus}/p$ bin, for the 2.3~GeV data collected on NH$_3$. 

For a specific $I_{torus}/p$ bin, the CC efficiency for a given $(\theta_{vtx},\phi_{DC1})$ bin was determined by taking calibration data that utilized EC alone for electron identification. For each bin, we denote the number of electrons recorded in EC-only calibration data to be $N_1$ and among those events, $N_2$ electrons also pass the CC selection criteria. The efficiency, and its uncertainty for $N_1\neq N_2\neq 0$ case, were calculated as
\begin{eqnarray}
  \eta = \frac{N_2}{N_1}~, &&
  \Delta\eta_{N_2\neq N_1} = \sqrt{\frac{N_2(1-N_2/N_1)}{N_1}}~.\label{eq:dcceff1}
\end{eqnarray}
For the case of $N_2=N_1$ or if $N_2=0$, we must consider the uncertainty due to $N_2$ being an integer (that is, we cannot detect half of an electron!). If $N_2=N_1$, we assign:
\begin{eqnarray}
  \eta_{N_2=N_1} = \frac{N_2-0.25}{N_1}~, && \Delta\eta_{N_2=N_1} = \frac{0.25}{N_1}~.\label{eq:dcceff2}
\end{eqnarray}
If $N_2=0$ we assign:
\begin{eqnarray}
  \eta_{N_2=0} = \frac{0.25}{N_1}~,&& 
  \Delta\eta_{N_2=0} = \frac{0.25}{N_1}~.\label{eq:dcceff3}
\end{eqnarray}
Note that the uncertainty due to $N_2$ being integer also exists for the general case, but was completely negligible compared with Eq.~(\ref{eq:dcceff1}) for $N_1$ as small as $2$. 

\begin{figure}[!ht]
\begin{center}
  \includegraphics[width=0.45\textwidth]{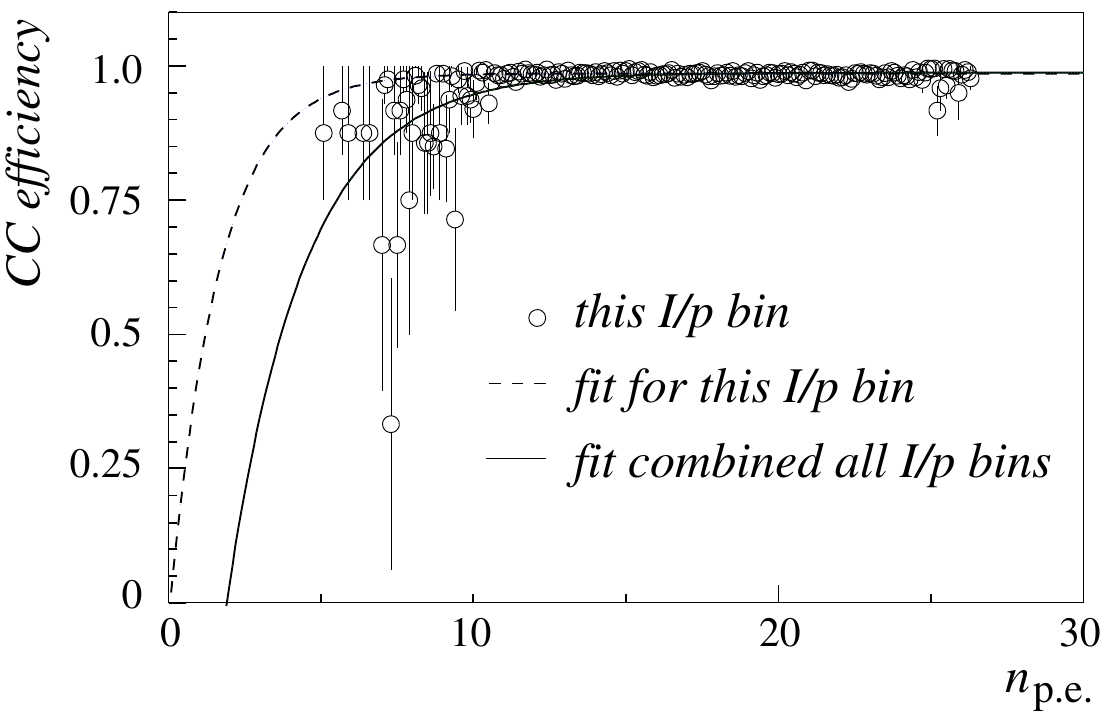}
  \includegraphics[width=0.45\textwidth]{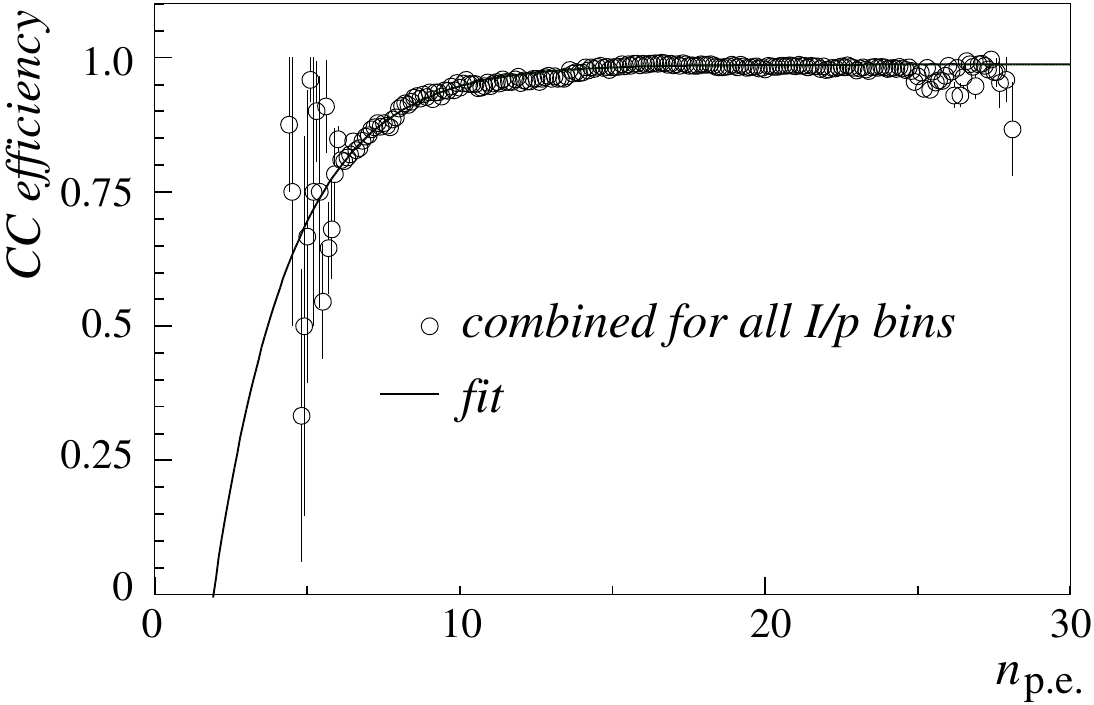}
\end{center}
\caption{CC efficiency (open circles) vs. the number of photoelectrons $n_{\rm p.e.}$ for the $I_{torus}/2250$A$/p$ (GeV)$=(0.6,0.7)$ bin (top) and for all $I_{torus}/p$ bins combined (bottom) for 2.3~GeV data taken on the NH$_3$ top long cell. 
The efficiencies and error bars were evaluated using Eqs.~(\ref{eq:dcceff1}-\ref{eq:dcceff3}). The fit results for the $I_{torus}/2250$A$/p$(GeV)$=(0.6,0.7)$ bin is $\eta_{cc}= 0.986-e^{-0.062(x-0.44)}$ (dashed curve), and for the combined results is $\eta_{cc}= 0.987-e^{-0.040(x-18.8)}$ (solid curve).
When making these plots, all good electron cuts were applied, thus there is no entry below $n_{\rm p.e.}=2$ by definition. Note that even after combining all bins, the region 
with $n_{\rm p.e.}\leqslant 6$ has low statistics and relatively large uncertainties. 
}\label{fig:cc_eff_fit}
\end{figure}

Once we extracted the CC efficiency and its uncertainty map for given $I_{torus}/p$ range and $(\theta_{vtx},\phi_{DC1})$ bin, we matched it to the $n_{\rm p.e.}$ map for the same bin. Then we fit the values of the efficiency as a function of $n_{\rm p.e.}$ using the Poisson function form, see Fig.~\ref{fig:cc_eff_fit} top panel. We studied the efficiency fit for both individual $I_{torus}/p$ bins as well as all bins combined. (When combining bins, we combined $N_1$ and $N_2$ first and then evaluated the efficiency and its uncertainty.) We found the fit to achieve good results for all cases, and the fit results for individual $I_{torus}/p$ bins were consistent with those achieved by combining all bins, see Fig.~\ref{fig:cc_eff_fit} bottom panel. These results were used in the simulation to account for the CC efficiency, see Section~\ref{sec:sim}.

\subsection{Background Contamination}\label{sec:ana_bg}
For inelastic scattering processes, charged pions can be misidentified as electrons. Similarly, neutral pion decay can produce photons and subsequently electrons via the pair production process. Contamination from these two background processes were studied using charged pion yields, and positron yields taken with the in-bending torus configuration. The pion contamination was found to be less than 1\%, and pair production yield was on the order of $10^{-3}$ for all kinematics. We did not correct these backgrounds, but instead incorporated them into the systematic uncertainties. An additional background is the electron scattering off the $^{15}$N in the target. The $^{15}$N is slightly polarized, and thus we discuss this background along with its correction in Section~\ref{sec:syst_n15} as part of the systematic uncertainties.

\section{Simulation and Extraction of Physics Results\label{sec:sim}}
As described in the previous section, this analysis was based on extraction of 
polarized yield differences $\left[\frac{N^+}{N_e^+}-\frac{N^-}{N_e^-}\right]$ 
from data. This was done for all $Q^2$ and $W$ ranges covered by the experiment, 
including both elastic and inelastic scattering.  By comparing (``normalizing'') 
the yield difference from simulation and data
in the region of the elastic (proton)
or quasi-elastic (deuteron) peak, factors such as beam and 
target polarizations, luminosity, average trigger efficiency and detector 
efficiencies of Eq.~(\ref{eq:analysis0}) were accounted for and did not need to be 
treated or corrected separately.  
Then, the same normalization was applied to the yield difference from simulation 
of inelastic scattering. By comparing these differences to the measured spectra, the
structure functions $g_1$ and $A_1F_1$ can be extracted.  In this section we 
describe the simulation procedure and the extraction of the structure functions.

\subsection{Simulation of (Quasi-)Elastic Polarized Yield Differences}\label{sec:elsim}
A simulation for $ep$ elastic and $ed$ 
quasi-elastic scattering 
events was performed using GSIM~\cite{GSIM:2002a,GSIM:2002b}, the standard CLAS simulation program, combined with an event generator for (quasi-)elastic scattering. 
Events were generated for ranges of $\theta=(5^\circ,45^\circ)$ and $\phi=(250^\circ,325^\circ)$. The $\theta$ and $\phi$ ranges ensured the whole Sector 6 was covered, while keeping a reasonable generator efficiency. 

Two functions were used to generate separately the proton elastic peak and its radiative 
tail with appropriate relative weights. The proton form factors were from Ref.~\cite{Arrington:2007ux}, and the radiative tail included both internal and external radiation effects following the prescription of Refs.~\cite{Mo:1968cg}\cite{Tsai:1973py}. The depolarization effect of the Bremsstrahlung photons was calculated following Ref.~\cite{Olsen:1959zz}.

{For the quasi-elastic events, we used a convolution prescription~\cite{Kahn:2008nq} to calculate the inclusive structure function of the deuteron stemming only from scattering elastically off a
moving proton or neutron. Radiative effects were included using the code RCSLACPOL~\cite{CLAS:2017qga}, which was
based on the treatments 
of Refs.~\cite{Mo:1968cg} and~\cite{Kukhto:1983pv}. For the proton contribution, we used the same form factors as for elastic $ep$ scattering; for the neutron
form factors, we used a continued fraction parameterization~\cite{Kubon:2001rj} for $G_M^n$
and the two-parameter Galster fit~\cite{E93-038:2003ixb} for $G_E^n$. 
}

The simulated data were then passed through the GSIM Post Processor (GPP)~\cite{GSIM:2002a} to account for smearing factors in the detectors, and RECSIS~\cite{RECSIS:1997}, the standard CLAS simulation program for event reconstruction. Detailed inputs to the simulation, GPP, and corrections applied in the reconstruction are given below.

\subsection{Energy Loss Correction}\label{sec:elsim_eloss}
The beam energy measured in the experiment (Section~\ref{sec:exp_beam}) was used in the simulations. 
The quality of the simulation was evaluated by comparing the reconstructed $W$ spectra of the elastic peak to data in all $Q^2$ bins. Special attention was paid to the position and the width of the elastic peak because both can affect the normalization.  Because kinematic corrections were applied to the data, the elastic peak position extracted from data was very close to the proton mass (within $\sim$1-2~MeV). 
However, the elastic peak position extracted from the reconstructed simulated events was shifted from the proton mass by several MeV due to energy loss of scattering electrons in the simulation that were not corrected completely in the reconstruction.  To correct for this effect, we calculated the correction $E_\q{corr}$ that should be added to $E'$ in order to align the $W$ value of the reconstructed simulated elastic peak to the proton mass.
A linear dependence on $E'$ was observed when combining simulation of all five beam
energies: $E_\mathrm{corr}$(MeV)$\times \cos\theta=1.812+1.399E'$ with $E'$ in~GeV, and the $\cos\theta$ factor was to account for the possible $\theta$-dependence due to most 
of material being oriented perpendicular to the beamline. 

\subsection{Smearing Factor for DC Resolution}\label{sec:elsim_gppsmear}
By comparing the $W$ peaks of the elastic simulation with those obtained from data, it was found necessary to apply DC smearing in the GPP stage of the simulation. For the ND$_3$ analysis, a single DC smearing factor for each beam energy was used. 
For the NH$_3$ analysis, we found it necessary to adjust the smearing factor separately for each beam energy.  This was done by adopting a wire-number dependence in the DC smearing factor of GPP. The functional form used for the DC smearing factor $y$ is
\begin{eqnarray}
 y=1.0+\frac{p_0}{1.0+e^{(x-p_1)/p_2}}~,\label{eq:DCfitform}
\end{eqnarray}
where $x$ is the DC wire number and $p_{0,1,2}$ are fit 
parameters, see Table~\ref{tab:DCsmear}. The smearing factor starts from a value greater than 1 for low wire numbers (small scattering angle) to approximately 1.0 for high wire numbers (large scattering angle). The large smearing factor for small scattering angle was possibly due to limits in the knowledge of the magnetic field of CLAS in that region. 
\begin{table}[!h]
  \begin{center}
\begin{tabular}{c|c|c|c} \hline
  $E_b$ (GeV) & $p_0$ & $p_1$ & $p_2$\\\hline
  1.1  &  $1.952\pm 0.007$ & $24.13\pm 0.12$ & $14.18\pm 0.04$\\
  1.3  &  $1.349\pm 0.014$ & $24.22\pm 0.35$ & $14.47\pm 0.14$\\
  2.0  &  $1.837\pm 0.030$ & $15.22\pm 0.37$ & $9.48\pm 0.14$\\
  2.23 &  $2.198\pm 0.021$ & $11.07\pm 0.18$ & $7.90\pm 0.06$\\
  3.0  &  $2.161\pm 0.040$ & $10.67\pm 0.30$ & $6.70\pm 0.11$\\\hline
\end{tabular}
\end{center}
\caption{Fitting results for the wire-dependent smearing factor used in simulation of
the NH$_3$ target. The functional form of Eq.~(\ref{eq:DCfitform}) was used for the fit. 
 All three layers of DC used the same smearing factor.}\label{tab:DCsmear}
\end{table}

\subsection{Comparison of Elastic Simulation with Data}\label{sec:elsim_comp1}

\begin{figure}[!h]
\begin{center}
  \includegraphics[width=0.45\textwidth]{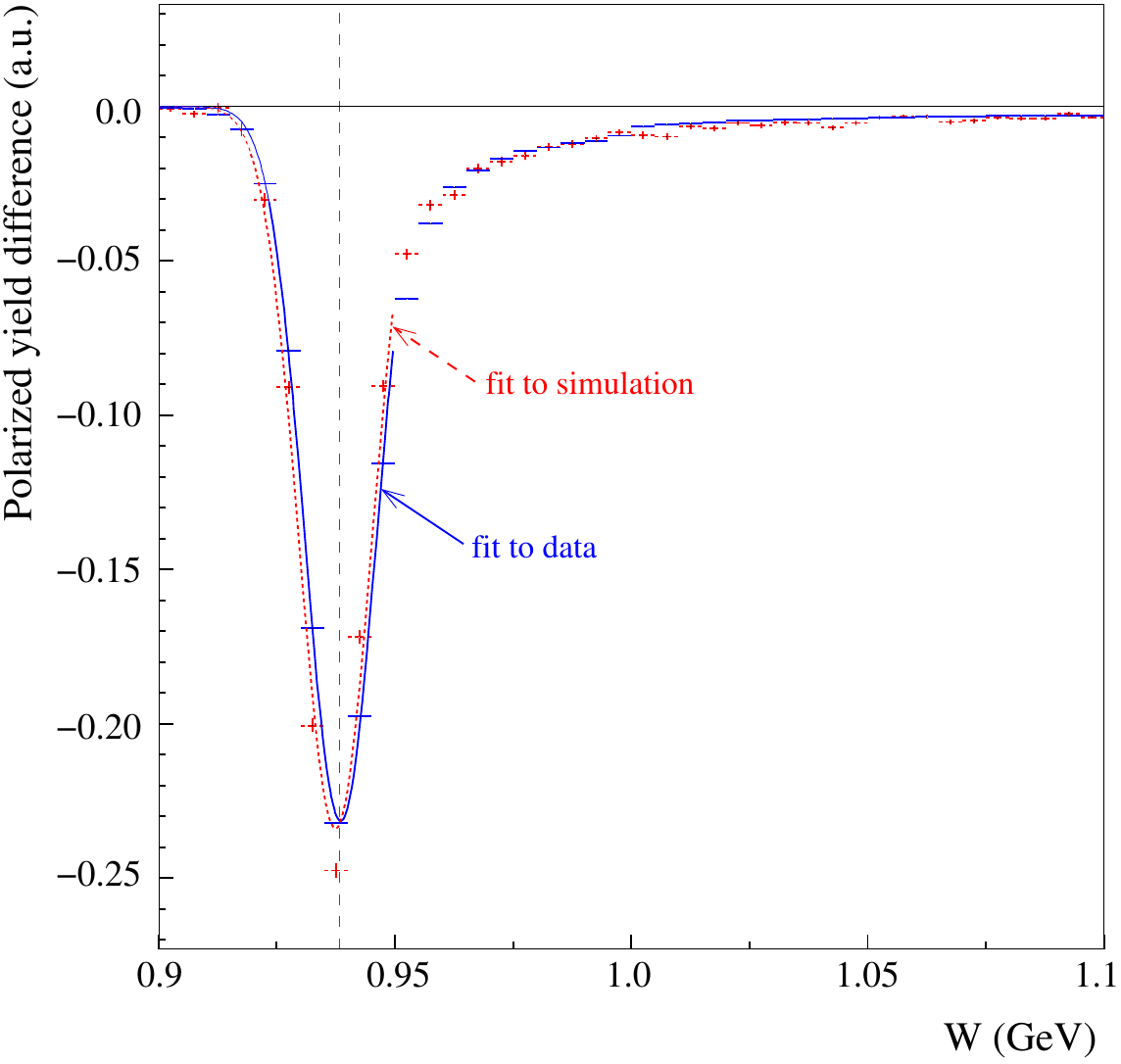}
\end{center}
\caption{[Color online] 
Comparison of the elastic peak summed over all good $Q^2$ bins, simulation (red dotted) vs. data (blue solid), for the 1.1~GeV data using the NH$_3$ bottom target cell. The vertical dashed (red) line shows the location of the proton mass. 
A single normalization factor was tuned until the integral under the simulated elastic peak matches that of data to within 0.2\% relative. The solid (blue) and the dashed (red) curves show fits to the proton peak of the data and simulation, which find mean values of 0.9385~GeV and 0.9374~GeV, and the peak widths ($\sigma$) of 0.0076~GeV and 0.0077~GeV, respectively. 
Because of the different beam polarization and target direction combinations, the sign of the elastic peak was arbitrary at this stage, but was corrected for in the extraction of the structure functions.
}\label{fig:el_ratio_sum}
\end{figure}

Figure~\ref{fig:el_ratio_sum} shows a representative bin for the comparison between data and simulation for elastic $ep$ scattering. Data from low $Q^2$ bins %
were not used for normalization because of the $^{15}$N elastic contamination (see Section~\ref{sec:syst_n15}). Both the energy loss correction and the DC smearing were applied to the simulation. %

Careful studies were done for the elastic simulation quality for individual $Q^2$ bins. Gaussian fits were used to compare the peak location and width of simulation with data to ensure that the simulated elastic peak agrees almost perfectly with the data (and the peak position agrees with the proton mass). Bins where the simulated peak position or shape deviate from data were excluded from the normalization input. This happened only at very low $Q^2$ bins due to $^{15}$N contamination (see Section~\ref{sec:syst_n15}), and at very high $Q^2$ bins due to sensitivity to the low-efficiency area of the DC as shown in Fig.~\ref{fig:cc_fid_nphe}. 
After rejecting the very low and very high $Q^2$ bins, we normalized the simulated elastic peak to that from data, summed over all $Q^2$ bins. 
The normalization for the 
deuteron data proceeded along the same lines,
using the simulation of the quasi-elastic peak
instead.

The uncertainty in the elastic peak normalization included the uncertainty in the proton form factor input~\cite{Arrington:2003qk}, 
which was found to be less than $2\%$ at 1.1~GeV and increased slightly for higher beam energies. 
The uncertainty due to radiative corrections was found to be $\approx 1\%$ for all energies. 
In addition, we compared~\cite{Arrington:2007ux} with a recent proton form factor fit~\cite{Griffioen:2015hta} that accounted for the uncertainty in the proton radius, and found the effect to be less than 1\%. 
Overall, we estimated the uncertainty in elastic normalization to be $(1+E_b)\%$ with $E_b$ the beam energy in GeV. 
For the deuteron normalization based
on the quasi-elastic peak, we estimated a total
uncertainty of 10\% for each beam energy, accounting for form factor uncertainties, the observed variation of the measured yield to simulated yield ratio with $Q^2$, and uncertainties in the folding and radiative correction procedure.
These uncertainties were taken into account in our estimate of the systematic uncertainties (see Section~\ref{sec:results_allsyst}). 

\subsection{Simulation of Inelastic Polarized Yield Differences}\label{sec:inelsim}
We used an event generator originally developed for Ref.~\cite{CLAS:2003iiq} and improved for EG4 to generate events distributed according to the polarized cross section model from RCSLACPOL~\cite{CLAS:2017qga}. 
The generator worked in two steps: first, it generated separate maps for radiated inclusive polarized cross section differences $$\Delta\sigma=\sigma_{\ua\Ua}-\sigma_{\da\Ua}$$ 
for polarized electron scattering off a longitudinally polarized proton or deuteron target. The map covered the 3D space of $(E',\sin\theta, \phi)$ for ranges of $\theta=(5^\circ,45^\circ)$, 
$\phi=(250^\circ,325^\circ)$ and $E'=(0.2~$GeV$,E_b)$. The $\theta$ and $\phi$ ranges were identical to the elastic scattering generator. Two such maps were generated for any given model of the double spin
asymmetries $A_1, A_2$; one for positive values of $\Delta\sigma$ and one for negative values. 
Next, events were thrown according to the two cross section maps. The events were given vertex coordinates that were uniformly distributed over the volume of a 1-cm-long cylinder with a radius of 0.01~cm around the beamline -- with the center of this volume being at the EG4 target position of $z=-100.93$~cm. Equal numbers of events were generated for each map of $\Delta\sigma$. The events generated were then fed into the GSIM/GPP/RECSIS chain to create the reconstructed $W, Q^2$ spectra from the simulation. 
These spectra were normalized according to the total integrated cross section of each map. The positive and negative $\Delta\sigma$ spectra were then combined to form the full spectra for inelastic scattering. Careful studies were carried out to ensure the radiative tail of elastic scattering was subtracted from the data before they were compared with the inelastic simulation. 

\subsection{Extraction of $g_1$ and $A_1F_1$}\label{sec:extraction}
Extraction of $g_1$ and $A_1F_1$ was done by varying the structure function model 
used in the inelastic simulation as follows: 
first, we constructed the {\it standard} simulation spectra using the best models for 
the unpolarized structure functions and asymmetries: 
\begin{itemize}
 \item The choice for $F_{1,2}$ and $R$
was based on the latest fit by Bosted/Christy/Kalantarians 
(dated 2014), interpolating to the real photon point at very low $Q^2$.  
The most up-to-date references are \cite{Bosted:2007xd} and \cite{Christy:2007ve}. 
\end{itemize}
For asymmetries, the latest fit performed by the EG1b collaboration~\cite{CLAS:2017qga} was used. The parameterization itself is unpublished. The asymmetry model includes three options:
\begin{itemize}
\item The asymmetries $A_1$ and $A_2$ in the resonance region are based on 
a combination of the deep inelastic scattering (DIS) fit with extensions into the resonance region. A parametrized fit 
from MAID is used to get the resonance behavior. All JLab data as of the early 2010’s, as well as some MIT BATES 
and NIKHEF data, were used in determining the parametrization.  Sum rules are obeyed as much as possible (such as the Burkhardt-Cottingham sum 
rule~\cite{Burkhardt:1970ti} and 
the Soffer inequality~\cite{Soffer:1994ww}).
\item The asymmetry $A_1$ in the DIS region was based on a fit to world data on inclusive spin structure functions in the DIS region (COMPASS, SLAC, HERMES, and JLab early 2010’s)~\cite{SpinMuon:1993gcv,SpinMuonSMC:1997voo,SpinMuonSMC:1994met,SpinMuon:1995svc,SpinMuonSMC:1997ixm,SpinMuonSMC:1997mkb,HERMES:1997hjr,HERMES:2006jyl,E142:1996thl,E143:1994vcg,E143:1995rkd,E143:1995clm,E143:1998hbs,E154:1997xfa,E155:1999pwm,E155:2000qdr,COMPASS:2010wkz,COMPASS:2017hef,CLAS:2017qga,CLAS:2008xos,CLAS:2015otq,JeffersonLabHallA:2004tea}. 
The measured $A_1(x,Q^2)$ values were fit, with simple Regge (powers of $x$) behavior assumed as $x\to 0$. 
The full uncertainty matrix of the fit 
was used to evaluate the systematic uncertainties.
 \item The asymmetry $A_2$ in the DIS region was determined from 
$g_2^{WW}$~\cite{Wandzura:1977qf}. 
To evaluate systematic uncertainties, we added an extra twist-3 term obtained by 
fitting to the SLAC E155x~\cite{E155:2002iec} data. 
\end{itemize}

After establishing the {\it standard} simulation as above, 
we varied $A_1$ by an arbitrary value, $+0.1$ or $-0.1$, for all inelastic $W$ values, and generated the difference in the polarized cross section: 
\begin{eqnarray}
\Delta\sigma_\mathrm{diff} &=& 
\Delta\sigma_{\mathrm{non-std}(A_1\mathrm{changed~by~}\pm 0.1) }- \Delta\sigma_\mathrm{std}.\label{eq:dsig_diff}
\end{eqnarray}
After applying the same normalization factor as the {\it standard} simulation, 
the resulting simulated event count was denoted $\Delta n^{\q{diff}}$ and was 
added to the {\it standard} simulation $\Delta n^{\q{sim0}}$ to form the {\it non-standard} 
spectra $\Delta n^{\q{sim1}}$, which corresponds to the model with $A_1$ changed by $\pm 0.1$:
\begin{eqnarray}
\Delta n_\mathrm{diff} &=& 
\Delta n_{\mathrm{non-std}(A_1\mathrm{changed~by~}\pm 0.1) }- \Delta n_\mathrm{std}.\label{eq:dn_diff}
\end{eqnarray}
The polarized yield obtained from data, $\Delta n^{\q{data}}$, would nominally agree with neither the 
{\it standard} $\Delta n^{\q{sim0}}$ nor the {\it non-standard} $\Delta n^{\q{sim1}}$, since neither model would match our data perfectly. 

By comparing the polarized yield from our data ($\Delta n^{\q{data}}$) 
with that from the {\it standard} simulation ($\Delta n^{\q{sim0}}$) and 
the {\it non-standard} simulation ($\Delta n^{\q{sim1}}$), 
the value for $g_1$ was extracted as: 
\begin{eqnarray}
  g_1^\q{data} &=& g_1^{\q{sim0}}+(g_1^{\q{sim1}}-g_1^{\q{sim0}})\frac{\Delta n^{\q{data}}-\Delta n^{\q{sim0}}}{\Delta n^{\q{sim1}}-\Delta n^{\q{sim0}}} \\
&=& g_1^{\q{sim0}}+(g_1^{\q{sim1}}-g_1^{\q{sim0}})\frac{\Delta n^{\q{data}}-\Delta n^{\q{sim0}}}{\Delta n^{\q{diff}}}~,\label{eq:g1_extract} 
  \end{eqnarray}
where $g_1^{\q{sim0}}$ and $g_1^{\q{sim1}}$ are the $g_1$ values used in the {\it standard} 
and the {\it non-standard} simulations, respectively. 
Similarly, the value for $A_1F_1$ was extracted as:
\begin{eqnarray}
  (A_1F_1)^\q{data} &=& (A_1F_1)^{\q{sim0}}+\nonumber\\
  &&\hspace*{-2cm}\left[(A_1F_1)^{\q{sim1}}-(A_1F_1)^{\q{sim0}}\right]\frac{\Delta n^{\q{data}}-\Delta n^{\q{sim0}}}{\Delta n^{\q{diff}}}~.\label{eq:a1f1_extract} 
  \end{eqnarray}

The ratios of ${\Delta n^{\q{diff}}}/(g_1^{\q{sim1}}-g_1^{\q{sim0}})$ and 
${\Delta n^{\q{diff}}}/[(A_1F_1)^{\q{sim1}}-(A_1F_1)^{\q{sim0}}]$ on the RHS of Eqs.~(\ref{eq:g1_extract})
and~(\ref{eq:a1f1_extract}) 
describe how the yield varies with the input $g_1$ or $A_1F_1$ of the model, and directly affect the 
extraction of these structure functions.  These ratios were sensitive to the statistical fluctuations 
in $\Delta n^{\q{diff}}$, which can be large if small $W$ bins are used.  In our analysis, these ratios were 
smoothed out by taking a 5-point average of adjacent $W$ bins.

We emphasize that our method of extracting $g_1$
and $A_1 F_1$ described above does not depend strongly on
the specific models used for the Monte Carlo generator, since a
larger deviation of the model from the data would simply lead to a larger ``correction'' in 
Eqs.~(\ref{eq:g1_extract}-\ref{eq:a1f1_extract}), 
bringing the result to the same final value.

\subsection{Systematic Uncertainty due to Models Used In the Extraction}\label{sec:extraction_syst}
In the extraction of $g_1$ or $A_1 F_1$ described above, there can be systematic uncertainties 
due to the models used for $F_{1,2}$, $R$, $A_1$, and in particular, for $A_2$, because only longitudinally polarized targets were used in this experiment. 
We evaluated the model dependence as follows: a simulation corresponding to a variation of the input structure function was performed, following the same method as the {\it non-standard} simulation (where $A_1$ is changed by $\pm 0.1$). This variation of simulation gave $\Delta n^{\q{var}}$, with ``var'' for ``variation in the input model''. The value for $g_1$ with this model variation was also 
calculated, denoted as $g_1^\q{var}$. 
The systematic uncertainty in the extracted $g_1$ was calculated as:
\begin{eqnarray}
  \Delta g_1^{\q{syst}} &=& \left[\frac{(g_1^{\q{sim1}}-g_1^{\q{sim0}})}{\Delta n^{\q{diff}}}\right] \Delta n^{\q{var}}
 - (g_1^{\q{var}}-g_1^{\q{sim0}})\label{eq:dg1_var}~, ~~
\end{eqnarray}
and similarly for $A_1F_1$: 
\begin{eqnarray}
  \Delta (A_1F_1)^{\q{syst}} &=& \left[\frac{((A_1F_1)^{\q{sim1}}-(A_1F_1)^{\q{sim0}})}{\Delta n^{\q{diff}}}\right] \Delta n^{\q{var}}
 \nonumber\\
  &&\hspace*{-1cm}- \left[(A_1F_1)^\q{var}-(A_1F_1)^{\q{sim0}}\right]\label{eq:da1f1_var}~, 
\end{eqnarray}
where $\left[\frac{(g_1^{\q{sim1}}-g_1^{\q{sim0}})}{\Delta n^{\q{diff}}}\right]$ 
and $\left[\frac{((A_1F_1)^{\q{sim1}}-(A_1F_1)^{\q{sim0}})}{\Delta n^{\q{diff}}}\right]$ are the same 
values as those on the RHS of Eqs.~(\ref{eq:g1_extract}-\ref{eq:a1f1_extract}), 
extracted from comparison between the standard vs. the non-standard simulations.
A total of six variations was performed:
\begin{itemize}
 \item Variation 1: Both $F_1$ and $F_2$ were increased by 3\%; 
 \item Variation 2: $R$ was changed by $R\to R+dR$ with $dR$ the uncertainty of $R$ given by Refs.~\cite{Bosted:2007xd} and \cite{Christy:2007ve};
 \item Variation 3: The value of $A_1$ in the resonance region was changed to the fit before including the EG1b data ($A_2$ unchanged);
 \item Variation 4: The value of $A_2$ in the resonance was changed to the fit before 2009 ($A_1$ unchanged);
 \item Variation 5: The value of $A_1$ in the DIS region was changed by the uncertainty of the DIS fit used in the the {\it standard} version. The full uncertainty matrix of the fit is used. 
 \item Variation 6: The value of $A_2$ was changed by adding an extra twist-3 term, obtained by fitting to the SLAC E155x data~\cite{E155:2002iec}.
\end{itemize}

Note that the systematic uncertainty evaluated this way can be either positive or negative.
When integrating the measured $g_1$ or $A_1F_1$ for the moments, the sign of 
the systematic uncertainty was retained and there can be cancellations. 
On the other hand, when combining these variations to form the total systematic uncertainty, 
the uncertainty from each variation was added in quadrature for $g_1$, $A_1F_1$, and for the moments. 
Also note that model uncertainties affect both the extraction of $g_1$ and $A_1F_1$ directly 
as well as indirectly through radiative corrections that are based on these models. Our method 
accounted properly for correlations between these two effects.

\subsection{Contamination from Polarized Nitrogen}\label{sec:syst_n15}
The $^{15}$N in the target was slightly polarized, and scattering from the $^{15}$N can contribute to the polarized count difference measured in the data.  The $^{15}$N contribution had three effects: contribution of $^{15}$N nuclear elastic scattering to the proton elastic peak; contribution of radiative tail from $^{15}$N nuclear elastic scattering to proton inelastic scattering; and contribution from $^{15}$N inelastic scattering to proton inelastic scattering. The first affected normalization of the elastic simulation to the data, while the latter two affected the extraction of the structure functions from the inelastic polarized yield.

To account for each contribution from the $^{15}$N background, one first needs to obtain the polarization of $^{15}$N as follows: the ratio of $^{15}$N polarization over that of the proton is about $16\%$~\cite{Crabb:1995xi}, and there are three protons for every $^{15}$N nucleus in NH$_3$. Therefore, the $^{15}$N contamination to either the inelastic or elastic polarized yield can be obtained by multiplying the calculated $^{15}$N unpolarized 
yield by the effective polarization ratio of $^{15}$N$/p = 16\%/3 \approx 5\%$. 

The contribution from $^{15}$N nuclear elastic scattering was calculated using the $^{15}$N form factors~\cite{Dally:1970lvj}, and was found to be problematic only in the low $Q^2$ bins. This is because the invariant mass $W$ for the $^{15}$N nuclear elastic scattering, calculated using Eq.~(\ref{eq:Wdef_fixedtarg}), is separated from that of the proton elastic peak ($W=M$) by:
\begin{eqnarray}
 \Delta W &=& \frac{1}{2}\frac{Q^2}{M^2}\left(\frac{M}{M_{15N}}-1\right)M <0~,
\end{eqnarray}
where $M_{15N}$ is the $^{15}$N nuclear mass. 
Therefore in the $W$ spectra, the separation the between $^{15}$N nuclear elastic and proton elastic peaks was large enough for the two peaks to be distinguished from each other in all data except the very low $Q^2$ bins. The contamination from $^{15}$N nuclear elastic scattering to the elastic normalization was kept below $1\%$ by excluding elastic data below $Q^2=0.0379$~GeV$^2$ from the elastic normalization.

The contribution of radiative tails from $^{15}$N nuclear elastic scattering to the inelastic polarized yield was calculated and found to be no more than 2\% of the proton elastic tail itself. Similarly, the relative contribution from $^{15}$N inelastic scattering to the proton inelastic polarized yield difference was found to be below 0.7\%. We included both effects in the systematic uncertainty evaluation.

In the case of the deuteron, the possible corrections due to other components in the target
were somewhat larger, since the deuteron polarization
is typically a factor 2 smaller than that of protons.
In addition to the contribution from $^{15}$N, there is
also the possibility of contamination from $^{14}$N 
and ordinary hydrogen in the target. A detailed study
of possible corrections from these contributions showed
a maximum effect of 4.5\% which we included in the
systematic uncertainty budget for the deuteron results.

\subsection{Summary of Systematic Uncertainties}~\label{sec:results_allsyst}
In the following, we list all systematic uncertainties that enter our final results:

\begin{figure*}[!htp]
\includegraphics[width=0.9\textwidth]{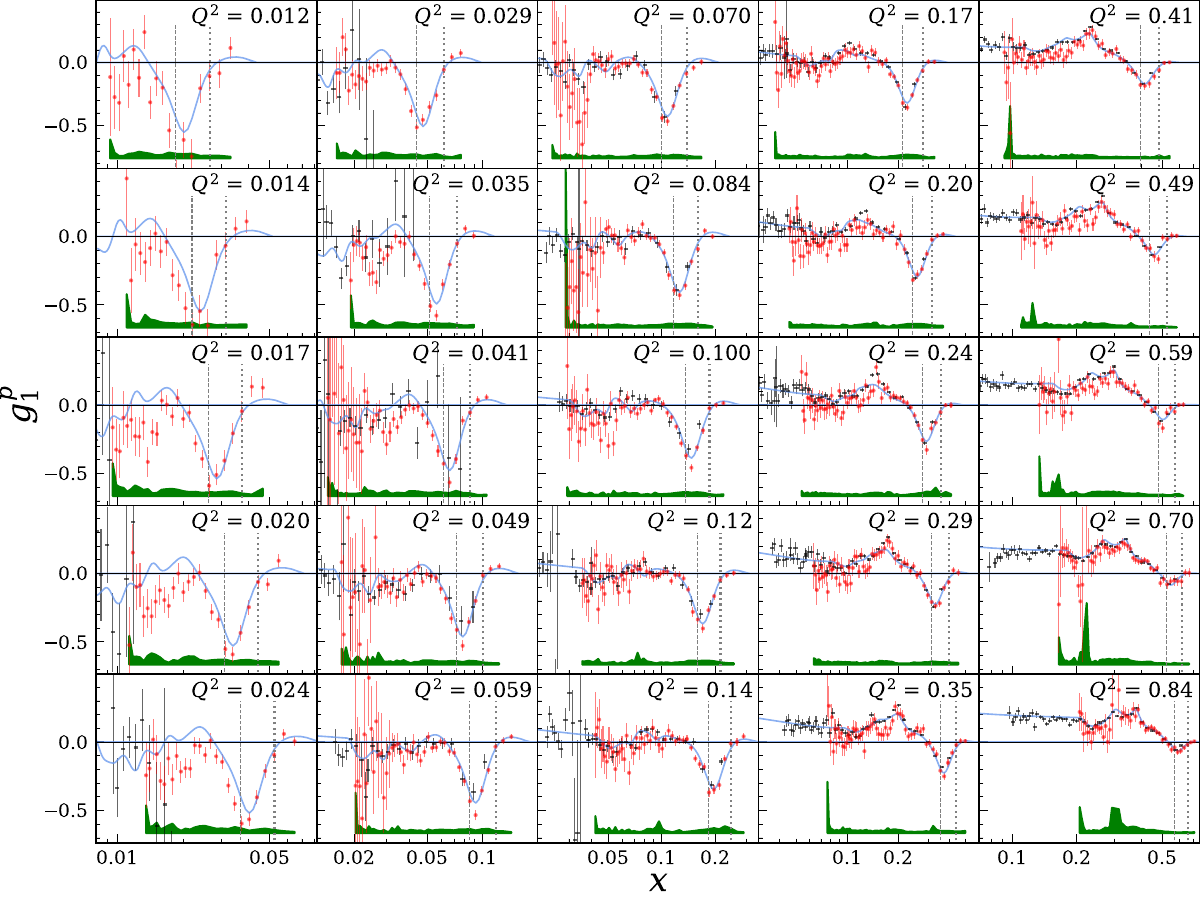}
\caption{[Color online] 
Results on $g_1$ of the proton (red solid circles) plotted vs. $x$,
compared with our standard parametrization 
(blue curves) and results from EG1b (black crossbars)~\cite{CLAS:2017qga}. 
The error bars are statistical only. The total systematic uncertainty is shown as the green band on the bottom edge of each panel and includes both experimental systematics and that from models used in the analysis. 
The size of the total systematic uncertainty varies from $0.013-0.15$ to $0.0045-0.19$, and the relative size of the total systematic to statistical uncertainties varies from $(8.0-33)\%$ to $(16-140)\%$, from the lowest to the highest $Q^2$ bins. 
The two vertical lines in each panel indicate the location of the $\Delta(1232)$ and the integration limit $x_\mathrm{hi}$ that corresponds to $W=1.15$~GeV, see Section~\ref{sec:results_moments_procedure}. Note that the $Q^2$ bins (in GeV$^2$) are ordered column-wise, first from top to bottom, then from left to right.  
}\label{fig:g1p_vs_W}  
\end{figure*}

\begin{enumerate}
   
 \item Statistical uncertainty from the simulation that enters 
Eqs.~(\ref{eq:g1_extract}) and~(\ref{eq:a1f1_extract}). This is typically negligible compared to 
data statistics and is not included in the total systematic uncertainty.
 
 \item Model uncertainties, analyzed by using the model variations described in 
Section~\ref{sec:extraction_syst}. 

\item Systematic uncertainty due to the radiative tail from proton elastic scattering and
 from radiative corrections included in our Monte Carlo simulation. There are two potential contributions to
 this uncertainty: the models for the form factors and structure functions that enter these tails 
 and radiative corrections, and the amount of target material traversed by the beam, which affects the {\em external} radiative tails. The former are ``automatically'' accounted for
 by our study of the overall systematic effect of using different structure function models, see
 Section~\ref{sec:extraction_syst}, and by trying different form factor parametrizations
 (see Section~\ref{sec:elsim_comp1}). Note that we modified those model inputs {\em simultaneously} in the entire analysis chain, and retained the sign and magnitude of all variations to propagate them to the final results for the structure functions and moments, to properly account for correlations. The second uncertainty comes mostly from the uncertainty
 of the overall areal density of the target material, which in turn is determined by the 
 packing fraction.
 The packing fraction (PF) is defined as the ratio of the actual quantity of ammonia target material contained in the target cell to the ideal quantity for a complete filling of the
 cell volume. While PF is not explicit in Eq.~(\ref{eq:analysis0}), it affects the total target thickness and thus the radiative effects in the simulation. 
 To obtain the uncertainty due to PF, we re-ran the full simulation using a set of ``alternate PF values'' that differ from the nominal PF values. The data were re-analyzed using the alternate PF simulations and $g_1$ and $A_1F_1$ were extracted. The differences in the structure function results and those obtained using nominal PF values were taken as the uncertainty due to PF values. The uncertainty in the PF values themselves is taken to be 0.1. 
The nominal PF values used in the  
analysis are shown in Table~\ref{tab:pf}.\\

\begin{table}[!h]
\begin{center}
\begin{tabular}{|c|c|c|}\hline
 Beam Energy & Target Cell & PF Value \\\hline
 \multirow{2}{*}
  {1.1~GeV} & long NH$_3$ top   &  0.717  \\
            & long NH$_3$ bottom & 0.625 \\ \hline
             
  \multirow{3}{*} 
  {1.3~GeV}  & long NH$_3$ bottom &  0.657 \\
             & short NH$_3$       &  0.602 \\
             & ND$_3$                & 0.624 \\\hline
 \multirow{2}{*}
  {2.0~GeV}   & long NH$_3$ top   & 0.716  \\
              & ND$_3$         & 0.764 \\ \hline
  \multirow{2}{*} 
  {2.3~GeV}   & long NH$_3$ top     & 0.682  \\
              & short NH$_3$         & 0.720  \\\hline
     3.0~GeV  & long NH$_3$ top    & 0.782  \\ \hline
\end{tabular}
\end{center}
\caption{Nominal packing factor (PF) values used in the analysis for each combination of target type and beam energy. See Fig.~\ref{fig:targinsert} for the detailed naming convention of the target cells. 
}\label{tab:pf}
\end{table}

 \item Systematic uncertainty due to normalizing the elastic peak of simulation to data. This includes: 
an uncertainty of $(1+E_b)\%$ (with $E_b$ the beam energy in GeV) (see Section~\ref{sec:elsim_comp1})
and the statistical uncertainty of the data elastic or quasi-elastic peak (Fig.~\ref{fig:el_ratio_sum}).

 \item Systematic uncertainty due to CC efficiency. This was obtained by re-running the analysis without
applying the CC efficiency to the simulated spectra. The difference in the extracted $g_1$ and $A_1F_1$ from 
the nominal analysis indicates the effect of the CC efficiency, and 20\% of this difference was taken 
as the systematic uncertainty due to the CC efficiency. Details of CC efficiency analysis were 
described in Section~\ref{sec:ana_pid}.

 \item Systematic uncertainty due to background contributions. This includes effect from 
pion and pair production background (1\% of the polarized yield, see Section~\ref{sec:ana_bg}); 
$^{15}$N elastic tail in the inelastic region (see Section~\ref{sec:syst_n15}); and the 
$^{15}$N inelastic scattering background in the inelastic polarized yield, which contributes 
0.7\% of the polarized yield estimated by calculating the $^{15}$N to proton inelastic polarized 
cross section ratio and applying the polarization ratio of the two.

 \item Systematic uncertainty due to event reconstruction. Even with momentum corrections applied to both data 
and simulation, the simulated elastic peak can differ from data in its $W$ position by a small amount 
(typically 1-2~MeV or smaller). To estimate the uncertainty due to this effect, we shifted the simulated $W$ spectra
by $+5$ or $-5$~MeV, and re-ran the analysis. The difference in the extracted $g_1$ and $A_1F_1$ from 
the nominal analysis was then scaled by the actual shift in the elastic peak, obtained from figures such 
as Fig.~\ref{fig:el_ratio_sum}. This uncertainty is typically very small except near the edge of the 
acceptance. When evaluating the integral of the structure functions, the data in the single $W$ bin on the edge are 
not used because of this reason. Another reason to exclude data from the edges is that they 
very likely do not cover the full $W$ range of the bin.

\end{enumerate}

\section{Results on the Structure Functions}\label{sec:results_g1}

We present in this section results on the proton and the deuteron structure functions $g_1$ and $A_1F_1$, which were reported previously in Refs.~\cite{CLAS:2017ozc,CLAS:2021apd}. We also present new results on the neutron $g_1^n$ as extracted from the proton and deuteron data, as well as their isovector combination $g_1^{p-n}$.

\subsection{Results on $g_1$ and $A_1F_1$ for the Proton and Deuteron }\label{sec:results_pd}

\begin{figure*}[!ht]
\centering
\includegraphics[width=0.9\textwidth]{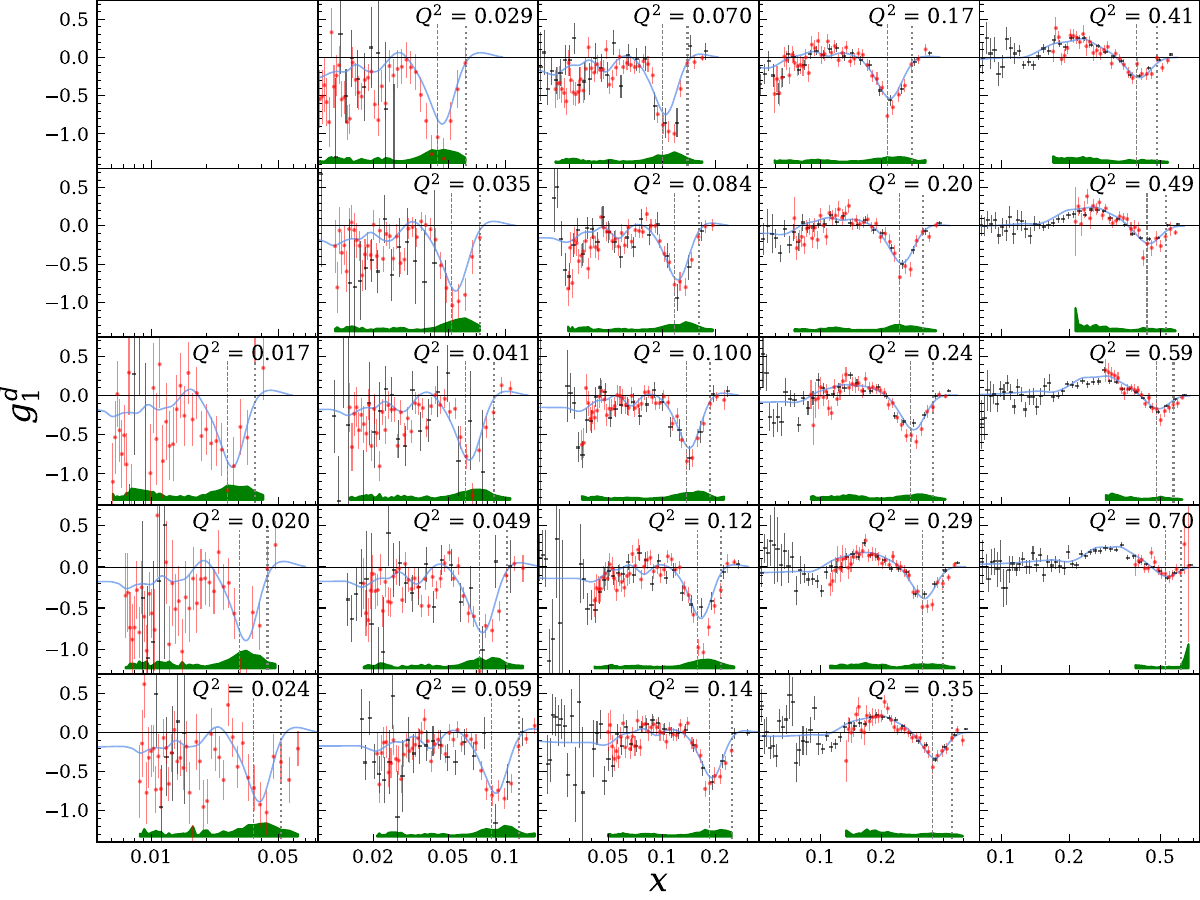}
  \caption{[Color online] Results on $g_1$ of the deuteron. The size of the total systematic uncertainty varies from $0.020-0.183$ to $0.010-0.294$, and the relative size of the systematic to statistical uncertainty varies from $(4.3-52)\%$ to $(16-55)\%$, from the lowest to the highest $Q^2$ bins. See Fig.~\ref{fig:g1p_vs_W} caption for more details.
}\label{fig:g1D_vs_W}  
\end{figure*}

In the previous section, we described the method of extracting $g_1$ and $A_1F_1$ and the determination of systematic uncertainties on our results. The $g_1$ and $A_1F_1$ results obtained from all target types and beam energies were then combined.
When doing so, we first checked whether any two data sets were compatible with each other, using Student's t-test to compare the distributions of both data sets in the region of common kinematics. The data were found to be consistent. 

When combining the five beam energies, the spin structure function and its statistical uncertainty were determined as
\begin{eqnarray}
  g_1^\mathrm{combined} &=& \frac{\sum_{i}\frac{g_{1,E_b^i}}{\delta g_{1,E_b^i,stat}^2}}
         {\sum_{i}\frac{1}{\delta g_{1,E_b^i,stat}^2}},\\
  \delta g_{1,stat}^\mathrm{combined} &=& \sqrt{\frac{1}{{\sum_{i}\frac{1}{\delta g_{1,E_b^i,stat}^2}}}}~,
\end{eqnarray}
where $g_{1,E_b^i}$ and $\delta g_{1,E_b^i,stat}$ are the $g_1$ value and its statistical uncertainty
obtained from beam energy $E_b^i$ ($i=1,2,3,4,5$). The systematic uncertainty was calculated as
\begin{eqnarray}
  \delta g_{1,syst}^\mathrm{combined} = \frac{\sum_{i}\frac{\delta g_{1,E_b^i,syst}}{\delta g_{1,E_b^i,stat}^2}}
         {\sum_{i}\frac{1}{\delta g_{1,E_b^i,stat}^2}}~, 
\end{eqnarray}
where $\delta g_{1,E_b^i,syst}$ is the systematic uncertainty 
obtained for the data taken with beam energy $E_b^i$. That is, both experimental systematic and model uncertainties were weighted by the relevant statistical uncertainties of each beam energy.  The same method was applied when combining data from different target cell types. 

\begin{figure*}[!hp]
  \includegraphics[width=0.76\textwidth]{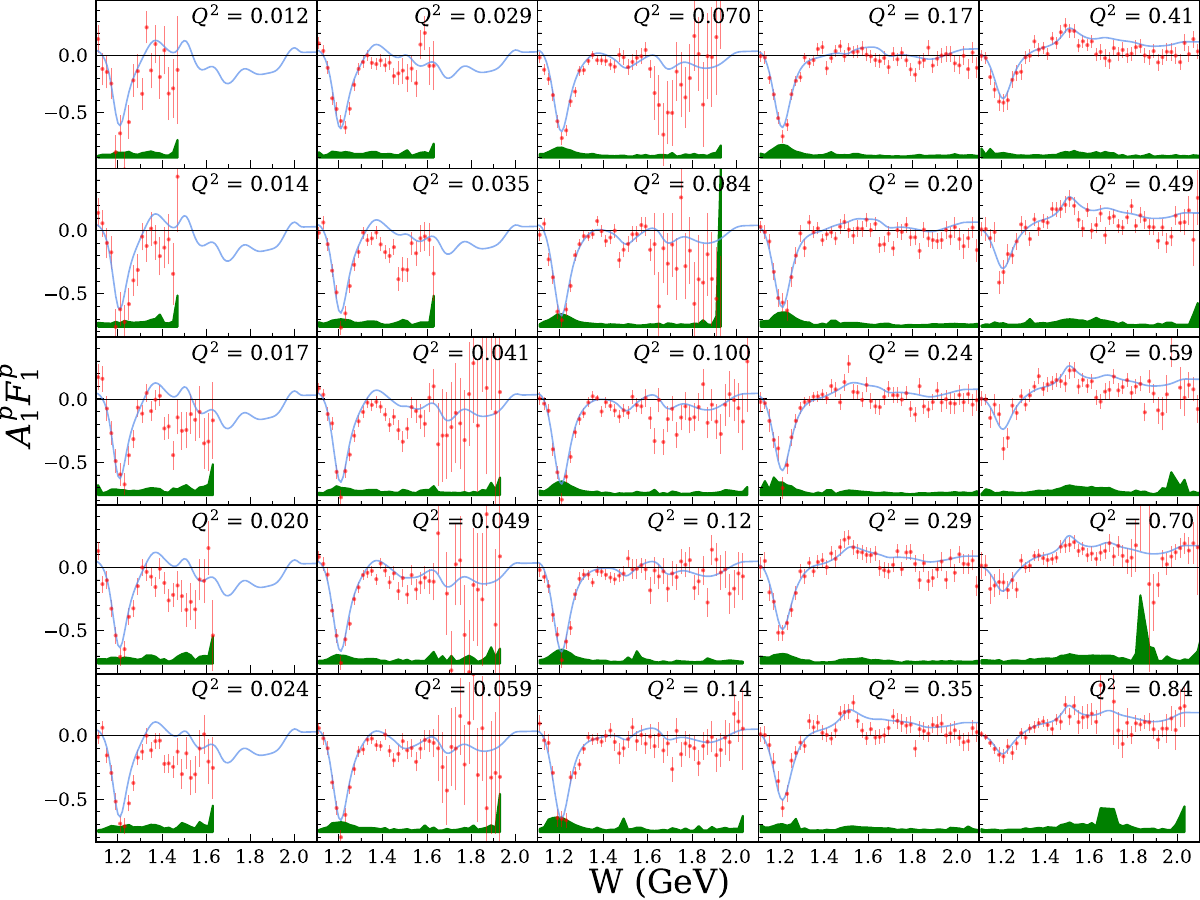}
\caption{[Color online] Results on $A_1F_1$ of the proton plotted vs. $W$ to emphasize the contribution of different resonance regions. The size of the total systematic uncertainty varies from $0.0134-0.151$ to $0.0045-0.190$, and the relative size of the systematic to statistical uncertainty varies from $(8.2-33)\%$ to $(8.0-118)\%$, from the lowest to the highest $Q^2$ bins. Note that earlier CLAS data~\cite{CLAS:2017qga} are not available for this quantity. See Fig.~\ref{fig:g1p_vs_W} caption for more details. 
}\label{fig:a1f1p_vs_W}  
  \vspace{\floatsep}
  \includegraphics[width=0.76\textwidth]{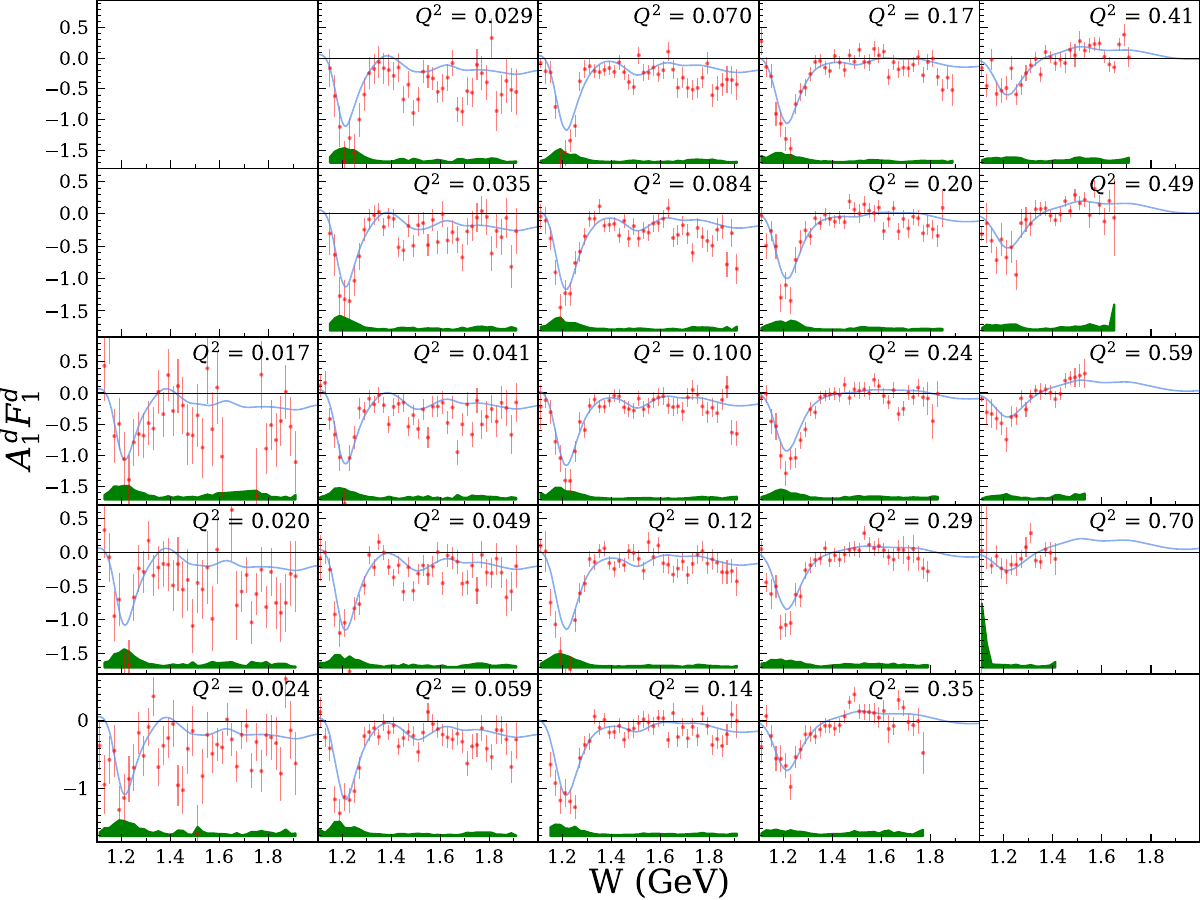}
  \caption{[Color online] Results on $A_1F_1$ of the deuteron plotted vs. $W$ to emphasize the contribution of different resonance regions. The size of the total systematic uncertainty varies from $0.020-0.22$ to $0.019-0.952$, and the relative size of the systematic to statistical uncertainty varies from $(4.3-55)\%$ in the lowest Q2 bin, to $(18-39)\%$, from the lowest to the highest $Q^2$ bins. Note that earlier CLAS data~\cite{CLAS:2017qga} are not available for this quantity. See Fig.~\ref{fig:g1p_vs_W} caption for more details. 
  }\label{fig:a1f1d_vs_W}
\end{figure*}

Our final results for $g_1$ and $A_1F_1$ of the
proton and the deuteron are shown in Figs.~\ref{fig:g1p_vs_W} through~\ref{fig:a1f1d_vs_W}. 
Our results are consistent with the previous data~\cite{CLAS:2015otq,CLAS:2017qga} 
where their coverage overlap, but extend the measured $Q^2$ range to three times smaller values, below the pion mass squared ($m_\pi^2$). This makes it possible to rigorously test $\chi$EFT calculations for the nucleon spin structure functions. Comparing to our existing $A_1$ model, which was based on existing data prior to EG4, one can see that it tends to be more positive than the data points, especially at lower $Q^2$. Our new data can help improve this model.

\subsection{Results on neutron $g_1$}\label{sec:results_g1n}

\begin{figure*}[!ht]
  \includegraphics[width=0.8\textwidth]
  {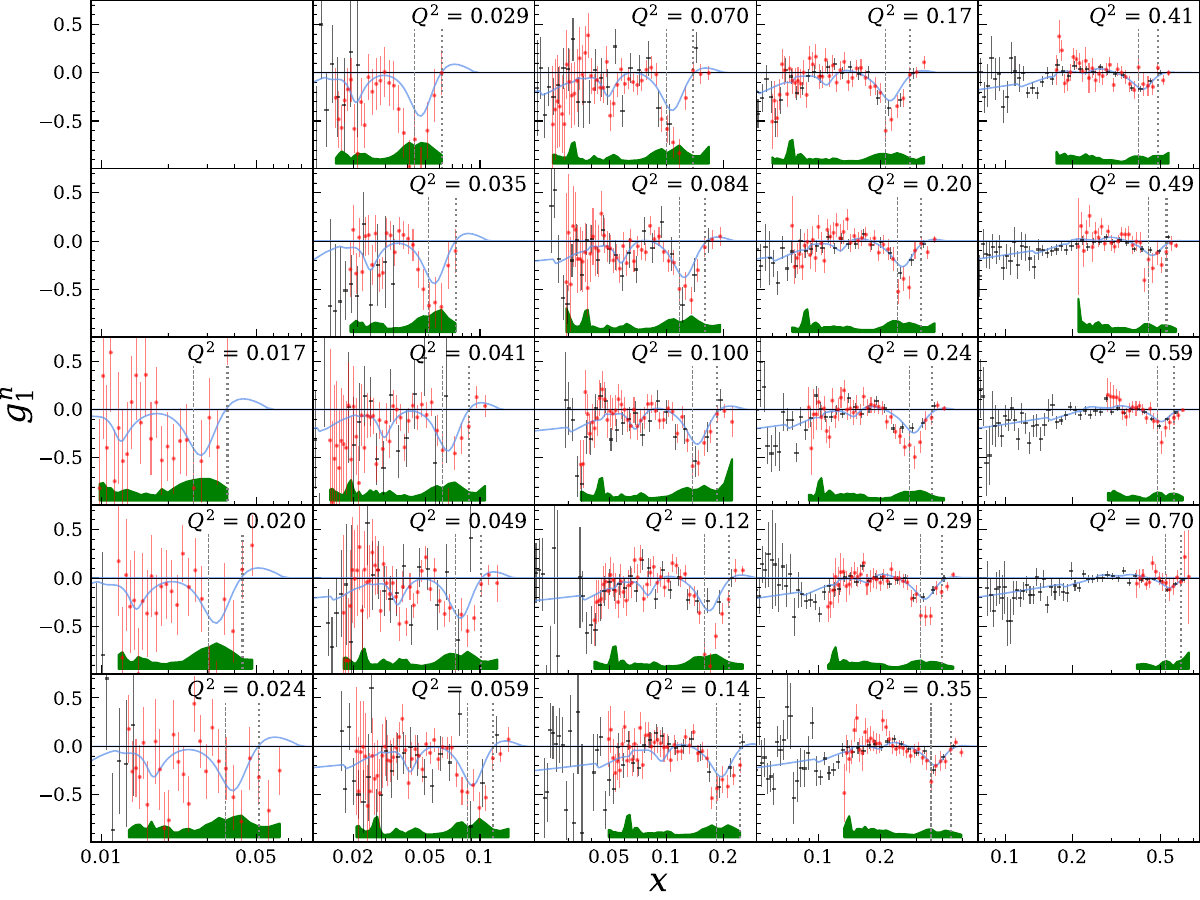}
\caption{[Color online] 
Results on $g_1$ of the neutron (red solid circles) vs. $x$ extracted from deuteron and proton data. These are 
compared with our standard parametrization with the best $A_1$ model (blue curves) and results from EG1b (black crossbars)~\cite{CLAS:2017qga}. 
The error bars are statistical only. The total systematic uncertainty is shown as the green band on the bottom edge of each panel that includes both experimental systematics and that from models used in the analysis. The size of the total systematic uncertainty varies from $0.040-0.305$ to $0.023-0.160$, and the relative size of the systematic to statistical uncertainty varies from $(10.7-55)\%$ to $(32-170)\%$, from the lowest to the highest $Q^2$ bins. 
}\label{fig:g1n_vs_W}  
\end{figure*}

From our results on $g_1$ of the proton and the deuteron, we further extracted values for the neutron spin structure function, $g_1^n$, which were not available previously. 
This extraction is based on the simplified
model of the deuteron structure functions described in Section~\ref{sec:intro_deuteron}, where the structure functions of the deuteron are expressed as sum of the two nucleon structure functions, convoluted with their spin and momentum distributions in the deuteron~\cite{Kahn:2008nq}. To extract the deconvoluted structure functions for a neutron at rest, we followed a similar procedure as for the extraction of $g_1^{p,d}$ from the data on the proton and deuteron, described in detail below. 
 
As a first step, we  took our $g_1^p$ results and converted them to the contribution from the bound proton to the deuteron spin structure function using the folding technique of Ref.~\cite{Kahn:2008nq}. The bound $g_1^p$ were then subtracted from our $g_1^d$ results to obtain $g_1$ data for the bound neutron $g_1^{n, \q{data, bound}}$. 
The data $g_1^{n, \q{data, bound}}$ were compared with expected values for the bound neutron, $g_1^{n, \q{model, bound}}$, produced using our model of the free neutron processed by the same folding technique, and any observed difference is attributed to how the input $g_1^n$ model of the free neutron differs from reality.  We then produced a {\it modification} of the base model $g_1^{n, \q{model-a(b), bound}}$ where $A_1$ was changed by $+0.1$ ($-0.1$) in the input, and the difference between the {\it modified} and the original model provides information on how much the $g_1^n$ model of the free neutron should be adjusted to match data.  More specifically, $g_1$ for the free neutron is extracted as: 
\begin{eqnarray}
 g_1^{n,\q{(data, free)}} &=& g_1^{n, \q{model, free}}\nonumber\\
 && + (g_1^{n, \q{model-a(b), free}}-g_1^{n, \q{model, free}}) \label{eq:g1n}\\
 &&\times\frac{g_1^{n, \q{data, bound}}-g_1^{n, \q{model, bound}}}{g_1^{n, \q{model-a(b), bound}}-g_1^{n, \q{model, bound}}}~.\nonumber
\end{eqnarray}
Note that the two versions ($a,b$) of the modification produce nearly identical results. The statistical and experimental systematic uncertainties of each of the input $g_1^p$ and $g_1^d$ were propagated to produce the experimental uncertainties in the extracted $g_1^n$. 
The uncertainty in this extraction was studied using a similar method as in Section~\ref{sec:extraction_syst}, but now by changing the initial model for the neutron. 

\begin{figure*}[!ht]
  \includegraphics[width=0.8\textwidth]%
  {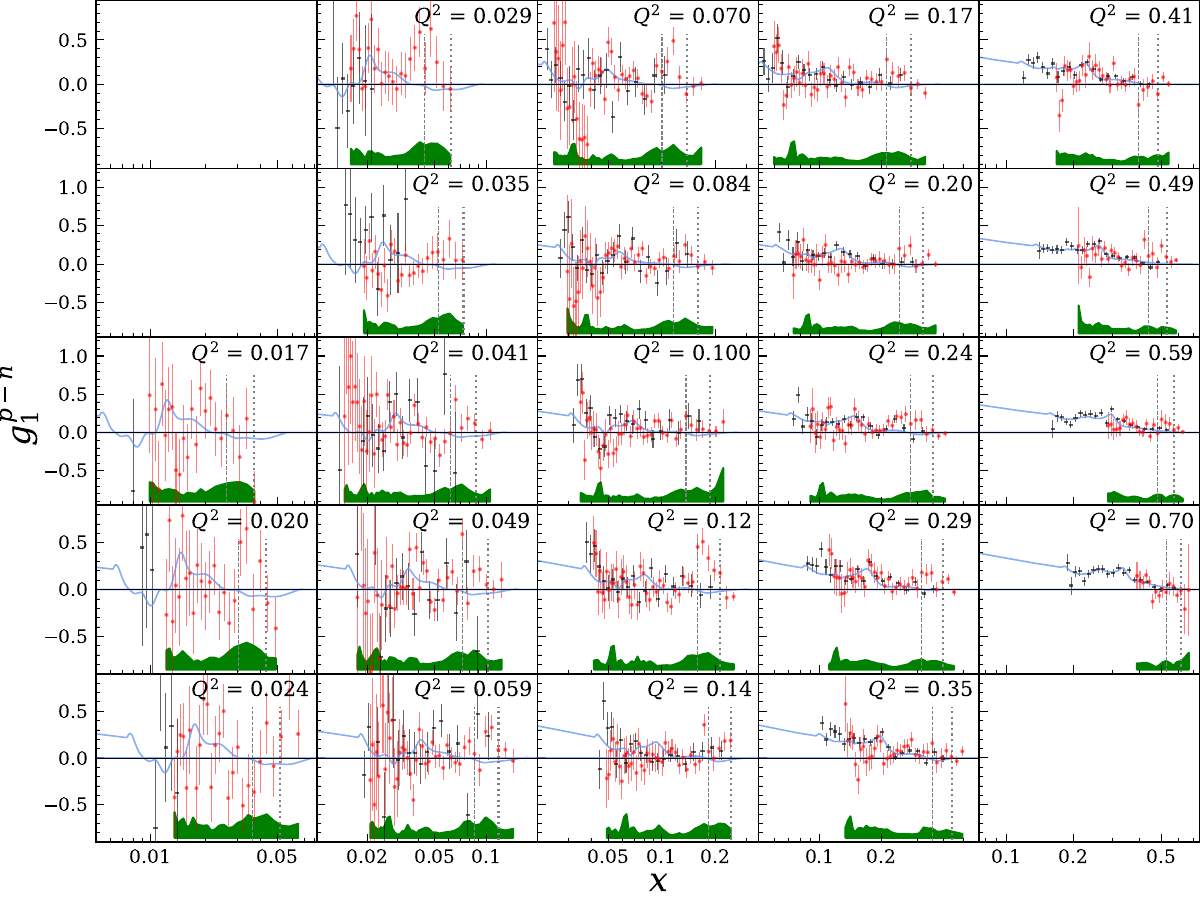}
\caption{[Color online] Results on the isovector component $g_1^{p-n}$. Across all $Q^2$ bins, the size of the total systematic uncertainty varies from $0.022$ to $1.013$, while the relative size of the total systematic to statistical uncertainty varies within $(7.7-173)\%$. See Fig.~\ref{fig:g1n_vs_W} caption for more details.
}\label{fig:g1pn_vs_W}  
\end{figure*}

Our results for $g_1^n$ are shown in Fig.~\ref{fig:g1n_vs_W}. 
Similar to the proton and the deuteron case, we see that the new results from EG4 are consistent with previous results in the region where they overlap, but the best $A_1$ model can be improved in the lowest $Q^2$ bins, below about $0.1$~GeV$^2$.

Finally, we show the isovector combination $g_1^{p-n}(x,Q^2)$ in Fig.~\ref{fig:g1pn_vs_W}. Comparing with Figs.\ref{fig:g1p_vs_W}, \ref{fig:g1D_vs_W}, and \ref{fig:g1n_vs_W}, one clearly observes the suppression in the isovector component of the $\Delta^{(1232)}$.  
This has important consequences for the Bjorken integral ({\it i.e.}, the first moment of $g_1^{p-n}(x,Q^2)$), see Section~\ref{isospin-moments}, and its comparison to $\chi$EFT~\cite{Burkert:2000qm}. 
Like the other EG4 results on $g_1$ and $A_1F_1$, Fig.~\ref{fig:g1pn_vs_W} also makes clear that in the $(x,Q^2)$ domain covered by EG4, the parameterization of the world data (continuous blue line) needs to be revised.

\section{Results on Moments of Structure Functions}\label{sec:results_moments}
Similar to the previous section(s), we summarize the procedure to form the moments of the proton and the deuteron structure functions, and present the results that were reported previously in Refs.~\cite{CLAS:2017ozc,CLAS:2021apd} but with updated $\chi$PT calculations. We also present new results on the moments of the neutron extracted directly from the moments of the proton and the deuteron, as well as their isospin combinations $\overline\Gamma_1^{p-n}$ and $\overline\gamma_0^{p\mp n}$.

\subsection{General Procedure Used to Form the Integrals}\label{sec:results_moments_procedure}
To form moments of $g_1$ and $A_1F_1$, we integrated these quantities over the full range of $10^{-3} \le x \le x_{th}$ ($x_{th}$ corresponds to the electroproduction threshold), using our model for the low and high $x$ region beyond the data coverage.  
In addition, for a few high $Q^2$ bins the coverage of our data has gaps at intermediate $W$ values because of disabled DC wires. The integrand over these gap regions was provided by the model as well. 
The full integral, {\it e.g.}, $\overline \Gamma_1 (Q^2)$ was therefore
evaluated as:
\begin{eqnarray}
  \overline \Gamma_1(Q^2) &=& \int_{0.001}^{x_\mathrm{lo}} g_1^\mathrm{mod}(x,Q^2)dx \nonumber\\
  &&\hspace*{-1cm}+ 
      \int_{x_\mathrm{lo}}^{x_\mathrm{hi}} g_1^\mathrm{EG4~data}(x,Q^2)dx + 
      \int_{x_\mathrm{hi}}^{x_{th}} g_1^\mathrm{mod}(x,Q^2)dx  \nonumber\\
    & &\hspace*{-1cm} +  \int_\mathrm{gaps~(when~applicable)} g_1^\mathrm{mod}(x,Q^2)dx~,\label{eq:G1int_full}
\end{eqnarray}
where $x_\mathrm{lo}$ and $x_\mathrm{hi}$ are the lower and upper limits of the data integration, $g_1^\mathrm{mod}$ is the estimate from our best model described in Section~\ref{sec:extraction} and $g_1^\mathrm{EG4~data}$ is our data. The low-$x$ integration starts at $x=10^{-3}$ and the step size was set to $dx=10^{-5}$. 

The range $(x_\mathrm{lo},x_\mathrm{hi})$ of the data integration in Eq.~(\ref{eq:G1int_full}) is slightly narrower than the actual kinematic  coverage of the experiment because the data uncertainties on the edge of the coverage are large: 
At large $x$, near the pion threshold $W\approx 1.07$~GeV, the polarized yield is dominated by the radiative tail from elastic scattering and the measured polarized yield is not sensitive to the value of $g_1$ or $A_1F_1$; 
while for the low $x$ or large $W$ edge of the data, the spectrometer acceptance drops and the statistics of the data is low, see Section~\ref{sec:results_allsyst}. 
Thus, the value of $x_\mathrm{hi}$ was chosen to be at $W=1.15$~GeV and $x_\mathrm{lo}$ was set at a few tens of MeV below the maximum $W$ covered by the data. 
The upper $x$-limits of the data integration ranges 
are shown in Figs.~\ref{fig:g1p_vs_W} and~\ref{fig:g1D_vs_W}.

Uncertainties in the model enters the integral in two ways. First, the model enters the extraction of the data itself, 
as described in Section~\ref{sec:extraction_syst}. 
Second, the uncertainty of the model integration below $x_\mathrm{lo}$ and above $x_\mathrm{hi}$ was evaluated by variation of the model parameters, similar to the method described in Section~\ref{sec:extraction_syst}. 
When evaluating model uncertainties
of the data integral, the sign of each model uncertainty of $g_1$ or $A_1F_1$ is retained to allow partial cancellation when integrating. The total uncertainty on the full integral was formed by adding all experimental and 
model uncertainties in quadrature.

\subsection{Latest $\chi$EFT Moment Predictions}
In the following sections, we will compare the EG4 result on the moments to theoretical predictions, in particular from $\chi$EFT. We will display only two recent predictions, from Ref.~\cite{Bernard:2012hb}  and Ref.~\cite{Alarcon:2020icz}, as they supersede earlier $\chi$EFT calculations~\cite{Bernard:1992nz, Ji:1999pd, Bernard:2002bs, Kao:2002cp}. The crucial advance of the latest calculations is that they account rigorously for the $\Delta^{(1232)}$ excitation. 
At the same time, the approaches by Refs.~\cite{Bernard:2012hb}  and \cite{Alarcon:2020icz} have significant differences. The chief reason is the effective accounting of higher-order contributions of the $\chi$EFT series by Ref.~\cite{Alarcon:2020icz}, which uses a phenomenological form factor. It softens the steep $Q^2$-dependence otherwise seen in Ref.~\cite{Bernard:2012hb}. The softening form factor of Ref.~\cite{Alarcon:2020icz} accounts for their typically better description of the $Q^2$-dependence of moments. 
The methodological differences between the two calculations are detailed in Section~\ref{sec:Discussion}.

\subsection{Moments of the Proton Spin Structure functions}\label{sec:results_moments_p}
\begin{figure}[!h]
  \includegraphics[width=0.48\textwidth]{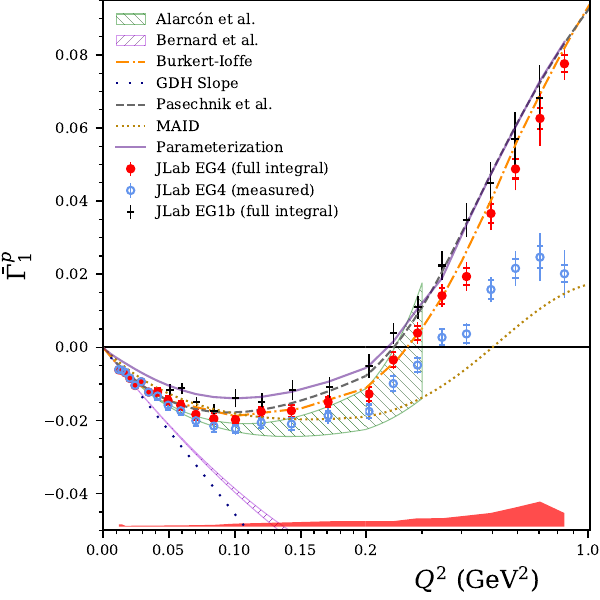}
\caption{[Color online] EG4 results on the proton $\overline \Gamma_1^p(Q^2)$. Integrals over the full (experimentally covered) $x$-range are shown as solid red (open blue) circles.
The inner and the outer error bars (sometimes too small to be seen) are for statistical and total uncertainties, respectively. 
To illustrate their relative contributions, we also show the systematic uncertainty as the (red colored) band at the bottom of the figure, which includes both experimental systematic and model uncertainties. 
Across all $Q^2$ bins, the size of the systematic uncertainty varies from $0.0002$ to $0.0068$, while the relative size of the systematic to statistical uncertainty varies within $(23-233)\%$. 
These results are compared with earlier data from CLAS~\cite{CLAS:2017qga} (solid black crossbars), $\chi$EFT predictions by Alarc\'on {\it et al.}~\cite{Alarcon:2020icz} (green backslash hatched band) and Bernard {\it et al.}~\cite{Bernard:2012hb} (magenta forward-slash hatched band), phenomenological models by Burkert and Ioffe~\cite{Burkert:1992tg,Burkert:1993ya} (orange dot-dashed curve) and Pasechnik {\it et al.}~\cite{Pasechnik:2010fg} (grey dashed curve), and our own spin structure function parameterization~\cite{CLAS:2017qga} (purple solid curve).  
The MAID model~\cite{Drechsel:1998hk} that includes only one-pion production contributions is shown by the (yellow) dotted curve. 
The GDH slope (black dotted line) is the GDH sum rule expectation for $Q^2 \to 0$. See texts for detailed discussions. 
}
\label{fig:Gamma1_p}
\end{figure}

Our results on $\overline \Gamma_1^p(Q^2)$ (Eq.~(\ref{eq:GDH_gen_ji})) are shown in Fig.~\ref{fig:Gamma1_p}. 
\begin{figure}[!ht]
  \includegraphics[width=0.48\textwidth]{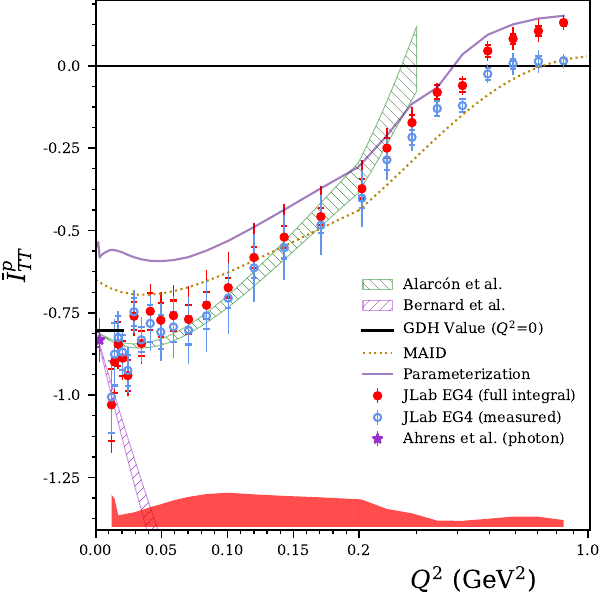}
\caption{[Color online] EG4 results on the proton $\overline I_{TT}^p(Q^2)$. Across all $Q^2$ bins, the size of the systematic uncertainty (red colored band) varies from $0.186$ to $0.104$, while the relative size of the systematic to statistical uncertainty varies within $(54-357)\%$. The GDH value is shown as the short horizontal line, at $\overline I_{TT}^p(0)=-0.804$. The result from the photoproduction experiment~\cite{GDH:2001zzk,GDH:2003xhc,GDH:2005noz,Ahrens:2006yx} 
  is shown as the solid purple star at $Q^2=0$. Note that earlier CLAS data~\cite{CLAS:2017qga} and the phenomenological model predictions~\cite{Burkert:1993ya, Pasechnik:2010fg} are not available for this integral. See Fig.~\ref{fig:Gamma1_p} caption for more details. }
\label{Fig:ITT_P}
\end{figure}
They agree with the most recent $\chi$EFT predictions~\cite{Bernard:2012hb, Alarcon:2020icz} up to $Q^2 \approx 0.03$~GeV$^2$, above which only that of Ref.~\cite{Alarcon:2020icz} agrees with the data.
The new $\overline \Gamma_1^p(Q^2)$ results  generally agree with a previous experiment~\cite{CLAS:2017qga} in the overlapping $Q^2$ region within their uncertainties. However, it is visible that the parameterization based on Ref.~\cite{CLAS:2017qga} (purple solid curve) can be improved by including the new results. The phenomenological models Burkert and Ioffe~\cite{Burkert:1992tg,Burkert:1993ya} and Pasechnik {\it et al.}~\cite{Pasechnik:2010fg} agree well with the new results for all $Q^2$ values. 
They predict the behavior of $\overline\Gamma_1$ in the non-perturbative domain of QCD and beyond: 
The Burkert and Ioffe model~\cite{Burkert:1992tg,Burkert:1993ya} extrapolates DIS data using vector meson dominance and a parameterization of resonance contributions; 
while the Pasechnik {\it et al.} model~\cite{Pasechnik:2010fg}, improved upon the earlier Soffer-Teryaev model~\cite{Soffer:2004ip}, employs the analytical perturbation theory (APT) approach that extrapolates DIS data to low $Q^{2}$, accounting for the mild $Q^{2}$-dependence of $g_1+g_2$. 
Lastly, the MAID unitary isobar phenomenological model~\cite{Drechsel:1998hk} is a 
parametrization of experimental pion production data based on
a partial wave analysis in the resonance region, and includes nonresonant background. MAID does not account for multi-particle final states and therefore provides a partial $\overline \Gamma_1(Q^2)$ in the resonance region, without the low-$x$ (high-$W$) contribution. Consequently, it can be roughly compared with the experimentally measured $\overline \Gamma_1(Q^2)$ (open blue circles),
although it does not necessarily have the same integration limits.
With these caveats in mind, MAID and the EG4 measured sum agree well up to $Q^2=0.2$~GeV$^2$, after which the model is below the data.

Our results on $\overline I_{TT}^p(Q^2)$ (Eq.~(\ref{eq:I_TT_2})) are shown in Fig.~\ref{Fig:ITT_P}. 
Compared with the $\chi$EFT predictions, similar conclusions as for $\overline \Gamma_1^p(Q^2)$ are reached. The MAID model still agrees with the experimentally measured integral within uncertainties.
Finally, the difference between the new results and the  parameterization of previous data~\cite{CLAS:2017qga} is more visible and the parameterization can clearly be improved. 

Extrapolating our $\overline I_{TT}^p(Q^2)$ 
results to $Q^2=0$ yields~\cite{CLAS:2021apd}
\begin{eqnarray}
  \overline I_{TT}^{p~\rm{EG4}}(Q^2\to 0)&=&-0.798\pm 0.042 \mathrm{~(tot)}~.
  \label{I_TT p extrapolation}
\end{eqnarray}
This agrees well with the GDH sum rule prediction $I^{p~\rm{theo}}_{TT}=-\kappa^2/4=-0.804(0)$ and the experimental photoproduction result 
$\overline I_{TT}^{p~\rm{exp}}(0)= -0.832\pm 0.023$ (stat)$\pm 0.063$ (syst)~\cite{GDH:2001zzk,GDH:2003xhc,GDH:2005noz,Ahrens:2006yx}.  
We note that the uncertainties from Refs.~\cite{GDH:2001zzk,GDH:2003xhc,GDH:2005noz,Ahrens:2006yx} and Eq.~(\ref{I_TT p extrapolation}) are comparable. For EG4, the uncertainty due to the $Q^2 \to 0$ extrapolation is compensated by the fact that 
inclusive electroproduction automatically sums over all reaction channels, removing uncertainties
due to the detection of all final states needed in photoproduction.
The EG4 data thus provide the first test of the GDH sum rule 
with a different technique than photoproduction. 

\begin{figure}[!ht]
  \includegraphics[width=0.48\textwidth]{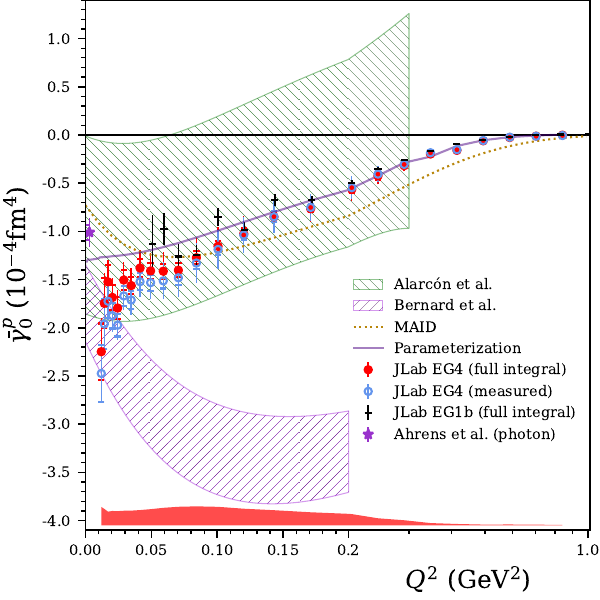}\\
  \vspace*{-0.2cm}
\caption{[Color online] EG4 results on the proton ${\overline \gamma_0}^p(Q^2)$. Across all $Q^2$ bins, the size of the systematic uncertainty (red colored band) varies from $0.0045$ to $0.198$ ($\times 10^{-4}$~fm$^4$), while the relative size of the systematic to statistical uncertainty varies within $(66-364)\%$. 
The photoproduction data point~\cite{Gurevich:2017cpt}
is slightly displaced from $Q^2=0$ (purple star). Note that phenomenological model predictions~\cite{Burkert:1993ya, Pasechnik:2010fg} are not available for this integral. See Fig.~\ref{fig:Gamma1_p} caption for more details. }
\label{fig:gamma0_P}
\end{figure}
One more moment that can be formed from the EG4 data 
is the generalized longitudinal spin polarizability $\overline \gamma_0(Q^2)$ (Eq.~(\ref{eq:gamma0})).
Compared to $\overline I_{TT}$ (Eq.~(\ref{eq:I_TT_2})), the $x^2$-weighting in Eq.~(\ref{eq:gamma0}) 
indicates that $\overline I_{TT}$ and $\overline \gamma_0$ have different contributions from both systematic and model uncertainties.
Results for $\overline \gamma_0^p(Q^2)$ are shown in Fig.~\ref{fig:gamma0_P}. 
They agree with $\chi$EFT calculation from Bernard {\it et al.}~\cite{Bernard:2012hb} only at very low $Q^2$ (below $\approx 0.03$~GeV$^2$) and agree marginally with Alarcon {\it et al.}~\cite{Alarcon:2020icz}, though both calculations have large uncertainties.
On the other hand, when combined with the photoproduction data point~\cite{Gurevich:2017cpt},
the EG4 results support the larger slope at very low $Q^2$ predicted by
Bernard {\it et al.}~\cite{Bernard:2012hb}. 
The agreement between the measured integrals and the MAID model is similar to both $\overline\Gamma_1^p$ and $\overline I_{TT}^p$.
The recent calculation by Bigazzi {\it et al.}~\cite{Bigazzi:2023odl} (not shown in the figure) 
based on the anti-de Sitter/conformal field theory (AdS/CFT) correspondence~\cite{Maldacena:1997re} applied to QCD~\cite{Brodsky:2014yha} recovers the sign and qualitative behavior of $\overline\gamma_0^p(Q^2)$, albeit up to a normalization factor.

\subsection{Moments of the Deuteron Spin Structure Functions}\label{sec:results_moments_d}
\begin{figure}[!h]
  \includegraphics[width=0.48\textwidth]{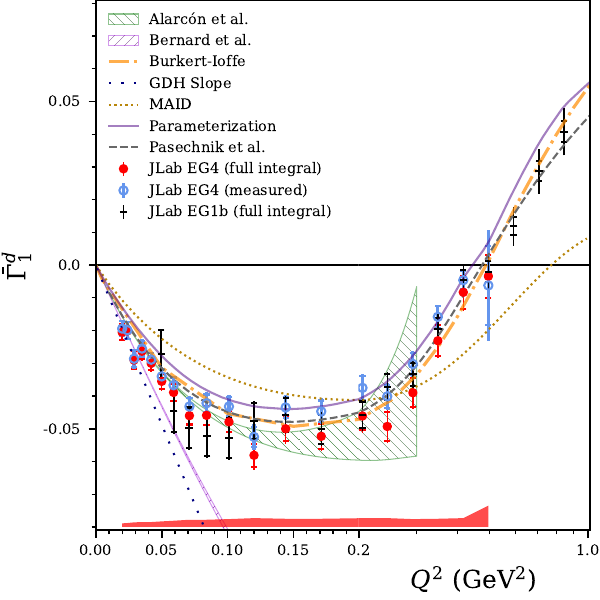}\\
  \vspace*{-0.2cm}
\caption{[Color online] EG4 results on the deuteron $\overline \Gamma_1^d(Q^2)$. 
The size of the systematic uncertainty (red colored band) varies from $0.0011$ to $0.0065$, while the relative size of the systematic to statistical uncertainty varies within $(44-98)\%$, from the lowest to the highest $Q^2$ bins. 
See Fig.~\ref{fig:Gamma1_p} caption for more details. 
}\label{fig:Gamma1_D}
\end{figure} 

\begin{figure}[!h]
  \includegraphics[width=0.48\textwidth]{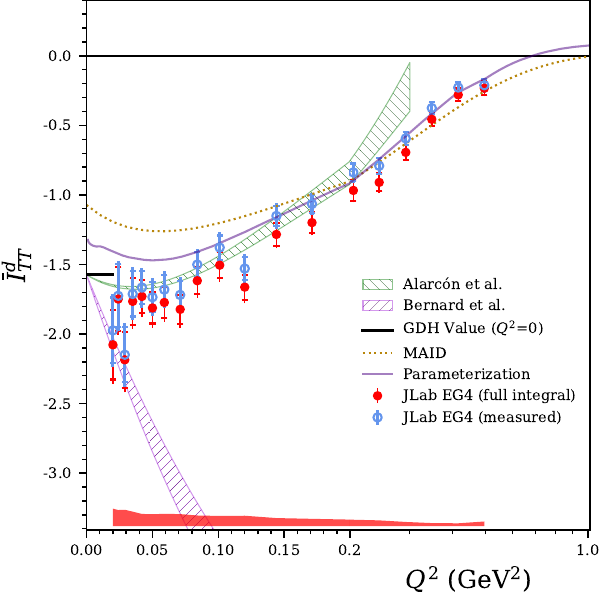}
\caption{[Color online] 
EG4 results for the deuteron $\overline I_{TT}^d(Q^2)$. 
Across all $Q^2$ bins, the size of the systematic uncertainty (red colored band) varies from $0.0198$ to $0.125$, while the relative size of the systematic to statistical uncertainty varies within $(42-83)\%$. 
The expected GDH value $\overline I_{TT}^d{(0)}=-1.574\pm0.026$ is shown by the short horizontal line.
Note that earlier CLAS data~\cite{CLAS:2017qga} and the phenomenological model predictions~\cite{Burkert:1993ya, Pasechnik:2010fg} are not available for this integral. See Fig.~\ref{fig:Gamma1_p} caption for more details.
}\label{fig:ITT_D}
\end{figure}

\begin{figure}[!h]
  \includegraphics[width=0.48\textwidth]{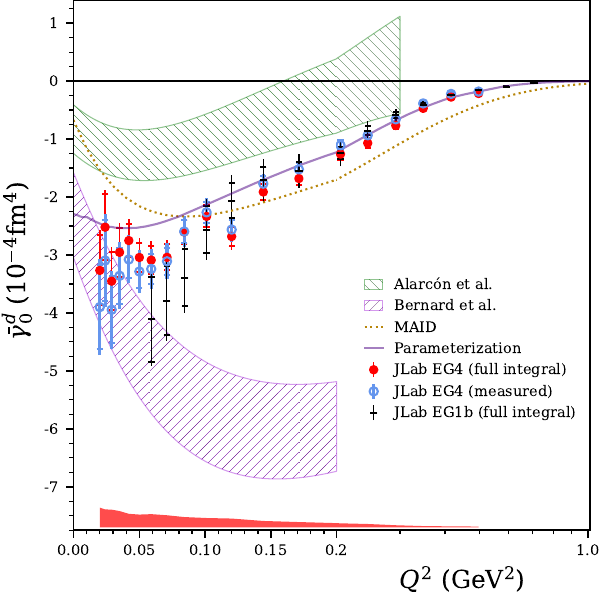}
\caption{[Color online] 
EG4 results on the deuteron $\overline \gamma_0^d(Q^2)$. Across all $Q^2$ bins, the size of the systematic uncertainty (red colored band) varies from $0.0123$ to $0.339$ ($\times 10^{-4}$~fm$^4$), while the relative size of the systematic to statistical uncertainty varies within $(37-93)\%$. Note that phenomenological model predictions~\cite{Burkert:1993ya, Pasechnik:2010fg} are not available for this integral. See Fig.~\ref{fig:Gamma1_p} caption for more details. }
\label{fig:gamma0_D}
\vspace{-0.2cm}
\end{figure}

Results for the deuteron moments $\overline I_{TT}^d $, $\overline \Gamma_1^d$, and $\overline \gamma_0^d $ are shown in Figs.~\ref{fig:Gamma1_D}, \ref{fig:ITT_D}, and \ref{fig:gamma0_D}, respectively. 
The bar now indicates that, in addition to the elastic reaction, the deuteron electro-disintegration contribution is also excluded from the integrals. Therefore, they approximately correspond to the sum of the proton and neutron moments rather than to the deuteron nucleus (see Section~\ref{sec:intro_deuteron}).


Similar observations as for the proton measurements can be drawn: 
the EG4 data agree with the previous data from CLAS~\cite{CLAS:2017qga} (and from SLAC E143~\cite{E143:1996vck}, not shown in the figure).
The measured $\overline \Gamma_1^d$ and $\overline{I}_{TT}^{d~\rm{exp}}(Q^2)$ agree with the $\chi$EFT results for the sum of proton and neutron
(multiplied with the D-state depolarization, see Section~\ref{sec:intro_deuteron}) of Alarc\'on {\it et al.}~\cite{Alarcon:2020icz}, and of Bernard {\it et al.}~\cite{Bernard:2012hb} for the lowest $Q^2$ points. 
The models of Pasechnik {\it et al.}~\cite{Pasechnik:2010fg} and 
Burkert-Ioffe 
\cite{Burkert:1992tg,Burkert:1993ya} agree well with the $\overline \Gamma_1^d$ data, while the parameterization~\cite{CLAS:2017qga} is systematically slightly larger. 
The comparison between the MAID model and the data is similar to that of the proton for all three integrals. 

Extrapolating the $\overline{I}_{TT}^{d}(Q^2)$ data to $Q^2=0$ 
yields~\cite{CLAS:2017ozc}
\begin{eqnarray}
 \overline{I}_{TT}^{d~\rm{EG4}}(Q^2\to 0)=-1.724 \pm 0.057~{\rm (tot)},
 \label{eq:ITT_D}
\end{eqnarray}
which agrees at the $1.5~\sigma$ level with the GDH sum rule prediction $-1.574 \pm 0.026$, obtained using Eq.~(\ref{eq:ITTd=p+n}), where the uncertainty is from the deuteron D-state correction. 
This can be compared with the data using real photons of energy between $0.2<\nu<1.8$~GeV,
$\overline{I}_{TT}^{d~\rm{exp}}(0)= -1.986\pm 0.008$ (stat)$\pm 0.010$ (syst)~\cite{GDH:2001zzk}. Note that the systematic uncertainty of~\cite{GDH:2001zzk} does not include that from the unmeasured low and large $\nu$ contributions. 

Regarding the generalized spin polarizability $\overline \gamma_0^d $, the EG4 results are clearly outside the range predicted by Alarc\'on {\it et al.}~\cite{Alarcon:2020icz}, 
but agree with the Bernard {\it et al.} calculation for the lowest $Q^2$ points. The parameterization~\cite{CLAS:2017qga} does not describe the data well below $Q^2=0.1$~GeV$^2$, similarly to the $\overline{I}_{TT}^d(Q^2)$ case.

\newpage
\subsection{Moments of the Neutron Spin Structure Functions}\label{sec:results_moments_n}
Moments of the neutron spin structure function can be extracted directly from those of the proton and neutron results, see Section~\ref{sec:intro_deuteron}. Understanding the deuteron moments as  ``per nucleus" (rather than ``per nucleon"), 
we rewrite Eqs.~(\ref{eq:Gamma1d=p+n}-\ref{eq:gamma0d=p+n}) as 
\begin{eqnarray}
\overline\Gamma_1^n&=\overline\Gamma_1^d/\left(1-1.5\omega_{D}\right)-\overline\Gamma_1^{p};\label{eq:Gamma1n=2d-p}\\
\overline I_{TT}^n&=\overline I_{TT}^d/\left(1-1.5\omega_{D}\right)-\overline I_{TT}^{p};\label{eq:ITTn=2d-p}\\
\overline\gamma_0^n&=\overline\gamma_0^d/\left(1-1.5\omega_{D}\right)-\overline\gamma_0^{p},
\label{eq:gamma0n=2d-p}
\end{eqnarray}
with $\omega_{D}$ given previously under Eq.~(\ref{eq:gamma0d=p+n}).
When extracting the neutron moments as above, we added in quadrature the proton and deuteron statistical uncertainties, as well as the systematic uncertainties.
\begin{figure}[!h]
  \includegraphics[width=0.48\textwidth]{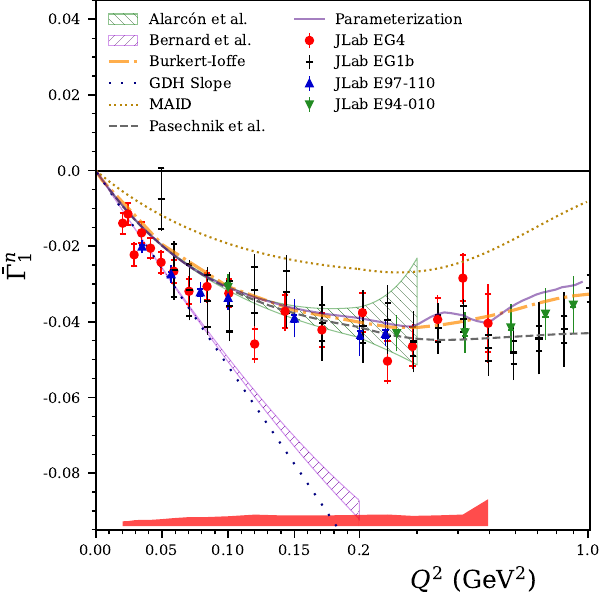}
\caption{[Color online] EG4 results on the neutron $\overline \Gamma_1^n(Q^2)$, compared with earlier results from 
 EG1b (d,p)~\cite{CLAS:2017qga,CLAS:2015otq} (black crossbars), 
JLab E97-110 (neutron extracted from $^3$He)~\cite{JeffersonLabE97-110:2019fsc} (blue triangles),  
 and E94-010 ($^3$He)~\cite{JeffersonLabE94-010:2003dvv} (green triangles). 
 The size of the systematic uncertainty (red colored band) varies from $0.0013$ to $0.0071$, while the relative size of the systematic to statistical uncertainty varies from $(45-92)\%$, from the lowest to the highest $Q^2$ bins. 
See Fig.~\ref{fig:Gamma1_p} caption for more details. 
}
\label{fig:Gamma1_n}
\end{figure}
\begin{figure}[!h]
  \includegraphics[width=0.48\textwidth]{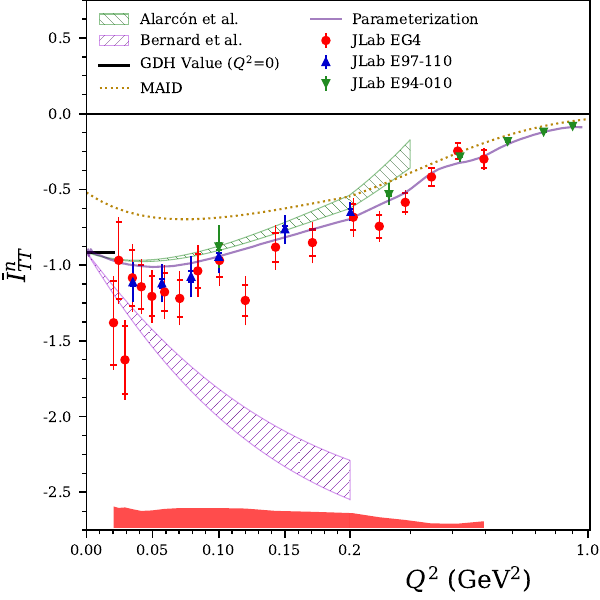}
\caption{[Color online] EG4 results on the neutron $\overline I_{TT}^n(Q^2)$, compared with results from JLab E94-010 ($^3$He)~\cite{Amarian:2002ar} (green triangles) and E97-110 ($^3$He)~\cite{JeffersonLabE97-110:2019fsc} (blue triangles). Across all $Q^2$ bins, the size of the systematic uncertainty varies from $0.0275$ to $0.141$, while the relative size of the systematic to statistical uncertainty varies within $(49-123)\%$. The GDH value is shown as the short horizontal line at $\overline I^n_{TT}(0)=-0.915$. 
Note that earlier CLAS data~\cite{CLAS:2017qga} and the phenomenological model predictions~\cite{Burkert:1993ya, Pasechnik:2010fg} are not available for this integral. See Fig.~\ref{fig:Gamma1_p} caption for more details. 
}
\label{fig:I_tt^n}
\end{figure}
The resulting $\overline\Gamma_1^{n}(Q^2)$ and $\overline I_{TT}^n(Q^2)$ 
are shown in Figs.~\ref{fig:Gamma1_n} and  \ref{fig:I_tt^n}, respectively.
They agree with previous data from E143~\cite{E143:1996vck}, EG1a~\cite{CLAS:2002knl,CLAS:2003rjt} and
EG1b~\cite{CLAS:2017qga,CLAS:2015otq} where the neutron information was extracted from the combined deuteron and proton data, as well as those from E94-010~\cite{JeffersonLabE94-010:2003dvv} and E97-110~\cite{JeffersonLabE97-110:2019fsc} for which
the neutron information was extracted from $^3$He data.
There is also agreement with the phenomenological 
models of Burkert and Ioffe~\cite{Burkert:1992tg,Burkert:1993ya} and Pasechnik {\it et al.}~\cite{Pasechnik:2010fg} (when available)
and with the $\chi$EFT predictions from Bernard {\it et al.}~\cite{Bernard:2012hb} (up to $Q^2 \simeq 0.08$~GeV$^2$) and, to a lesser extent, with Alarc\'on {\it et al.}~\cite{Alarcon:2020icz}. The MAID model prediction is systematically more positive than the data.
Like in the case of $\overline \gamma_0^n $, the calculation of~\cite{Bigazzi:2023odl} (not shown in the figure) agrees in sign and behavior with the $\overline \gamma_0^n (Q^2)$, up to a similar normalization factor as for $\overline \gamma_0^p(Q^2)$.

The real photon GDH sum for the neutron can be deduced from Eqs.~(\ref{I_TT p extrapolation}) and (\ref{eq:ITT_D}):
 \begin{eqnarray}
\overline I_{TT}^{n~\rm{EG4}}(Q^2\to 0)= -1.084\pm 0.130~\mathrm{(tot)}, 
\end{eqnarray}
where the uncertainty arises from the uncertainties of $p$, $d$ and the D-state component added linearly. This agrees within $\approx 1.3 \sigma$ from the sum rule expectation, $\overline I_{TT}^{n~\rm{theo}}(0)=-0.915(0)$. 

\begin{figure}[!h]
  \includegraphics[width=0.48\textwidth]{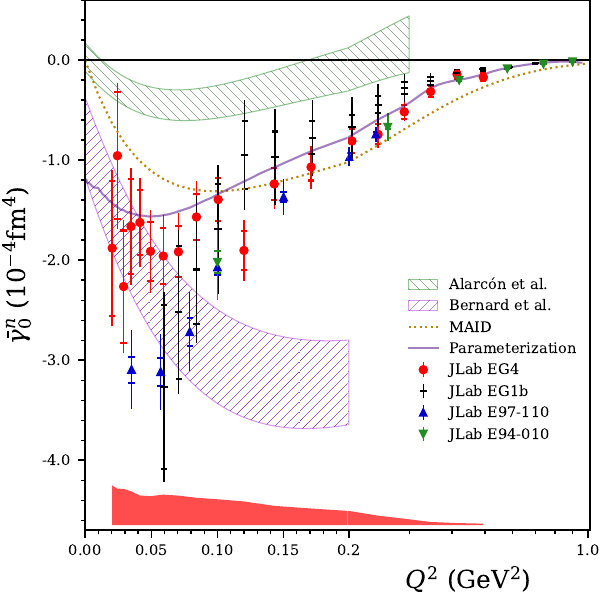}
\caption{[Color online] EG4 results on the generalized spin polarizability $\overline \gamma_0^n$, compared with earlier results from JLab 
 EG1b (proton, deuteron)~\cite{CLAS:2017qga,CLAS:2015otq} (black crossbars), 
E97-110 ($^3$He)~\cite{JeffersonLabE97-110:2019fsc} (blue triangles),  
 and E94-010 ($^3$He)~\cite{JeffersonLabE94-010:2003dvv} (green triangles). Across all $Q^2$ bins, the size of the systematic uncertainty varies from $0.0158$ to $0.397$ ($\times 10^{-4}$~fm$^4$), while the relative size of the systematic to statistical uncertainty varies from $(40-124)\%$. 
 Note that phenomenological model predictions~\cite{Burkert:1993ya, Pasechnik:2010fg} are not available for this integral. See Fig.~\ref{fig:Gamma1_p} caption for more details. 
  }
\label{fig:gam0_n}
\end{figure}

Results on the generalized longitudinal spin polarizability on the neutron are shown in Fig.~\ref{fig:gam0_n}, along with earlier results from JLab EG1b~\cite{CLAS:2017qga,CLAS:2015otq}), E97-110~\cite{JeffersonLabE97-110:2019fsc} and E94-010~\cite{JeffersonLabE94-010:2003dvv}).
The EG4 neutron results generally agree with the earlier data, though there is a tension between EG4 and E97-110 for $Q^2 \lesssim 0.1$~GeV$^2$. Adding the EG4 deuteron and proton systematic uncertainties linearly, rather than in quadrature, does not remove the tension. 
It contrasts with the agreement between EG4 and E97-110 for the neutron data on $\overline \Gamma_1^n$ or $\overline I_{TT}^{n}$, suggesting the discrepancy arises from the high-$x$ (near pion threshold) contribution.
The EG4 results agree with $\chi$EFT calculation from Ref.~\cite{Bernard:2012hb} up to 
$Q^2 \approx 0.1$~GeV$^2$ but not with that of  Alarc\'on {\it et al.}~\cite{Alarcon:2020icz}. 

\subsection{Isospin Analysis on the Moments of Spin Structure Functions} \label{isospin-moments}
Results on the proton and deuteron moments presented in the previous sections can be used  
to form various isospin combinations based on the description in Section~\ref{sec:intro_deuteron}.  
The isoscalar ($p+n$) combination would be similar to the deuteron results, up to a factor $\approx 1/0.92$ from the deuteron D-state. 
For isovector ($p-n$) moments, the proton and deuteron results were combined as:
\begin{eqnarray}
\overline\Gamma_1^{p-n}&=2\overline\Gamma_1^{p}-\overline\Gamma_1^d/\left(1-1.5\omega_{D}\right);\\
\overline\gamma_0^{p-n}&=2\overline\gamma_0^{p}-\overline\gamma_0^d/\left(1-1.5\omega_{D}\right),
\end{eqnarray}
where $\omega_D$ is given under Eq.~(\ref{eq:gamma0d=p+n}) and with the deuteron moments understood as ``per nucleus". 
Alternatively to using the EG4 proton and deuteron data, one can use the proton results from EG4 and the neutron results obtained from $^3$He data such as those from JLab  E97-110~\cite{JeffersonLabE97-110:2019fsc} to perform the isospin separation. 

%

The isovector moments are expected to be simpler to compute than the isoscalar or nucleon moments. For lattice QCD simulations, this is because the disconnected diagrams  
are suppressed~\cite{Deur:2018roz}. 
For $\chi$EFT calculations, {\it e.g.}, Refs.~\cite{Burkert:2000qm, Alarcon:2020icz}, the inclusion of the $\Delta^{(1232)}~3/2^+$ excitation is often difficult, which would cancel in the isovector combination.  
However, the consequence of the expectation that the $\chi$EFT predictions are more robust has been upheld only with data on the Bjorken sum $\overline \Gamma_1^{p-n}$~\cite{Deur:2004ti, Deur:2008ej, Deur:2014vea}, but not with 
$\overline \gamma_0^{p-n}$~\cite{Deur:2008ej}, nor with the longitudinal-transverse interference polarizability $\delta_{LT}$~\cite{JeffersonLabE94010:2004ekh, E97-110:2021mxm} 
for which the $\Delta^{(1232)}$ contribution is also expected to cancel. 
Our results provide improved tests of this expectation thanks to the lower $Q^2$ coverage and higher precision of EG4 data.

\subsubsection{Results on the Bjorken Sum $\overline\Gamma_1^{p-n}(Q^2)$ \label{sssec:BJSR}}
\begin{figure}[!h]
  \includegraphics[width=0.48\textwidth]{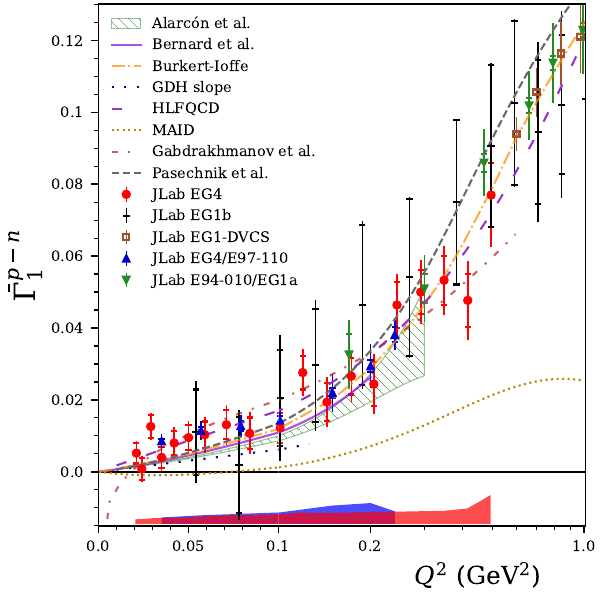}
\caption{[Color online] 
The Bjorken sum $\overline\Gamma_1^{p-n}(Q^2)$ from EG4 (solid red circles)
and from the EG4 proton data combined with the JLab E97-110 neutron data~\cite{JeffersonLabE97-110:2019fsc} 
(blue triangles).  
Also shown are earlier data from 
E94-010/EG1a~\cite{Deur:2004ti} (green triangles), 
CLAS EG1b~\cite{Deur:2008ej} (black crossbars), 
and EG1dvcs~\cite{Deur:2014vea} (brown open squares).
The inner (outer) error bars on the data points give their statistical (total) 
uncertainties, while the systematic uncertainties of EG4 and EG4/E97-110 are also given by the red and the blue bands, respectively.  Across all $Q^2$ bins, the size of the systematic uncertainty of the EG4 results (red colored band) varies from $0.0017$ to $0.0115$, while the relative size of the systematic to statistical uncertainty varies from $(58-131)\%$. 
The $\chi$EFT predictions from Alarc\'on {\it et al.}~\cite{Alarcon:2020icz} and Bernard {\it et al.}~\cite{Bernard:2012hb} (the latter without error band as given in Ref.~\cite{Bernard:2012hb}) are 
shown along with phenomenological model predictions from Burkert-Ioffe~\cite{Burkert:1992tg,Burkert:1993ya}, Pasechnik {\it et al.}~\cite{Pasechnik:2010fg}, MAID~\cite{Drechsel:1998hk}, HLFQCD~\cite{Brodsky:2010ur}, and Gabdrakhmanov 
{\it et al.}~\cite{Gabdrakhmanov:2023rjt}.}
\label{fig:bjsr} 
\end{figure}

The isovector combination $\overline \Gamma_1(Q^2)$ is the Bjorken sum~\cite{Bjorken:1966jh,Bjorken:1969mm}, see Eqs.~(\ref{eq:generalized Bjorken SR})-(\ref{eq:bj-GDH}).  
Our results on $\overline\Gamma_1^{p-n}(Q^2)$ are shown in Fig.~\ref{fig:bjsr}, compared with data from earlier experiments~\cite{E143:1994vcg,Deur:2004ti,Deur:2008ej,Deur:2014vea}. 
Besides the models already discussed in connection with Figs.~\ref{fig:Gamma1_p}-\ref{fig:gam0_n}, we also show predictions from holographic light-front QCD (HLFQCD)~\cite{Brodsky:2014yha}. 
In HLFQCD, the QCD coupling $\alpha_s$ is derived in the $g_1$ scheme following the effective charge 
prescription~\cite{Grunberg:1980ja, Deur:2016tte, Deur:2023dzc}, from which the  Bjorken sum can be obtained~\cite{Deur:2005cf, Deur:2008rf, Deur:2022msf}.

Note the comparison between the $\overline\Gamma_1^{p-n}$ results obtained from EG4's proton and deuteron data and those extracted from the EG4 proton and E97-110 neutron ($^3$He) data. 
An agreement between the two different approaches  
was not assured since the nuclear structures of the deuteron and $^3$He are quite different: $^3$He is strongly bound compared to the deuteron, therefore more sensitive to potential nuclear modification of the proton. 
Our treatment of the deuteron is minimal and consists of accounting for its D-state~\cite{Lacombe:1980dr, Machleidt:1987hj, Zuilhof:1980ae, Kotthoff:1976fg, Desplanques:1988mp}, while treatment of the $^3$He data accounted for the nucleon effective polarizations~\cite{CiofidegliAtti:1996cg}. 
As shown in Fig.~\ref{fig:bjsr}, the two approaches agree well with each other.  This suggests that, at least for $\overline\Gamma_1^{p-n}$, the nuclear corrections applied to the deuteron and $^3$He data to obtain the neutron information appear sufficient at these low $Q^2$ values.

Both the EG4 and EG4/E97-110 results on $\overline\Gamma_1^{p-n}$ are systematically above $\chi$EFT and model predictions, except that of HLFQCD~\cite{Brodsky:2010ur} and~\cite{Gabdrakhmanov:2023rjt}, which agree with the data. 
To allow a quantitative comparison between the data and theory, we 
employed a fit of the form $bQ^2+cQ^4$ to the entire world data set, based on Eq.~(\ref{eq:bj-GDH})~\cite{Deur:2021klh}, where $b$ is expected to be the GDH slope. 
    Our results are shown in Table~\ref{table_fit2}. 
The “$uncor$” uncertainty designates the point-to-point uncorrelated uncertainty. It is
the quadratic sum of the statistical uncertainty and a fraction of the systematic uncertainty determined so that $\chi^2$/n.d.f = 1
for the best fit. The “$cor$” uncertainty is the correlated uncertainty estimated from the remaining fraction of
the systematic uncertainty.

We note that our value for $b$ deviates significantly from that
expected from the GDH sum rule. This is partially due to the 
large cancellation for the GDH sum rule in the $p-n$ combination, which
is only 1/4 of the value for the individual nucleons. Experimentally, because of the nearly perfect cancellation of the large contribution from the $\Delta^{(1232)}$ resonance in the
$p-n$ difference, the isovector integral is much more strongly affected by small-$x$
contributions than the other moments, with relatively large uncertainties for
$g_1^p-g_1^n$. Of course, our results could also indicate a much stronger $Q^2$--dependence near the photon point than can be captured by a 2-parameter fit. In any case, they do not indicate a violation of the GDH sum rule since, as shown in the previous section, our results for the proton and neutron are individually consistent with the sum rule within (1 - 1.3)$\sigma$. Further discussions on the possible discrepancy between the $b$ parameter and the GDH slope can be found in Ref.~\cite{Deur:2021klh}.

\renewcommand{\arraystretch}{1.3} 
\begin{table}[!ht]
\resizebox{0.48\textwidth}{!}{%
\begin{tabular}{|c|c|c|} \hline
Data set                 &   $b $      [GeV$^{-2}$]        &  $c$    [GeV$^{-4}$]        \\  \hline 
World data              &  $0.182$\begin{tabular}{@{}c@{}} $\pm0.016~(uncor)$ \\ $\pm0.040~(cor)$ \end{tabular}    &   $-0.124$\begin{tabular}{@{}c@{}}$ \pm0.089~(uncor)$\\ $\pm0.112~(cor)$  \end{tabular} \\  \hline 
GDH sum rule~\cite{Gerasimov:1965et}      &  0.0618   &     -            \\ \hline
$\chi$EFT Bernard {\it et al.}~\cite{Bernard:2012hb}  &  0.07  &   0.3        \\ 
$\chi$EFT Alarc\'on {\it et al.}~\cite{Alarcon:2020icz}  &   0.066(4)  &   0.25(12)                 \\
\hline
Burkert-Ioffe~\cite{Burkert:1992tg,Burkert:1993ya} &   0.09   &     0.3           \\
Pasechnik {\it et al.}~\cite{Pasechnik:2010fg} &   0.09  &   0.4              \\ 
HLFQCD~\cite{Brodsky:2010ur} &  0.177  &   -0.067              \\ \hline 	
\end{tabular}}
\vspace{-0.2cm}
\caption{Best fit of the world data
on $ \overline\Gamma_1^{p-n}$ over the range $0.021 \leq Q^2 \leq 0.244$~GeV$^2$. The fit form used is $bQ^2+cQ^4$.
The meaning of the ``$uncor$'' and ``$cor$'' uncertainties is explained in the main text.
Also shown are the GDH sum rule expectation and the results of the fit applied to the theoretical predictions. 
}
\label{table_fit2}
\end{table}

\subsubsection{Isospin Study of $\overline \gamma_0(Q^2)$}

\begin{figure}[!h]
  \includegraphics[width=0.48\textwidth]{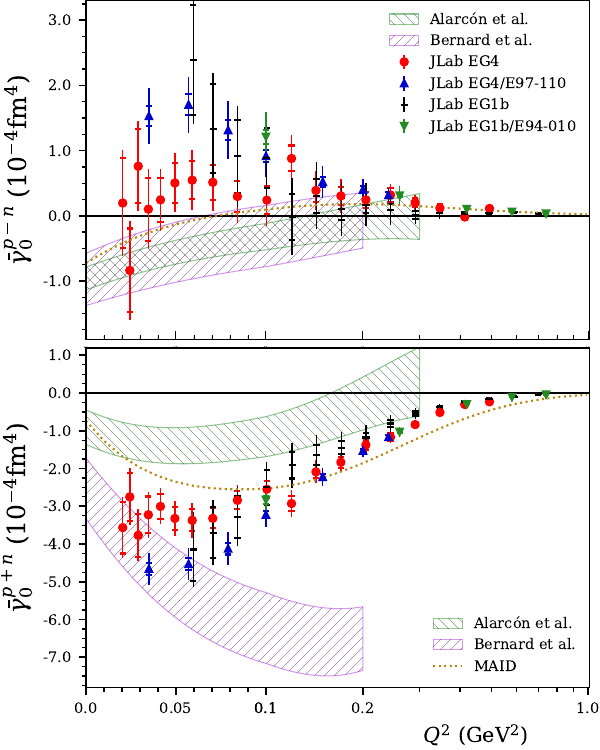}
\caption{[Color online]
EG4 results on the isovector (top) and isoscalar (bottom) combinations of the generalized longitudinal spin polarizability $\overline \gamma_0(Q^2)$.
The EG4-only data are shown by the red circles, compared with the earlier EG1b-only (black crossbars), the combination of EG4 $p$ and E97-110 $n$($^3$He) data (blue triangles), and the combination of EG1b $p$ and E94-010 $n$($^3$He) data (green triangles).
For $\overline \gamma_0^{p-n}(Q^2)$, the size of the systematic uncertainty varies from $0.026$ to $0.522$ ($\times 10^{-4}$~fm$^4$), while the relative size of the systematic to statistical uncertainty varies from $(58-216)\%$. 
For $\overline \gamma_0^{p+n}(Q^2)$, the size of the systematic uncertainty varies from $0.013$ to $0.367$ ($\times 10^{-4}$~fm$^4$), while the relative size of the systematic to statistical uncertainty varies from $(36-93)\%$. 
The results are compared with $\chi$EFT predictions from Alarc\'on {\it et al.}~\cite{Alarcon:2020icz} 
and Bernard {\it et al.},
as well as the MAID model. The prediction by Bernard {\it et al.}~\cite{Bernard:2012hb} does not provide uncertainty bands for the proton-neutron difference; for illustration, we show a band whose width is the larger of the proton and the neutron uncertainties.
}\label{fig:gamma0_decomp}
\vspace{-0.2cm}
\end{figure}

Our results on the isovector and isoscalar combination of $\overline \gamma_0(Q^2)$ from EG4 are shown in Fig.~\ref{fig:gamma0_decomp}. We also provide results based on EG4 $p$ data and E97-110 $n$($^3$He) data~\cite{E97-110:2021mxm}. 
The total experimental systematic uncertainties on the proton and deuteron/neutron($^3$He) are combined quadratically, as are the 
statistical uncertainties. The low-$x$ uncertainties 
are added linearly. 
The results shown in Fig.~\ref{fig:gamma0_decomp} agree reasonably well with those from the previous experiments EG1b and EG1b/E94-010~\cite{Deur:2008ej}. 
However, unlike for $\overline\Gamma_1^{p-n}$, results from the EG4-only and EG4/E97-110 approaches agree only at the higher $Q^2$ points. 
On the other hand, if we combine linearly the total systematic uncertainties of the two experiments, the tension at low $Q^2$ vanishes. 

For $\overline \gamma_0^{p-n}(Q^2)$, both the EG4 and EG4/E97-110 
results indicate that it stays positive in the measured $Q^2$ range, conversely to the $\chi$EFT and MAID predictions. 
In the case of 
$\overline \gamma_0(Q^2)^{p+n}$, both the EG4 and EG4/E97-110 combinations agree with Bernard {\it et al.} in the lower $Q^2$
range, but disagree with the Alarcón {\it et al.} prediction.

\subsection{Discussions \label{sec:Discussion}}
As discussed in previous sections, the success of $\chi$EFT in describing our results on the moments of the spin structure function remains limited. 
One reason for this limited success may be coming from the difficulty to fully account for the $\Delta^{(1232)}$ 
degree of freedom in the calculation.  In fact, early $\chi$EFT calculations~\cite{Bernard:1992nz,Ji:1999pd} did not include explicitly the $\Delta^{(1232)}$ excitation, which slows the convergence of the $\chi$EFT perturbation series, or they included it phenomenologically~\cite{Bernard:2002bs, Kao:2002cp}.
This was thought to not be important for some observables for which the $\Delta^{(1232)}$ contribution was expected to be suppressed, such as the longitudinal-transverse spin polarizability $\delta_{LT}$ and the isovector quantities $\overline \Gamma_1^{p-n}$ 
~\cite{Burkert:2000qm} or $\overline \gamma_0^{p-n}$.
It therefore came as a surprise that much of the early nucleon spin structure function data~\cite{Amarian:2002ar} disagreed with calculations~\cite{Bernard:1992nz, Ji:1999pd, Bernard:2002bs, Kao:2002cp}, noticeably $\delta_{LT}^n$ and $\overline \gamma_0^{p-n}$. 
This puzzle prompted refined $\chi$EFT calculations~\cite{Bernard:2012hb, Alarcon:2020icz} and a new experimental program at JLab optimized to cover the $\chi$EFT domain~\cite{CLAS:2017ozc, JeffersonLabE97-110:2019fsc, JeffersonLabHallAg2p:2022qap}, including the EG4 experiment reported here. 

The latest $\chi$EFT calculations by Bernard {\it et al.}~\cite{Bernard:2012hb} and Alarc\'on {\it et al.}~\cite{Alarcon:2020icz} 
both included the $\Delta^{(1232)}$ but differ in their expansion method for the $\pi$(pion)-$\Delta^{(1232)}$ corrections. 
In $\chi$EFT, the general expansion parameter is $m_\pi / \Lambda_\mathrm{\chi SB}$, where $m_\pi\approx 0.1$~GeV and  
$\Lambda_\mathrm{\chi SB} \approx 1$~GeV is the chiral symmetry breaking scale. 
To explicitly account for the $\Delta^{(1232)}$, 
the nucleon-$\Delta^{(1232)}$ mass difference 
$m_{N\Delta}\approx 0.3$~GeV must 
be included in the chiral expansion.
Bernard {\it et al.}~\cite{Bernard:2012hb} treated $m_{N\Delta}$ as a small parameter of the same order as $m_\pi$. 
Alarc\'on {\it et al.}~\cite{Alarcon:2020icz} used $m_{N\Delta}$ as an intermediate scale so that 
$m_{N\Delta} / \Lambda_\mathrm{\chi SB} \approx m_\pi / m_{N\Delta}$
is the expansion parameter for the $\pi$-$\Delta^{(1232)}$ corrections. 
This is referred to as the ``$\delta$-expansion scheme".
Additionally, Alarc\'on {\it et al.} included
a phenomenological form factor, viz not derived within $\chi$EFT, that suppresses the $Q^2$-dependence of observables at large $Q^2$. This clearly extends the range of applicability of the Alarc\'on {\it et al.} prediction~\cite{Alarcon:2020icz} to higher $Q^2$ compared to that of Bernard {\it et al.}~\cite{Bernard:2012hb}, albeit by introducing an extra scale factor. 

An important outcome that can be drawn from the results shown in this section is that $\chi$EFT, although successful in many instances, remains challenged by results from dedicated polarized experiments at low $Q^2$. This includes not only EG4, but also other experiments that used different detectors, methods, and spin observables~\cite{JeffersonLabE97-110:2019fsc, E97-110:2021mxm, JeffersonLabHallAg2p:2022qap}. 
To address this problem, further improvements on $\chi$EFT would be required, although it is very difficult to upgrade the predictions to the next order.
Since $\chi$EFT is the leading QCD-rooted approach to 
calculating the effects of the strong force at large distances,  
this difficulty  poses a problem in our pursuit of a complete description of nature. 
Alternatively or complementarily, it would be helpful to have predictions from other non-perturbative approaches to QCD, such as lattice QCD, the Dyson-Schwinger Equations, or the AdS/CFT methods. Recent progress in calculating doubly-virtual Compton scattering (VVCS) amplitudes in lattice QCD now makes moments involved in sum rules within reach of this method, see, {\it e.g.}, Refs.~\cite{Lee:2023lnx, Wang:2023omf} for recent development. At present, only calculations for spin-independent VVCS amplitudes are available; these
calculations are difficult because they involve computing 4-point correlation functions, in contrast to usual Lattice QCD calculations that require less computing-extensive 2- and 3-point correlations functions. 
Only recently, advances in methods and computer power have allowed such a computation. 
There is no difficulty in extending these calculations to computing the spin-flip VVCS amplitudes that enter our sum rules. 
Our results and their mixed level of agreement with $\chi$EFT predictions are strong incentives to extend these calculations to spin-dependent VVCS amplitudes, allowing lattice QCD predictions to be extensively tested.

Apart from tests of fundamental predictions in the hadronic spin sector, our data will also provide  input for calculations of two-photon effects in the hyperfine splitting of
the hydrogen (or muonic hydrogen) ground state. The 
lack of proton spin data at low enough $Q^2$ has made this contribution the dominant source of uncertainty in that calculation. The EG4 longitudinal spin data combined with the transverse spin data from the recent JLab Hall A experiment E08-027~\cite{JeffersonLabHallAg2p:2022qap} will help to remedy this situation~\cite{Ruth:2024bsl}.

\FloatBarrier

 \section{Summary and Conclusions \label{sec:conclusion}}

In this archival paper, we present the details of the Jefferson Lab EG4 experiment and our final results for proton and the deuteron spin structure function data for $Q^2$ values as low as 0.012 and 0.017~GeV$^2$, respectively.
We include results on the proton and the deuteron $g_1$,  $A_1F_1$, and their moments, reported previously in Refs.~\cite{CLAS:2017ozc,CLAS:2021apd}, along with details of the simulation and data analysis procedure that led to these results. 
Our results agree well with previous JLab experiments, but the significantly improved precision of the new data reveal that the world knowledge on low $Q^2$ spin structure functions, mostly encapsulated in the parameterization~\cite{CLAS:2017qga}, must be updated.

We also present new results on the neutron $g_1^n$ and various moments, extracted by combining the corresponding observable of the proton and deuteron and using a simple deuteron model to correct for Fermi smearing and the deuteron D-state component. Our results on the neutron moments agree with earlier results that used polarized $^3$He targets.

Extrapolating the lowest $Q^2$ EG4 data $\overline I_{TT}(Q^2)$ to $Q^2=0$ provides a check of the GDH sum rule validity, independent from exclusive photoproduction ($Q^2=0$) method. 
This novel technique using quasi-real photons tests the sum rule with a competitive accuracy compared to real photon experiments. It provides results in excellent agreement with the GDH expectation for the proton, and reasonable agreement for the neutron and the deuteron.
By combining the proton with neutron results, isovector or isoscalar moments can be formed, which provide complementary tests of theory predictions at low $Q^2$.  
We also compared the new results on moments to several phenomenological models, and found a mixed level of success.

Finally, our results test extensively $\chi$EFT predictions using 8 distinct moments representing diverse $x$-weightings and isospin components. The new data are of high precision and are at $Q^2$ values well into domain of 
$\chi$EFT. Despite these advantages and the recent improvements on the $\chi$EFT calculation,  
there is mixed level of agreement between data and $\chi$EFT depending on the observable, $Q^2$ range, and type of calculations. Clearly, $\chi$EFT remains challenged by the experimental data at low $Q^2$. 
We hope other methods, such as lattice QCD, Dyson-Schwinger Equations, and AdS/CFT, will soon provide predictions that can be compared with experimental results at low $Q^2$, advancing the theoretical non-perturbative approaches to the strong interaction. 

\acknowledgments{
We would like to acknowledge the outstanding efforts of
the staff of the Accelerator, the Target Group, and the Physics Divisions at Jefferson Lab that made this experiment possible.
This work was supported in part by the U.S. Department of Energy, the U.S. National Science Foundation, the U.S. Jeffress Memorial Trust; 
the United Kingdom Science and Technology Facilities Council (STFC), the Italian Istituto Nazionale 
di Fisica Nucleare; the French Institut National de Physique Nucl\'eaire et de 
Physique des Particules, the French Centre National de la Recherche Scientifique; 
and the National Research Foundation of Korea. 
J.~Zhang and X.~Zheng are supported by the U.S. Department of Energy, Office of Science, Office of Nuclear Physics under contract number DE–SC0014434.
This material is based upon work supported by the U.S. Department of Energy, Office of Science, Office of Nuclear Physics under contract DE-AC05-06OR23177. 
We are grateful to J.M. Alarc\'on, A. Kotikov, R. Pasechnik, and D. Volkova for sending us the predictions from their models. 

\bibliographystyle{JHEP}
\bibliography{EG4_PRC.bib}


\appendix

\section{Table for Results on $g_1^p$ and $A_1^p F_1^p$}

\noindent 
Results for $g_1^p$ and $A_1^pF_1^p$ as shown in Figs.~10 and 12, respectively.

\begin{longtable}{|c|c|c|c|c|c|c|c|c|} \hline
$Q^{2}$ & W  & $x_{bj}$ & $g_{1}^p$ & Stat. & Syst. & $A_{1}^pF_{1}^p$ & Stat. & Syst.\\
(GeV$^{2}$) & (GeV) &  & & &  & & & \tabularnewline
\hline
  0.012 &  1.11& 0.0330 &   0.115 &  0.090 & 0.013 &    0.145 &   0.120 &   0.013 \tabularnewline
  0.012 &  1.13& 0.0294 &  -0.085 &  0.115 & 0.021 &   -0.119 &   0.144 &   0.021 \tabularnewline
  0.012 &  1.15& 0.0264 &  -0.108 &  0.120 & 0.022 &   -0.146 &   0.144 &   0.022 \tabularnewline
  0.012 &  1.17& 0.0240 &  -0.205 &  0.117 & 0.021 &   -0.248 &   0.137 &   0.021 \tabularnewline
  0.012 &  1.19& 0.0219 &  -0.745 &  0.136 & 0.038 &   -0.854 &   0.155 &   0.038 \tabularnewline
  0.012 &  1.21& 0.0202 &  -0.613 &  0.144 & 0.034 &   -0.687 &   0.161 &   0.034 \tabularnewline
\end{longtable}
  \label{Tab:g1p}

\section{Table for Results on $g_1^d$ and $A_1^d F_1^d$}
\noindent 
Results for $g_1^d$ and $A_1^dF_1^d$ as shown in Figs.~11 and 13, respectively.

\begin{longtable}{|c|c|c|c|c|c|c|c|c|} \hline

$Q^{2}$ & W  & $x_{bj}$ & $g_{1}^D$ & Stat. & Syst. & $A_{1}^DF_{1}^D$ & Stat. & Syst.\\
(GeV$^{2}$) & (GeV) &  & & &  & & & \tabularnewline
\hline
            0.0171 &      1.13 &  0.0413 &    0.353 &  0.413 &  0.060 &         0.436 &  0.558 &  0.077 \tabularnewline
            0.0171 &      1.15 &  0.0372 &    0.760 &  0.362 &  0.102 &         0.936 &  0.464 &  0.129 \tabularnewline
            0.0171 &      1.17 &  0.0338 &   -0.536 &  0.348 &  0.170 &        -0.692 &  0.430 &  0.214 \tabularnewline
            0.0171 &      1.19 &  0.0309 &   -0.400 &  0.353 &  0.160 &        -0.496 &  0.422 &  0.208 \tabularnewline
            0.0171 &      1.21 &  0.0285 &   -0.900 &  0.369 &  0.174 &        -1.060 &  0.431 &  0.221 \tabularnewline
            0.0171 &      1.23 &  0.0263 &   -1.209 &  0.349 &  0.183 &        -1.391 &  0.398 &  0.220 \tabularnewline
\end{longtable}

\section{Table for Results on $g_1^n$}
\noindent 
Results for $g_1^n$ as shown in Fig.~14.
\begin{longtable}{|c|c|c|c|c|c|} \hline

$Q^{2}$ & W  & $x_{bj}$ & $g_{1}^n$ & Stat. & Syst. \\
(GeV$^{2}$) & (GeV) &  & & &  \tabularnewline
\hline
                                                             
 0.0171 &  1.15 &  0.0372 &   0.8620 &  0.3999 &  0.1226 \tabularnewline   
 0.0171 &  1.17 &  0.0338 &  -0.3898 &  0.3905 &  0.1905 \tabularnewline   
 0.0171 &  1.19 &  0.0309 &  -0.0851 &  0.4073 &  0.2234 \tabularnewline   
 0.0171 &  1.21 &  0.0285 &  -0.5346 &  0.4303 &  0.2233 \tabularnewline   
 0.0171 &  1.23 &  0.0263 &  -0.8129 &  0.3972 &  0.2110 \tabularnewline   
 0.0171 &  1.25 &  0.0245 &  -0.3160 &  0.3701 &  0.1938 \tabularnewline
\end{longtable}

\section{Table for Results on Moments}
\begin{table*}
\begin{center}
\begin{tabular}{|c|c|c|c|c|c|c|c|c|c|c|c|c|}
\hline 
$Q^{2}$ (GeV$^{2}$) & $\overline\Gamma_1^{p,\mbox{data}}$ & $\overline\Gamma_1^{p,\mbox{full}}$ & Stat. & Syst. & $\overline I_{TT}^{p,\mbox{data}}$ & $\overline I_{TT}^{p,\mbox{full}}$ & Stat. & Syst.& $\overline\gamma_0^{p,\mbox{data}}$ & $\overline\gamma_0^{p,\mbox{full}}$ & Stat. & Syst.\tabularnewline

\hline
0.0120 & -0.0061 & -0.0062 & 0.0007 & 0.0006 & -1.0058 & -1.0291 & 0.1088 & 0.0978 & -2.4741 & -2.2482 & 0.2926 & 0.1917\tabularnewline
0.0143 & -0.0062 & -0.0063 & 0.0007 & 0.0006 & -0.8755 & -0.8991 & 0.0950 & 0.0863 & -1.9606 & -1.7464 & 0.2603 & 0.1710\tabularnewline
0.0171 & -0.0068 & -0.0070 & 0.0006 & 0.0002 & -0.8247 & -0.8453 & 0.0663 & 0.0357 & -1.7245 & -1.5237 & 0.1782 & 0.1429\tabularnewline
0.0204 & -0.0084 & -0.0085 & 0.0005 & 0.0002 & -0.8692 & -0.8875 & 0.0532 & 0.0375 & -1.8753 & -1.6853 & 0.1323 & 0.1497\tabularnewline
0.0244 & -0.0103 & -0.0104 & 0.0006 & 0.0003 & -0.9246 & -0.9405 & 0.0479 & 0.0410 & -1.9733 & -1.7951 & 0.1176 & 0.1481\tabularnewline
0.0290 & -0.0094 & -0.0094 & 0.0006 & 0.0002 & -0.7466 & -0.7600 & 0.0424 & 0.0442 & -1.6696 & -1.5045 & 0.1026 & 0.1522\tabularnewline
0.0347 & -0.0122 & -0.0122 & 0.0006 & 0.0003 & -0.8319 & -0.8429 & 0.0379 & 0.0511 & -1.7124 & -1.5622 & 0.0904 & 0.1532\tabularnewline
\hline
\end{tabular}
\caption{Results for the proton moments $\overline\Gamma_1^p$, $\overline I_{TT}^p$ and $\overline\gamma_0^p$ in $10^{-4}$fm$^4$, for the measured and the full ranges, along with their statistical and systematic uncertainties, as shown in Figs.14-16 and reported in~\cite{CLAS:2021apd}. 
For Figs.~4 and 5 of~\cite{CLAS:2021apd}, the four lowest $Q^2$ bins were combined into two for visual clarity.}
\end{center}
\end{table*}
 
\begin{table*}
\begin{center}
{
\begin{tabular}{|c|c|c|c|c|c|c|c|c|c|c|c|c|}

\hline 

$Q^{2}$ (GeV$^{2}$) & $\overline\Gamma_1^{d,\mbox{data}}$ & $\overline\Gamma_1^{d,\mbox{full}}$ & Stat. & Syst. & $\overline I_{TT}^{d,\mbox{data}}$ & $\overline I_{TT}^{d,\mbox{full}}$ & Stat. & Syst.& $\overline\gamma_0^{d,\mbox{data}}$ & $\overline\gamma_0^{d,\mbox{full}}$ & Stat. & Syst.\tabularnewline

\hline
0.0204 & -0.0194 & -0.0205 & 0.0025 & 0.0011 & -1.9738 & -2.0776 & 0.2489 & 0.1252 & -3.8999 & -3.2665 & 0.6048 & 0.3394\tabularnewline
0.0244 & -0.0199 & -0.0201 & 0.0026 & 0.0012 & -1.7253 & -1.7485 & 0.2295 & 0.1151 & -3.0963 & -2.5206 & 0.5697 & 0.3085\tabularnewline
0.0290 & -0.0286 & -0.0290 & 0.0027 & 0.0014 & -2.1487 & -2.1853 & 0.2021 & 0.1171 & -3.9462 & -3.4522 & 0.5043 & 0.3041\tabularnewline
0.0347 & -0.0255 & -0.0263 & 0.0025 & 0.0015 & -1.7120 & -1.7651 & 0.1670 & 0.1035 & -3.3601 & -2.9548 & 0.4277 & 0.2799\tabularnewline
0.0414 & -0.0289 & -0.0300 & 0.0021 & 0.0015 & -1.6658 & -1.7296 & 0.1212 & 0.0872 & -3.0780 & -2.7518 & 0.2878 & 0.2322\tabularnewline
0.0494 & -0.0340 & -0.0355 & 0.0023 & 0.0017 & -1.7351 & -1.8125 & 0.1119 & 0.0849 & -3.2859 & -3.0435 & 0.2565 & 0.2191\tabularnewline
0.0590 & -0.0367 & -0.0389 & 0.0025 & 0.0019 & -1.6804 & -1.7729 & 0.1087 & 0.0880 & -3.2418 & -3.0891 & 0.2441 & 0.2266\tabularnewline
\hline
\end{tabular}
}
\caption{Results for the (per nucleus) deuteron moments $\overline\Gamma_1^d$, $\overline I_{TT}^d$ and $\overline\gamma_0^d$ in $10^{-4}$fm$^4$, for the measured and the full ranges, along with their statistical and systematic uncertainties, as shown in Figs.~19-21 and reported in~\cite{CLAS:2017ozc}.
}
\end{center}
\end{table*}
 
\begin{table*}
\begin{center}
\begin{tabular}{|c|c|c|c|c|c|c|c|c|c|c|c|c|}
\hline 
$Q^{2}$ (GeV$^{2}$) & $\overline\Gamma_1^{n}$ & Stat. & Syst. & $\overline I_{TT}^{n}$ & Stat. & Syst.& $\overline\gamma_0^{n}$ & Stat. & Syst.\tabularnewline

\hline
0.0204 & -0.0139 & 0.0028 & 0.0013 & -1.3806 & 0.2769 & 0.1409 & -1.8808 & 0.6734 & 0.3974\tabularnewline
0.0244 & -0.0115 & 0.0029 & 0.0013 & -0.9684 & 0.2551 & 0.1313 & -0.9566 & 0.6329 & 0.3657\tabularnewline
0.0290 & -0.0223 & 0.0030 & 0.0015 & -1.6257 & 0.2247 & 0.1346 & -2.2643 & 0.5600 & 0.3633\tabularnewline
0.0347 & -0.0165 & 0.0028 & 0.0017 & -1.0840 & 0.1862 & 0.1233 & -1.6636 & 0.4756 & 0.3399\tabularnewline
0.0414 & -0.0205 & 0.0026 & 0.0016 & -1.1431 & 0.1439 & 0.1115 & -1.6240 & 0.3265 & 0.2967\tabularnewline
0.0494 & -0.0243 & 0.0028 & 0.0018 & -1.2063 & 0.1325 & 0.1148 & -1.9128 & 0.2920 & 0.2908\tabularnewline
0.0590 & -0.0266 & 0.0030 & 0.0021 & -1.1772 & 0.1272 & 0.1249 & -1.9592 & 0.2779 & 0.3057\tabularnewline
\hline
\end{tabular}
\caption{Results for the neutron moments $\overline\Gamma_1^n$, $\overline I_{TT}^n$ and $\overline\gamma_0^n$ in $10^{-4}$fm$^4$, extracted from the proton and the deuteron moments results as shown in Figs.~22-24 and reported in~\cite{CLAS:2021apd,CLAS:2017ozc}, along with their statistical and systematic uncertainties. The moments shown are for the full $x$ range. 
}
\end{center}
\end{table*}

 \begin{table}[!h]
\begin{tabular}{|c|c|c|c|c|c|} \hline
$Q^2$ (GeV$^2$) & $\overline\Gamma_1^{p-n}$ (meas.) & $\overline\Gamma_1^{p-n}$ (full) & Stat. & Syst.  \\  \hline 
0.021& 0.0042  & 0.0052  & $ 0.0029$   &  $ 0.0017$   \\
0.024& 0.0008  & 0.0008  & $ 0.0031$   &  $ 0.0018$   \\ 
0.029& 0.0122  & 0.0126  & $ 0.0031$   &  $ 0.0020$   \\
0.035& 0.0031  & 0.0040  & $ 0.0030$   &  $ 0.0021$   \\
0.042& 0.0046  & 0.0080  & $ 0.0033$   &  $ 0.0030$   \\
0.050& 0.0051  & 0.0095  & $ 0.0035$   &  $ 0.0031$   \\
0.059& 0.0050 & 0.0103  & $ 0.0037$   &  $ 0.0033$   \\
0.071& 0.0065  & 0.0131  & $ 0.0041$   &  $ 0.0037$   \\
0.084& 0.0026  & 0.0107  & $ 0.0044$   &  $ 0.0039$   \\
0.101& 0.0022  & 0.0122  & $ 0.0042$   &  $ 0.0038$   \\
0.120& 0.0156  & 0.0276  & $ 0.0047$   &  $ 0.0045$   \\
0.144& 0.0051  & 0.0194  & $ 0.0052$   &  $ 0.0048$   \\
0.173& 0.0109  & 0.0266  & $ 0.0051$   &  $ 0.0049$   \\
0.205& 0.0055  & 0.0244  & $ 0.0061$   &  $ 0.0056$   \\
0.244& 0.0236  & 0.0464  & $ 0.0063$   &  $ 0.0059$   \\
0.292& 0.0230  & 0.0500  & $ 0.0062$   &  $ 0.0065$   \\
0.348& 0.0226  & 0.0533  & $ 0.0068$   &  $ 0.0068$   \\
0.416& 0.0123  & 0.0477  & $ 0.0073$   &  $ 0.0082$   \\
0.496& 0.0384  &  0.0770  & $ 0.0088$   &  $ 0.0115$   \\
\hline 	
\end{tabular}
\caption{Results on the Bjorken integral  $\overline\Gamma_1^{p}-\overline\Gamma_1^{n}$ as shown in Fig.~25: partial integral evaluated over the measured $x$-ranges, the near-full integral after adding an estimate of the unmeasured low-$x$ contributions down to $x = 0.001$, and the uncertainties. The statistical uncertainty is from the measured data only, while the systematic uncertainty includes that from the low $x$ contribution.
}
\label{table_data}
\end{table}
 \begin{table}[!h]
\begin{tabular}{|c|c|c|c|c|c|c|c|} \hline
$Q^2$ (GeV$^2$)& $\overline\gamma_0^{p-n}$  & Stat. & Syst.  & $\overline\gamma_0^{p+n}$  & Stat. & Syst. \\  \hline 
0.020 &  0.161 & $ 0.705$   &  $ 0.522$  & -3.531  & $ 0.654$   &  $ 0.367$    \\
0.024 & -0.865 & $ 0.659$   &  $ 0.495$  & -2.725  & $ 0.616$   &  $ 0.334$    \\
0.029 &  0.723 & $ 0.583$   &  $ 0.494$  & -3.732  & $ 0.545$   &  $ 0.329$    \\
0.035 &  0.070 & $ 0.496$   &  $ 0.478$  & -3.194  & $ 0.462$   &  $ 0.303$    \\
0.042 &  0.215 & $ 0.358$   &  $ 0.462$  & -2.975  & $ 0.311$   &  $ 0.251$    \\
0.050 &  0.471 & $ 0.323$   &  $ 0.461$  & -3.290  & $ 0.277$   &  $ 0.237$    \\
0.059 &  0.513 & $ 0.308$   &  $ 0.488$  & -3.340  & $ 0.264$   &  $ 0.245$    \\
0.071 &  0.481 & $ 0.283$   &  $ 0.495$  & -3.288  & $ 0.243$   &  $ 0.224$    \\
0.084 &  0.269 & $ 0.255$   &  $ 0.485$  & -2.814  & $ 0.218$   &  $ 0.191$    \\
0.101 &  0.215 & $ 0.231$   &  $ 0.460$  & -2.524  & $ 0.203$   &  $ 0.173$    \\
0.120 &  0.850 & $ 0.212$   &  $ 0.419$  & -2.899  & $ 0.181$   &  $ 0.163$    \\
0.144 &  0.368 & $ 0.187$   &  $ 0.400$  & -2.068  & $ 0.152$   &  $ 0.118$    \\
0.171 &  0.287 & $ 0.148$   &  $ 0.320$  & -1.817  & $ 0.128$   &  $ 0.097$    \\
0.205 &  0.231 & $ 0.138$   &  $ 0.272$  & -1.361  & $ 0.112$   &  $ 0.074$    \\
0.244 &  0.308 & $ 0.122$   &  $ 0.183$  & -1.156  & $ 0.091$   &  $ 0.061$    \\
0.292 &  0.192 & $ 0.080$   &  $ 0.129$  & -0.828  & $ 0.063$   &  $ 0.038$    \\
0.348 &  0.113 & $ 0.067$   &  $ 0.066$  & -0.508  & $ 0.052$   &  $ 0.023$    \\
0.416 & -0.021 & $ 0.052$   &  $ 0.044$  & -0.298  & $ 0.041$   &  $ 0.017$    \\
0.496 &  0.110 & $ 0.045$   &  $ 0.026$  & -0.231  & $ 0.036$   &  $ 0.013$    \\
\hline 	
\end{tabular}
\caption{Results on the isovector and isoscalar combinations $\overline\gamma_0^{p\mp n}$, in $10^{-4}$~fm$^4$, along with their uncertainties, as shown in Fig.~26. 
}
\label{table_data}
\end{table}

\end{document}